\documentclass[a4paper,11pt]{article}
\pdfoutput=1 
\usepackage{jheppub} 
\usepackage[T1]{fontenc} 
\usepackage[dvipsnames]{xcolor}
\usepackage{graphicx}
\usepackage{subcaption} 
\usepackage{adjustbox}  
\usepackage{comment}
\usepackage{multirow}
\usepackage{tabularx}
\newcolumntype{Y}{>{\centering\arraybackslash}X}
\usepackage{hyperref}
\usepackage{doi} 
\newcommand{\be}{\begin{equation}}
\newcommand{\ee}{\end{equation}}
\newcommand{\bea}{\begin{eqnarray}}
\newcommand{\eea}{\end{eqnarray}}
\newcommand{\ba}{\begin{align}}
\newcommand{\ea}{\end{align}}
\newcommand{\lsim}{\raisebox{-0.13cm}{~\shortstack{$<$ \\[-0.07cm]
      $\sim$}}~}
\newcommand{\mg}{{\tt MadGraph5\_aMC@NLO}}
\newcommand{\nn}{\nonumber}

\newcommand{\s}{\hat{s}}
\newcommand{\T}{\hat{t}}
\newcommand{\U}{\hat{u}}

\newcommand{\MG}{ {\tt Madgraph5\_aMC@NLO} }

\title{\boldmath 
\boldmath Collider sensitivity to SMEFT heavy-quark operators at one-loop in top-quark processes}

\author[a]{C. Degrande}
\author[b]{, R. Rosenfeld}
\author[c]{, A. Vasquez}
\affiliation[a]{Centre for Cosmology, Particle Physics and Phenomenology (CP3), Universit\'e Catholique de Louvain,\\ B-1348 Louvain-la-Neuve, Belgium}
\affiliation[b]{ICTP South American Institute for Fundamental Research \& 
Instituto de F\'{\i}sica Te\'orica \\
UNESP - Universidade Estadual Paulista \\
Rua Dr.~Bento T.~Ferraz 271  -  01140-070  S\~ao Paulo, SP, Brazil}
\affiliation[c]{Bethe Center for Theoretical Physics, Universit\"at Bonn, \\ D-53115 Bonn, Germany}
\emailAdd{celine.degrande@uclouvain.be}
\emailAdd{rogerio.rosenfeld@unesp.br}
\emailAdd{avasquez@uni-bonn.de}

\abstract{
We study the effects of four-heavy-quark operators in the production of top quarks in the framework 
of the Standard Model Effective Field Theory (SMEFT) at the LHC. 
In particular, we compute for the first time the total contribution of the 
four-top-quark operator which enters only at the one-loop level in the top-quark pair production process.
Analytical results at one-loop are presented for the gluon- and quark-initiated sub-processes, which allowed a first complete validation of the {\tt SMEFT@NLO} in  \mg. The 95\% CL bounds on four-heavy-quark operators from the available top-quark pair and four-top-quark production data are provided, which are complementary to other bounds found in the literature. We focus on the comparison of the sensitivities of the top-quark pair and the four-top-quark production processes, where in the latter case the four-top-quark operator contributes at tree-level. We conclude that the sensitivities of the two processes to four-heavy-quark operators are comparable.  The projected sensitivities of both processes at HL-LHC are also presented. 
}

\begin{document} 
\begin{flushright}
 BONN-TH-2024-02
/IRMP-CP3-24-04 
\end{flushright}
\maketitle
\flushbottom

\section{Introduction}
\label{sec:intro}

A decade has passed since the discovery of the Higgs boson~\cite{CMSHiggsDisc,ATLASHiggsDisc}, the last fundamental particle predicted by the Standard Model (SM) to be found  and whose properties are being measured with an increasing precision. After many measurements, the SM has been established as the most precise theory to describe collider experiments. Nevertheless, indirect observations such as the matter-antimatter asymmetry and the dark matter content of the universe reveal the shortcomings  of the SM and hint towards a more complete theory. Furthermore, a direct measurement of the triple and quartic Higgs self-interactions is yet to be performed and it is possible that the full story about the Higgs sector is not completely told. 

These are some of the issues that motivate the search for physics beyond the Standard Model. In recent years, new physics searches have been approached in two ways. One method is to extend  the particle content of the SM and enlarge its symmetries, as  it is carried out in extended Higgs sectors and in SUSY models. Another method is to consider the SM as the renormalizable sector of an Effective Field Theory (EFT), in which case one keeps the symmetries and particle content of the SM but modifies the interactions among the known particles via higher dimensional operators, with the so-called Wilson coefficients resulting from integrating-out heavy degrees of freedom that appear at a high-energy scale $\Lambda$. There are several realizations of the latter approach, and the so-called Standard Model Effective Field Theory (SMEFT) \cite{Buchmuller1986,Grzadkowski2010,Brivio2019}, in which the Higgs boson is part of an $SU(2)_L$ doublet in concordance with the SM,  is widely used.  Similarly widespread, the Higgs Effective Field Theory (HEFT) \cite{Falkowski2019,Alonso2016,Cohen2021,Contino2013} assumes that the Higgs boson resonance corresponds to a singlet scalar, so that no assumption about the origin of  spontaneous symmetry breaking (SSB) mechanism is made. Both approaches provide a complete and linearly independent basis of operators that can be used in precision analysis because of their renormalizability order by order in the $1/\Lambda$ expansion.
The HEFT is a more general approach than the SMEFT, in the sense that $\mathcal{L}_{\mathrm{HEFT}} \supset \mathcal{L}_\mathrm{SMEFT}$, but this comes at the cost of inserting more parameters that have to be fitted, which complicates global analyses. Moreover, current measurements point towards a SM-like Higgs boson. In this paper, we work in the SMEFT framework in order to directly compare our findings with previous results in the literature \footnote{In fact, the heavy-four-quark operators are the same in both frameworks.}.

Studies dedicated to the search for indirect traces of new physics in experimental observations are becoming more important since no new resonances at the current colliders have been measured. Following the precision approach for finding deviations, it is clear that the highest accuracy in the experimental measurements and theoretical predictions from our models are required. Experimentally,  the recently started Run III at the Large Hadron Collider (LHC) is expected to measure new observables and significantly improve several of the successful measurements of the Run II. Furthermore, an upgrade in luminosity, the HL-LHC, is planned to become operational in 2029. On the theoretical side, the SMEFT presents the appropriate framework to look for deviations from the SM: it is model independent, gauge invariant and, being a consistent QFT order by order in $1/\Lambda$, it is possible to find accurate predictions at higher orders. Thus, improving the SMEFT predictions enhances our sensitivity to new physics.

Large efforts have concentrated on understanding the structure of the SMEFT as well as on consistently and efficiently constraining the Wilson coefficients. Several of these constraints include one-loop corrections, showing that some processes are sensitive to operators that do not contribute at the Born level. Moreover, Next-to-Leading-Order (NLO) corrections have proven to break degeneracies and flat directions present in the tree-level contributions of the effective operators. Finally, considering the SMEFT at higher loops systematically improves its predictions not only by significantly reducing the scale uncertainties, but also by making more accurate computations of the shapes of differential distributions \cite{Maltoni2016}.

In addition, the treatment of effective operators in one-loop processes requires a high level of consistency. First, the running Wilson coefficients associated with the renormalization scheme should be included to achieve full NLO accuracy. Second, these operators might introduce spurious gauge-anomalous contributions, which have to be treated in a consistent scheme. Third, evanescent operators might change our results depending on the basis used to write down the four-fermion operators in the Lagrangian.  In our analysis, we employ the package {\tt SMEFT@NLO}, which provides a consistent automation of the SMEFT at one-loop. 

When considering the different sectors of the SMEFT, the top quark sector is presented as a good prospect in the search for BSM physics. The largest coupling of the Higgs boson is the Yukawa interaction with the top quark, hence it is expected that the  properties of the SSB can be further studied through the top, one of the least experimentally constrained quarks. The leading top quark production mechanisms at hadron colliders are the top-pair and single-top production processes. In addition, multiple  processes involving the top quark have been measured in the recent years, from which we highlight the four-top process for which,  since the reports by ATLAS and CMS \cite{ATLAStttt2020, CMStttt2020, CMStttt2018}, its relevance in global fits has grown. The interest on this process has led to the more recent measurements found in \cite{ATLAStttt2023,CMStttt2023}. The first measurements of the four-top process lead to constrain  several Wilson coefficients at leading-order (LO) in global analysis using top quark data \cite{Rojo2019,Saavedra2018,Moore2016,Brivio2020}. The operators modifying the couplings of the top quark to gauge bosons and operators modifying the self-interactions of gluons have been constrained in dedicated studies \cite{Martini2020,Bylund2016,Franzosi2015}. Furthermore, four-fermion operators have also been studied in such global analyses, which, after identifying the relevant operators in top production processes at the LHC, can be organized into two groups:

\begin{itemize}
\item Two-light-two-heavy operators (2L2H): color-singlet and color-octet operators formed by one current with quarks from the 3rd generation and one current with light quarks. 
\item Four-heavy-quark operators (4H): color-singlet and color-octet operators with purely 3rd generation quarks. 
\end{itemize}

The first set of operators above has been extensively studied since they can be accessed at tree-level by several processes. In this work, we will be interested in the less constrained four-fermion operators containing four top interactions, \textit{i.e.} the set of four-heavy-quark operators. At tree-level, those operators can be constrained in the four-top production process, as shown in Ref. \cite{Zhang2018}, while in the top-pair production they receive almost no constraints, as they only contribute via the PDF suppressed bottom pair annihilation channel. At the same perturbative order, no insertions are possible in the single-top production. However, one-loop corrections involving four-heavy-quark operators in the top-pair production have not been explored, which turn out to be non-negligible and lead to sizable constraints, as we will show in this paper. The sensitivity of single-top production to the four-heavy-quark operators is expected to be smaller than top pair production as this process has a lower cross-section due to its electroweak origin. Even though top-pair production has the disadvantage of receiving suppressed one-loop or bottom-quark-initiated contributions from the four-heavy-quark operators, the center-of-mass energy distributions peak close to the threshold. On the contrary, the invariant-mass distributions for four top quarks present peaks around 1.3 TeV that fall gradually ranging along some few TeV's (see Fig. \ref{fig:4t_sq_Minv}), and therefore is expected to receive larger higher-order correction in the EFT expansion.

In this paper, we present for the first time a comparison between the sensitivity of top-pair and four-top production to four-fermion operators involving purely the 3rd generation of quarks. These operators have been addressed by the SMEFiT collaboration~\cite{smefit2021}, with bounds coming exclusively from $pp\rightarrow t\bar{t}t\bar{t}$ and $pp\rightarrow t\bar{t}b\bar{b}$ processes. Recently, it has been suggested that the contributions of these operators to electroweak precision observables (EWPO) through loop corrections lead to sensitive effects~\cite{Dawson2022}, raising the question whether a thorough study in the top-pair production could provide more information about their NLO effects.  Additionally,  this analysis is motivated by the fact that both theoretical and experimental uncertainties are better controlled in top-pair production. As a matter of fact, in the Standard Model, the top-pair cross-section is known at NNLO-QCD and  NLO-EW~\cite{Tsinikos2017} while four-top predictions are known at NLO QCD+EW~\cite{Pagani2018}. The relative theoretical uncertainties both at LO and at NLO are larger for four-top than top-pair production, suggesting that even if NNLO four-top predictions became available, they would not be  as accurate as the predictions for top-pair production. At the experimental level, results have been presented with much larger precision for the cross-section of top-pair production and several reports with differential distributions have been published by CMS and ATLAS~\cite{CMSttdiff2018a,CMSttdiff2018b,CMSttdiff2019,ATLASttdiff2019,CMSttInclusive2023,ATLAStt2023}. We show that with the current level of precision, the top-pair and four-top productions have similar sensitivities when only the interference between new physics  and the SM (linear terms in the SMEFT expansion) is taken into account. However, most of the sensitivity in the four-top process comes from the square of the new physics effects (quadratic terms in the SMEFT expansion). Thus, questions arise regarding the validity of the dimension-6 truncation in the four-top process. Since a full description regarding dimension-8 contributions to four-top has not been accomplished, the bounds obtained from the top-pair production appear to be under better theoretical control. Moreover, the current measurements lead to loose and thus questionable bounds in the top-pair production regarding the EFT validity. In this work we focus on the bounds obtained at the interference level and consider the quadratic terms as an indicator for the stability of the EFT expansion. Using this indicator, we find that the top-pair and four-top productions are in the same ballpark in terms of the EFT  validity. Beyond the EFT validity issues associated to the four-heavy-quark operators, the top-pair production probes new physics scales of the same order as those bounded via the four-top production at the linear order in the SMEFT expansion, and constrains different parameter-space directions. Including the top-pair production could break possible existing degeneracies in global analyses. We also notice that the EWPO observables lead to bounds of roughly the same order. Although they do not cover all the five four-heavy-quark operators, the EFT expansion is under better theoretical control due to the lower energy probed.  In addition, we revisit the subtleties that arise in the study of the process $p p \rightarrow t\bar{t}$ at one-loop in the SMEFT. The analytical expressions presented in this paper for the one-loop gluon- and quark-initiated sub-processes are used to validate the implementation of the {\tt SMEFT@NLO} in \mg, which lead to improvements in the code\footnote{The analytical expressions for the one-loop 
gluon-initiated sub-processes were also computed in \cite{Muller2021}.}.

This paper is organized as follows: in section \ref{sec:smeft}  we present a short overview of the effective operators  relevant for the top physics and show the selected operators considered in our study. Subsequently, the theoretical computations of the $pp \rightarrow t\bar{t}$ and $pp \rightarrow t\bar{t}t\bar{t}$ processes in the SMEFT are discussed in section \ref{sec:calculation_Frame}. In  section \ref{sec:pheno} we present our analysis strategy and the 95\% confidence level (CL) bounds on the selected Wilson coefficients from the current data. In section \ref{sec:HLLHC} we report on the projected sensitivity at the HL-LHC to the selected operators. We conclude in section \ref{sec:conclusions}.

\section{Effective operators}
\label{sec:smeft} 

In general, the SMEFT Lagrangian can be written as
\be 
{\cal L}_{\rm SMEFT}={\cal L}_{\rm SM}+\sum_{i,d>4}\frac{c_i^{(d)}(\mu)}{\Lambda^{d-4}}{\cal O}_i^{(d)},
\ee
where the coefficients $c_i$ are the Wilson coefficients of the dimension-$d$ operators and $\Lambda$ the energy scale at which we expect to find direct new physic effects. The gauge invariant ${\cal O}_i$ are the effective operators  built of SM fields. The Lagrangian above can be interpreted as an expansion around the SM theory, with the $c_i$ constituting a basis that parametrizes possible deviations from the SM in the observable ${\rm O}_n$ as                                                               
\be
\Delta {\rm O}_n = {\rm O}_n^{\rm EXP} - {\rm O}_n^{\rm SM} = \sum_i \frac{a_{n,i}^{(6)} (\mu) c_i^{(6)}(\mu)}{\Lambda^2}  + \sum_{ij} \frac{b_{n,ij}^{(6)} (\mu) c_i^{(6)}(\mu)c_j^{(6)}(\mu)}{\Lambda^4} +\sum_i \frac{a_{n,i}^{(8)} (\mu) c_i^{(8)}(\mu)}{\Lambda^4} +\dots,
\ee
with the coefficients $a_i$ and $b_{ij}$ determining the size of the effects of the operators $\mathcal{O}_i$ and which are obtained by computing the contributions of the operators to each observable\footnote{The coefficients $a_i$ and $b_{ij}$ contributing to our results are obtained with \mg. }.  In this approach, if we want trustworthy predictions of the new physics effects, the experiments and the theoretical computations from the SM should be performed with high accuracy. In addition, the parametrization of such deviations must also be accurate and consistent. The SMEFT provides the framework required to parametrize new physics effects in such a way that its predictions can be systematically improved by including loop corrections. Hence, to enhance our sensitivity to new physics, we can improve the predictions from the SMEFT by going at NLO.

We are interested in the top-pair and four-top production processes and so we focus on the operators that are relevant in the top sector. In what follows, we classify those operators and provide an overview of the current status of their constraints. Only dimension-six operators are considered ($d\leq 6$) and the Warsaw basis is used, following the notation in Ref. \cite{Grzadkowski2010}. When considering the top production in the LHC via strong interactions, there are several classes of operators that should be considered. Bounds found in the literature  on several of the operators that we discuss in this section are collected in Table \ref{tab:boundsLit}. In a first class of effective operators, we have the coupling of a top quark current with bosons (2FB). In a second class are the purely bosonic operators (B). These two classes of effective interactions involving gluons, the top and the Higgs, are relevant in gluon-initiated processes,
\begin{align}
    \mathcal{O}_{t\varphi} & =  (\varphi^{\dagger} \varphi)  (\bar{Q} t_R \widetilde{\varphi}), \nonumber \\
    \mathcal{O}_{tG} & = (\Bar{Q}\sigma_{\mu\nu}T^A t) \Tilde{\varphi} G_{\mu}^{A\mu\nu}, \nonumber \\
    {\cal O}_{\varphi G} &= \varphi^{\dagger} \varphi G_{\mu\nu}^{A}G^{A\mu\nu}, \nonumber \\
    \mathcal{O}_G & = f^{ABC} G_{\mu}^{A\nu}G_{\nu}^{B\rho}G_{\rho}^{C\mu}, \nonumber \\
    \mathcal{O}_{\tilde{G}} & = f^{ABC} \tilde{G}_{\mu}^{A\nu}G_{\nu}^{B\rho}G_{\rho}^{C\mu}, \nonumber \\
    {\cal O}_{\varphi \Tilde{G}} &= \varphi^{\dagger} \varphi \tilde{G}_{\mu\nu}^{A}G^{A\mu\nu}. \label{eq:bosonop}
\end{align}
The first three operators in Eq. \eqref{eq:bosonop} receive constraints from the Higgs sector and have been studied at LO in \cite{Maltoni2016} and at NLO in the gluon-fusion Higgs production~\cite{Deutschmann2017}. The chromomagnetic operator ${\cal O}_{tG}$, affects at tree-level  $t\bar{t}$ production but flips the chirality of the top lines, which introduces an overall factor of $m_t v$ in its interference with the SM. As a consequence, this operator is suppressed in the $t\bar{t}$ production, for which the interference goes as $m_t v/\Lambda^2$ at high energies, instead of growing with the center-of-mass energy. Corrections of order QCD-NLO on this operator lead to increments to the $t\bar{t}$ up to 50\% compared to the LO at the LHC~\cite{Franzosi2015}. The operator $\mathcal{O}_{t\varphi}$ enters through loop corrections in the sub-process $gg\rightarrow t\bar{t}$. The operator $\mathcal{O}_G$ has been extensively studied in global fits using $t\bar{t}$ and $t\bar{t}V$ data and dedicated multijet studies~\cite{Ellis2021}. Finally, the interactions in Eq. \eqref{eq:bosonop} involving the dual field strength $\tilde{G}_{\mu\nu}$ do not contribute at the order $\mathcal{O}(1/\Lambda^2)$ when studying unpolarized cross-sections~\cite{Moore2016}, given their CP-violating nature. Hence, even though all those operators are relevant in the $t\bar{t}$ process, their implications are well known and we do not consider them in the analysis below (see Table \ref{tab:boundsLit}). 

As a third class of operators, are the four-fermion interactions involving two light and two heavy quarks (2L2H) 
\begin{align}
    \mathcal{O}_{Qq}^{(8,3)}  = (\Bar{Q}_L \gamma_{\mu} T^A \tau^{i} Q_L ) (\Bar{q}_L \gamma^{\mu} T^A \tau^{i} q_L), & \qquad \mathcal{O}_{Qq}^{(1,3)}  = (\Bar{Q}_L \gamma_{\mu} \tau^{i} Q_L ) (\Bar{q}_L \gamma^{\mu} \tau^{i} q_L), \nonumber \\
    \mathcal{O}_{Qq}^{(8,1)}  = (\Bar{Q}_L \gamma_{\mu} T^A Q_L ) (\Bar{q}_L \gamma^{\mu} T^A  q_L),   \quad\:\:\:  & \qquad \mathcal{O}_{Qq}^{(1,1)}  = (\Bar{Q}_L \gamma_{\mu}  Q_L ) (\Bar{q}_L \gamma^{\mu}   q_L), \nonumber \\
    \mathcal{O}_{td}^{(8)}  = (\Bar{t}_R \gamma_{\mu} T^A  t_R ) (\Bar{d}_R \gamma^{\mu} T^A d_R), \qquad\, & \qquad \:\:\, \mathcal{O}_{td}^{(1)}  = (\Bar{t}_R \gamma_{\mu}  t_R ) (\Bar{d}_R \gamma^{\mu} d_R), \nonumber \\
    \mathcal{O}_{tu}^{(8)}  = (\Bar{t}_R \gamma_{\mu} T^A  t_R ) (\Bar{u}_R \gamma^{\mu} T^A u_R), \qquad & \qquad \:\:\, \mathcal{O}_{tu}^{(1)}  = (\Bar{t}_R \gamma_{\mu} t_R ) (\Bar{u}_R \gamma^{\mu} u_R), \nonumber \\
    \mathcal{O}_{tq}^{(8)}  = (\Bar{t}_R \gamma_{\mu} T^A  t_R ) (\Bar{q}_L \gamma^{\mu} T^A q_L), \qquad\:\, & \qquad  \:\:\, \mathcal{O}_{tq}^{(1)}  = (\Bar{t}_R \gamma_{\mu}  t_R ) (\Bar{q}_L \gamma^{\mu} q_L), \nonumber \\
    \mathcal{O}_{Qd}^{(8)}  = (\Bar{Q}_L \gamma_{\mu} T^A  Q_L ) (\Bar{d}_R \gamma^{\mu} T^A d_R), \quad\:\, & \qquad \:\:\, \mathcal{O}_{Qd}^{(1)}  = (\Bar{Q}_L \gamma_{\mu}  Q_L ) (\Bar{d}_R \gamma^{\mu}  d_R), \nonumber \\
    \mathcal{O}_{Qu}^{(8)}  = (\Bar{Q}_L \gamma_{\mu} T^A  Q_L ) (\Bar{u}_R \gamma^{\mu} T^A u_R), \quad\:\,  & \qquad  \:\:\, \mathcal{O}_{Qu}^{(1)}  = (\Bar{Q}_L \gamma_{\mu} Q_L ) (\Bar{u}_R \gamma^{\mu} u_R). \label{eq:2light2heavy}
\end{align}
The operators to the left of Eq. \eqref{eq:2light2heavy} are composed by color-octet heavy quark currents, while the ones on the right are composed by color-singlet currents. Hence, the upper indices shown in parentheses in the names given to the operators in Eq. \eqref{eq:2light2heavy} indicate the type of currents composing the effective operator, explicitly $\blacklozenge^{(8)}$ stands for color-octet, $\blacklozenge^{(3)}$ stands for $SU(2)_L$ triplet and $\blacklozenge^{(1)}$ stands for color-singlet operators. Because of the color structure, when we consider the $t\bar{t}$ process at tree-level, color-octet operators generate diagrams that interfere with the QCD-SM amplitude, while the color-singlet ones only interfere with the EW-SM amplitude. Therefore, the contributions of the order ${\cal O} \left(\Lambda^{-2}\right)$ for the color singlets, obtained only from the interference with the EW-SM amplitude, are suppressed by the electroweak coupling constant. This class of operators can also be constrained via the single-top, top-pair in association with jets and four-top processes\cite{smefit2021}. These operators are well studied and since they can easily be constrained at tree-level we do not consider them in the following sections.

As a fourth class of operators, we have the four-heavy-quark operators (4H) defined as follows
\begin{align}
\mathcal{O}^{(1)}_{Qt} = \left( \bar{Q}_L \gamma_{\mu} Q_L \right)\left( \bar{t}_R \gamma^{\mu} t_R \right), \qquad  & \quad
\mathcal{O}^{(8)}_{Qt} = \left( \bar{Q}_L \gamma_{\mu} T^A Q_L \right)\left( \bar{t}_R \gamma^{\mu} T^A t_R \right), \nonumber \\
\mathcal{O}^{(1)}_{QQ} = \frac{1}{2} \left( \bar{Q}_L \gamma_{\mu} Q_L \right) \left( \bar{Q}_L \gamma^{\mu} Q_L \right), \: & \quad
\mathcal{O}^{(8)}_{QQ} = \frac{1}{2}\left( \bar{Q}_L \gamma_{\mu} T^A Q_L \right)\left( \bar{Q}_L \gamma^{\mu} T^A Q_L \right),\,\,\, \nonumber \\
\mathcal{O}^{(1)}_{tt} = \left( \bar{t}_R \gamma_{\mu} t_R \right)\left( \bar{t}_R \gamma^{\mu} t_R \right). \quad \quad \:\:\:\, & \label{eq:4heavy}
\end{align}
The color-octet operator involving only right-handed top quarks is equivalent to the operator $\mathcal{O}^{(1)}_{tt}$ after using Fierz identities, and so we do not listed it in Eq. \eqref{eq:4heavy}. Since they are all hermitian for a single generation, the five operators in Eq. \eqref{eq:4heavy} constitute a maximal set of possible operators that can be written as consisting of the third generation of quarks. The Wilson coefficients corresponding to these four-heavy-quark operators must be non-zero if the NP couples to the top quark, hence their importance. Four-heavy-quark operators appear in several BSM scenarios, among these we find two-Higgs-doublet models \cite{Craig2012} and composite models of the top quark \cite{Pomarol2008,Banelli2021}. In composite models, the four-top effective operators have coefficients larger than those corresponding to other operators. Top-philic scenarios where vector or scalar resonances mainly couple to the top quarks, but interact weakly with the rest of the SM fermions, are easily relatable to these effective operators \cite{Darme2021,Greiner2015,Kim2016}. The four-heavy-quark operators are constrained via the four-top production at tree-level \cite{Zhang2018,ElFaham2022} and the top-pair production at one-loop, as will be presented. Additionally, subsets of those operators receive constraints from the top-pair production in association with a bottom pair, the top-pair production in association with a Higgs boson, Higgs production and EWPO~\cite{smefit2021, Dawson2022,Alasfar2022}. Other four-heavy-quark operators are
\begin{align}
    \mathcal{O}^{(1)}_{QtQb} = \epsilon_{jk} (\bar{Q}^j t)  (\bar{Q}^k b),& \quad\quad \mathcal{O}^{(8)}_{QtQb}=\epsilon_{jk} (\bar{Q}^j T^A t)  (\bar{Q}^k T^A b),
\end{align}
which can contribute to several top production mechanisms, but their interference with SM amplitudes is suppressed by factors of the bottom mass. These factors arise from the flip in chirality of the bottom quark. Hence, we do not consider them in our analysis.

For completeness, we present the anomalous electroweak couplings that have been studied at NLO in the $t\bar{t}$ production. They impose constraints on effective operators that modify SM-like vertices \cite{Martini2020}. The operators to consider in this case are of the type 2FB, and in the Warsaw basis, they are given by
\begin{align}
    \mathcal{O}_{\varphi t} & = (\varphi^{\dagger} i \overset{\leftrightarrow}{D}_{\mu} \varphi ) (\bar{t}\gamma^{\mu}t), \nn \\
    \mathcal{O}_{\varphi Q}^{(1)} & = (\varphi^{\dagger} i \overset{\leftrightarrow}{D}_{\mu} \varphi ) (\bar{Q}\gamma^{\mu}Q), \nn \\
    \mathcal{O}_{\varphi Q}^{(3)} & = (\varphi^{\dagger} i \overset{\leftrightarrow}{D}_{\mu} \tau^I \varphi ) (\bar{Q}\gamma^{\mu} \tau^I Q). \label{eq:5.1}
\end{align}

{ 
\setlength{\tabcolsep}{3pt}
\renewcommand{\arraystretch}{1.2}
\begin{table}
\begin{center}
{\fontsize{7.5}{9.2}\selectfont 
\begin{tabular}{|c|c|c|c|c|c|c|c|}
\hline
    \multirow{2}{*}{Class} &\multirow{2}{*}{$c_i$} & \multirow{2}{*}{Ref.} & \multicolumn{2}{c}{ Individual } & \multicolumn{2}{c|}{ Marginalized }   \tabularnewline

    & & & \multicolumn{1}{c}{ $\mathcal{O}(\Lambda^{-2})$ } & \multicolumn{1}{c}{ $\mathcal{O}(\Lambda^{-4})$ } & \multicolumn{1}{c}{ $\mathcal{O}(\Lambda^{-2})$} & \multicolumn{1}{c|}{ $\mathcal{O}(\Lambda^{-4})$ } \tabularnewline
\hline 
\hline 
    \multirow{13}{*}{4H} & \multirow{3}{*}{$c_{QQ}^{1}$} & \cite{smefit2021} & \multicolumn{1}{c}{\multirow{1}{*}{ $ [-6.13,23.3] $ }}& \multicolumn{1}{c}{\multirow{1}{*}{ $ [-2.23,2.02]  $ }}   & \multicolumn{1}{c}{ $[-190,189]  $ } & \multicolumn{1}{c|}{ $[-2.99,3.71]  $ }   \tabularnewline
    
     & & \cite{Dawson2022} & \multicolumn{1}{c}{\multirow{1}{*}{ $[- 1.61, 2.68]$ }}  & \multicolumn{1}{c}{\multirow{1}{*}{ - }} & \multicolumn{1}{c}{\multirow{1}{*}{ - }} & \multicolumn{1}{c|}{\multirow{1}{*}{ - }} \tabularnewline
     
    & & \cite{ElFaham2022} & \multicolumn{1}{c}{\multirow{1}{*}{ - }}  & \multicolumn{1}{c}{\multirow{1}{*}{ $[-2.2,3]$ }} & \multicolumn{1}{c}{\multirow{1}{*}{ - }} & \multicolumn{1}{c|}{\multirow{1}{*}{ - }} \tabularnewline
    \cline{2-7}
   
    &  \multirow{3}{*}{$c_{QQ}^{8}$} & \cite{smefit2021} & \multicolumn{1}{c}{\multirow{1}{*}{ $ [ -26.5,57.8] $ }}& \multicolumn{1}{c}{\multirow{1}{*}{ $ [-6.81,5.83] $ }}   & \multicolumn{1}{c}{ $[-190,170]  $ } & \multicolumn{1}{c|}{ $[-11.2,8.17]  $ }  \tabularnewline
    
     & & \cite{Dawson2022} & \multicolumn{1}{c}{\multirow{1}{*}{ $[- 15.23, 25.41]$ }}  & \multicolumn{1}{c}{\multirow{1}{*}{ - }} & \multicolumn{1}{c}{\multirow{1}{*}{ - }} & \multicolumn{1}{c|}{\multirow{1}{*}{ - }} \tabularnewline
     
     & & \cite{ElFaham2022} & \multicolumn{1}{c}{\multirow{1}{*}{ - }}  & \multicolumn{1}{c}{\multirow{1}{*}{ $[-6.75,9]$ }} & \multicolumn{1}{c}{\multirow{1}{*}{ - }} & \multicolumn{1}{c|}{\multirow{1}{*}{ - }} \tabularnewline
    \cline{2-7}
    
    &  \multirow{4}{*}{$c_{Qt}^{1}$} & \cite{smefit2021} & \multicolumn{1}{c}{\multirow{1}{*}{ $ [-195,159] $ }} & \multicolumn{1}{c}{\multirow{1}{*}{ $ [-1.83,1.86] $ }}   & \multicolumn{1}{c}{ $[-190,189]  $ }  & \multicolumn{1}{c|}{ $[-1.39,1.25]  $ } \tabularnewline
    
    &  & \cite{Dawson2022} & \multicolumn{1}{c}{\multirow{1}{*}{ $[- 2.24, 1.35]$ }}  & \multicolumn{1}{c}{\multirow{1}{*}{ - }} & \multicolumn{1}{c}{\multirow{1}{*}{ - }} & \multicolumn{1}{c|}{\multirow{1}{*}{ - }} \tabularnewline
    
    & & \cite{ElFaham2022} & \multicolumn{1}{c}{\multirow{1}{*}{ - }}  & \multicolumn{1}{c}{\multirow{1}{*}{ $[-2.6,2]$ }} & \multicolumn{1}{c}{\multirow{1}{*}{ - }} & \multicolumn{1}{c|}{\multirow{1}{*}{ - }} \tabularnewline

    & & \cite{Alasfar2022} & \multicolumn{1}{c}{\multirow{1}{*}{ $[-1.1,1.2]$ }}  & \multicolumn{1}{c}{\multirow{1}{*}{ $[-1,0.6]$ }} & \multicolumn{1}{c}{\multirow{1}{*}{ - }} & \multicolumn{1}{c|}{\multirow{1}{*}{ - }} \tabularnewline
    \cline{2-7}
    
    &  \multirow{3}{*}{$c_{Qt}^{8}$} & \cite{smefit2021}  & \multicolumn{1}{c}{\multirow{1}{*}{ $ [-5.72,20.1] $ }} & \multicolumn{1}{c}{\multirow{1}{*}{ $ [-4.21,3.35]  $ }}    & \multicolumn{1}{c}{ $[-190,162]  $ } & \multicolumn{1}{c|}{ $[-3.04,2.20]  $ }  \tabularnewline
    
    & & \cite{ElFaham2022} & \multicolumn{1}{c}{\multirow{1}{*}{ - }}  & \multicolumn{1}{c}{\multirow{1}{*}{ $[-4.2,5.3]$ }} & \multicolumn{1}{c}{\multirow{1}{*}{ - }} & \multicolumn{1}{c|}{\multirow{1}{*}{ - }} \tabularnewline

    & & \cite{Alasfar2022} & \multicolumn{1}{c}{\multirow{1}{*}{ $[-4.6,4.9]$ }}  & \multicolumn{1}{c}{\multirow{1}{*}{ $[-3.2,4.7]$ }} & \multicolumn{1}{c}{\multirow{1}{*}{ - }} & \multicolumn{1}{c|}{\multirow{1}{*}{ - }} \tabularnewline
    \cline{2-7}
    
     & \multirow{2}{*}{$c_{tt}^{1}$} & \cite{smefit2021}  & \multicolumn{1}{c}{\multirow{1}{*}{ $ [-2.78,12.1] $ }} & \multicolumn{1}{c}{\multirow{1}{*}{ $ [-1.15,1.02] $ }}   & \multicolumn{1}{c}{ $[-115,153]  $ } & \multicolumn{1}{c|}{ $[-0.79,0.71]  $ } \tabularnewline
     
     & & \cite{ElFaham2022} & \multicolumn{1}{c}{\multirow{1}{*}{ - }}  & \multicolumn{1}{c}{\multirow{1}{*}{ $[-1.2,1.4]$ }} & \multicolumn{1}{c}{\multirow{1}{*}{ - }} & \multicolumn{1}{c|}{\multirow{1}{*}{ - }} \tabularnewline
\hline 
\hline    
    \multirow{14}{*}{2L2H} & $c_{Qq}^{8,1}$ & \cite{smefit2021} & \multicolumn{1}{c}{\multirow{1}{*}{ $ [ -0.273,0.509 ] $ }} & \multicolumn{1}{c}{\multirow{1}{*}{ $ [-0.373,0.309 ] $ }}    & \multicolumn{1}{c}{ $[ -2.26,4.82 ]  $ } & \multicolumn{1}{c|}{ $[-0.555,0.236 ]  $ }  \tabularnewline
    
    & $c_{Qq}^{1,1}$ &\cite{smefit2021} &  \multicolumn{1}{c}{\multirow{1}{*}{ $ [-3.60,0.307 ] $ }} & \multicolumn{1}{c}{\multirow{1}{*}{ $ [-0.303,0.225 ] $ }}    & \multicolumn{1}{c}{ $[ -8.05,9.40]  $ } & \multicolumn{1}{c|}{ $[-0.354,0.249 ]  $ }  \tabularnewline
    
    & $c_{Qq}^{8,3}$ &\cite{smefit2021} & \multicolumn{1}{c}{\multirow{1}{*}{ $ [-1.81,0.625 ] $ }} & \multicolumn{1}{c}{\multirow{1}{*}{ $ [-0.470,0.439 ] $ }}    & \multicolumn{1}{c}{ $[-3.01,7.36 ]  $ } & \multicolumn{1}{c|}{ $[-0.462,0.497 ]  $ }  \tabularnewline
    
    & $c_{Qq}^{1,3}$ & \cite{smefit2021} & \multicolumn{1}{c}{\multirow{1}{*}{ $ [-0.099,0.155 ] $ }} & \multicolumn{1}{c}{\multirow{1}{*}{ $ [-0.088,0.166] $ }}    & \multicolumn{1}{c}{ $[-0.163,0.296 ]  $ } & \multicolumn{1}{c|}{ $[-0.167,0.197 ]  $ }  \tabularnewline
    
    & $c_{tq}^{8}$ & \cite{smefit2021} & \multicolumn{1}{c}{\multirow{1}{*}{ $ [-0.396,0.612 ] $ }} & \multicolumn{1}{c}{\multirow{1}{*}{ $ [-0.483,0.393 ] $ }}    & \multicolumn{1}{c}{ $[ -4.03,4.39 ]  $ } & \multicolumn{1}{c|}{ $[-0.687,0.186 ]  $ }  \tabularnewline
    
    & $c_{tq}^{1}$ & \cite{smefit2021} & \multicolumn{1}{c}{\multirow{1}{*}{ $ [-0.784,2.77 ] $ }} & \multicolumn{1}{c}{\multirow{1}{*}{ $ [-0.205,0.271 ] $ }}    & \multicolumn{1}{c}{ $[-12.4,6.63 ]  $ } & \multicolumn{1}{c|}{ $[-0.222,0.226 ]  $ }  \tabularnewline
    
    & $c_{tu}^{8}$ & \cite{smefit2021} & \multicolumn{1}{c}{\multirow{1}{*}{ $ [-0.774,0.607 ] $ }} & \multicolumn{1}{c}{\multirow{1}{*}{ $ [ -0.911,0.347 ] $ }}    & \multicolumn{1}{c}{ $[-16.9,0.368 ]  $ } & \multicolumn{1}{c|}{ $[-1.12,0.260 ]  $ }  \tabularnewline
    
    & $c_{tu}^{1}$ & \cite{smefit2021} & \multicolumn{1}{c}{\multirow{1}{*}{ $ [ -6.05,0.424] $ }} & \multicolumn{1}{c}{\multirow{1}{*}{ $ [-0.380,0.293 ] $ }}    & \multicolumn{1}{c}{ $[-15.6,15.4  ]  $ } & \multicolumn{1}{c|}{ $[-0.383,0.331 ]  $ }  \tabularnewline  
    
    & $c_{Qu}^{8}$ & \cite{smefit2021} & \multicolumn{1}{c}{\multirow{1}{*}{ $ [-1.50,1.02] $ }} & \multicolumn{1}{c}{\multirow{1}{*}{ $ [ -1.007,0.521 ] $ }}    & \multicolumn{1}{c}{ $[ -12.7,13.8 ]  $ } & \multicolumn{1}{c|}{ $[-1.00,0.312 ]  $ }  \tabularnewline
    
    & $c_{Qu}^{1}$ & \cite{smefit2021} & \multicolumn{1}{c}{\multirow{1}{*}{ $ [-0.938,2.46 ] $ }} & \multicolumn{1}{c}{\multirow{1}{*}{ $ [ -0.281,0.371 ] $ }}    & \multicolumn{1}{c}{ $[-17.0,1.07 ]  $ } & \multicolumn{1}{c|}{ $[-0.207,0.339 ]  $ }  \tabularnewline
    
    & $c_{td}^{8}$ & \cite{smefit2021} & \multicolumn{1}{c}{\multirow{1}{*}{ $ [-1.46,1.36 ] $ }} & \multicolumn{1}{c}{\multirow{1}{*}{ $ [-1.31,0.638 ] $ }}    & \multicolumn{1}{c}{ $[-5.49,25.4 ]  $ } & \multicolumn{1}{c|}{ $[-1.33,0.643 ]  $ }  \tabularnewline
    
    & $c_{td}^{1}$ & \cite{smefit2021} & \multicolumn{1}{c}{\multirow{1}{*}{ $ [-9.50,-0.086 ] $ }} & \multicolumn{1}{c}{\multirow{1}{*}{ $ [-0.449,0.371 ] $ }}    & \multicolumn{1}{c}{ $[-27.7,11.4 ]  $ } & \multicolumn{1}{c|}{ $[-0.474,0.347 ]  $ }  \tabularnewline

    & $c_{Qd}^{8}$ & \cite{smefit2021} & \multicolumn{1}{c}{\multirow{1}{*}{ $ [ -2.39,2.04] $ }} & \multicolumn{1}{c}{\multirow{1}{*}{ $ [-1.61,0.888 ] $ }}    & \multicolumn{1}{c}{ $[-24.5,11.2 ]  $ } & \multicolumn{1}{c|}{ $[-1.26,0.715]  $ }  \tabularnewline
    
    & $c_{Qd}^{1}$ & \cite{smefit2021} & \multicolumn{1}{c}{\multirow{1}{*}{ $ [-0.889,6.46 ] $ }} & \multicolumn{1}{c}{\multirow{1}{*}{ $ [ -0.332,0.436 ] $ }}    & \multicolumn{1}{c}{ $[-3.24,34.6 ]  $ } & \multicolumn{1}{c|}{ $[-0.370,0.384 ]  $ }  \tabularnewline
\hline 
\hline 
    \multirow{9}{*}{2FB} & $c_{t\varphi} $  & \cite{smefit2021} & \multicolumn{1}{c}{\multirow{1}{*}{ $ [-1.33,0.355 ] $ }}  & \multicolumn{1}{c}{\multirow{1}{*}{ $ [-1.29,0.348  ] $ }}  & \multicolumn{1}{c}{ $[-5.74,3.43 ]  $ } & \multicolumn{1}{c|}{ $[-2.32,2.80 ]  $ }   \tabularnewline
    \cline{2-7}
    
    & \multirow{3}{*}{$c_{tG}$} & \cite{smefit2021} & \multicolumn{1}{c}{\multirow{1}{*}{ $ [0.007,0.111 ] $ }}  & \multicolumn{1}{c}{\multirow{1}{*}{ $ [0.006,0.107] $ }}  & \multicolumn{1}{c}{ $[ -0.127,0.403]  $ } & \multicolumn{1}{c|}{ $[ 0.062,0.243]  $ }  \tabularnewline
    
    &  & \cite{Franzosi2015} & \multicolumn{1}{c}{\multirow{1}{*}{ $[-0.42, 0.30]$}}  & \multicolumn{1}{c}{\multirow{1}{*}{ - }} & \multicolumn{1}{c}{\multirow{1}{*}{ - }} & \multicolumn{1}{c|}{\multirow{1}{*}{ - }} \tabularnewline
    
    &  & \cite{Moore2016} & \multicolumn{1}{c}{\multirow{1}{*}{ - }}  & \multicolumn{1}{c}{\multirow{1}{*}{ $[-0.300, 0.650]$ }} & \multicolumn{1}{c}{\multirow{1}{*}{ - }} & \multicolumn{1}{c|}{\multirow{1}{*}{ $[-1.32, 1.22]$}} \tabularnewline
    \cline{2-7}
    
    & \multirow{2}{*}{ $c_{tW}$ }& \cite{smefit2021} & \multicolumn{1}{c}{\multirow{1}{*}{ $ [-0.093,0.026] $ }}  & \multicolumn{1}{c}{\multirow{1}{*}{ $ [-0.084,0.029] $ }}  & \multicolumn{1}{c}{ $[ -0.313,0.123 ]  $ } & \multicolumn{1}{c|}{ $[ -0.241,0.086]  $ }  \tabularnewline
    
    & & \cite{Moore2016} & \multicolumn{1}{c}{\multirow{1}{*}{ - }}  & \multicolumn{1}{c}{\multirow{1}{*}{ $[ 1.32, 1.82]$ }}  & \multicolumn{1}{c}{ -} & \multicolumn{1}{c|}{ $ [-4.03, 3.43] $ }  \tabularnewline
    \cline{2-7}
    
    & $c_{tZ}$ & \cite{smefit2021} & \multicolumn{1}{c}{\multirow{1}{*}{ $ [-0.039,0.099 ] $ }}  & \multicolumn{1}{c}{\multirow{1}{*}{ $ [-0.044,0.094 ] $ }}  & \multicolumn{1}{c}{ $[ -15.9,5.64 ]  $ } & \multicolumn{1}{c|}{ $[ -1.13,0.856 ]  $ }  \tabularnewline
    
    & $c_{\varphi Q}^{-}$ &\cite{smefit2021} & \multicolumn{1}{c}{\multirow{1}{*}{ $ [-0.998,1.44 ] $ }}  & \multicolumn{1}{c}{\multirow{1}{*}{ $ [-1.15,1.58] $ }}  & \multicolumn{1}{c}{ $[ -1.69,11.6 ]  $ } & \multicolumn{1}{c|}{ $[-2.25,2.85]  $ }  \tabularnewline
    
    & $c_{\varphi t}$ & \cite{smefit2021}  & \multicolumn{1}{c}{\multirow{1}{*}{ $ [-2.09,2.46 ] $ }}  & \multicolumn{1}{c}{\multirow{1}{*}{ $ [-3.03,2.19
    ] $ }}  & \multicolumn{1}{c}{ $[ -3.27,18.3 ]  $ } & \multicolumn{1}{c|}{ $[-13.3,3.95 ]  $ }  \tabularnewline

\hline 
\hline     
    \multirow{3}{*}{B} & \multirow{2}{*}{ $c_{G} $ } & \cite{Rojo2019,Hirschi2018} & \multicolumn{1}{c}{\multirow{1}{*}{-}}  & \multicolumn{1}{c}{\multirow{1}{*}{ $ [-0.04, 0.04 ] $ }} & \multicolumn{1}{c}{-} & \multicolumn{1}{c|}{ -} \tabularnewline
    
     &  & \cite{Moore2016} & \multicolumn{1}{c}{\multirow{1}{*}{ - }}  & \multicolumn{1}{c}{\multirow{1}{*}{ $[-0.300, 0.450]$ }} & \multicolumn{1}{c}{\multirow{1}{*}{ - }} & \multicolumn{1}{c|}{\multirow{1}{*}{ $[-1.62, 1.42]$}} \tabularnewline
    \cline{2-7}
    
    & $c_{\varphi G} $  & \cite{smefit2021} & \multicolumn{1}{c}{\multirow{1}{*}{ $ [-0.002,0.005 ]  $ }}  & \multicolumn{1}{c}{\multirow{1}{*}{ $ [-0.002,0.005 ] $ }} & \multicolumn{1}{c}{ $[-0.043,0.012 ]  $ } & \multicolumn{1}{c|}{ $[ -0.019,0.003 ]  $ }  \tabularnewline
\hline
\end{tabular}
}
\caption{ Compendium of 95\% confidence level bounds  ($\Lambda$ = 1 TeV) from the literature for CP-even Wilson coefficients relevant in the top sector. Individual bounds correspond to results obtained from allowing only one coefficient to vary and marginalized indicates bounds from allowing a subset of coefficients to be non-zero at the same time (See text).  } 
\label{tab:boundsLit}
\end{center}
\end{table}
}

The Wilson coefficients corresponding to the last two operators normally receive bounds in the combination $ c_{\varphi Q}^- = c_{\varphi Q}^{1} - c_{\varphi Q}^{3} $. Bounds on the operators in Eq. \eqref{eq:5.1} from measurements of the top and W boson masses have been recently reported \cite{deBlas2022}. Other operators modifying the electroweak interactions of the top are
\begin{align}
    \mathcal{O}_{tW} = i(\bar{Q} \sigma^{\mu\nu} \tau^I t) \tilde{\varphi} W_{\mu\nu}^I,  & \quad\quad \mathcal{O}_{tB}= i(\bar{Q} \sigma^{\mu\nu} t) \tilde{\varphi} B_{\mu\nu},
\end{align}
where the latter is often constrained through the combination $c_{tZ}=-\sin \theta_W \, c_{tB} + \cos \theta_W  \, c_{tW}$. The neutral couplings of the top at NLO have been constrained in \cite{Bylund2016}. More recently, it has been shown that neural networks can improve the sensitivity to NP from operators that modify the electroweak couplings of the top \cite{Atkinson2021}.

Table \ref{tab:boundsLit} presents a summary of the current state of the bounds on the operators presented in this section. The results from the TopFitter collaboration \cite{Moore2016} marginalize over a subset of 12 operators entering at tree-level in top-pair, single-top and associated-top processes, as well as top decays. Results from the SMEFiT collaboration \cite{smefit2021} marginalize over 36 independent and 14 dependent directions in the space of operators entering Higgs, diboson and top quark processes. A global fit of the top sector in the SMEFT should include all the operators aforementioned (see  \cite{Englert2015} for early attempts to achieve this).  The constraints on four-heavy-quark operators reported by the SMEFiT collaboration were obtained only from four-top production and top-pair production in association with a bottom-pair,  for which the tree-level contributions are dominant. Let us notice that the top-pair production in association with a bottom-pair constrains four of the five operators in Eq. \eqref{eq:4heavy} at tree-level. The four-heavy-quark operators induce sizable  new physics effects to such process (around 0.5\% of the corresponding SM cross-section with $c_i/\Lambda^2 = 1 $ TeV$^{-2}$, which is of the same relative order as those operator effects in the top-pair production). However, the three measurements reported to date by the CMS and ATLAS collaborations do not lead to  strong bounds. Because of this, we do not consider the top-pair production in association with a bottom-pair. Finally,  simultaneous determination of parton distribution functions (PDF) and BSM effects has been explored in the literature~\cite{Carrazza2019,Gao2023}, showing cases where the bounds on the Wilson coefficients are modified by considering that new physics can also take place inside the proton. In Ref.~\cite{Costantini2024simunet} a simultaneous global fit of the PDFs and SMEFT coefficients is presented, thus setting bounds on several of the coefficients presented in this section. However, the impact of combining PDFs with the SMEFT on the bounds of four-heavy-quark operators has not been considered. Since we are interested in a first estimate of the sensitivity of the top-pair production to these operators, we focus only on the hard scattering and assume that the PDFs of the SM provide a good enough approximation for this purpose.

In this paper, we are interested in the sensitivity of the top-pair production to the four-heavy-quark operators in Eq. \eqref{eq:4heavy},  which induce one-loop contributions that have not been taken into account previously in any of the literature cited above. We compare such sensitivity to that of the tree-level contributions of these higher-dimensional operators to four-top production.

\section{Calculation framework for the top-pair and four-top processes }
\label{sec:calculation_Frame}

The simulations of the $pp$-collisions are obtained with \mg\footnote{In particular, the \mg~ version 3.4.1 is known to correctly handle the rational terms. Previous versions suffered from a bug in the indexing of the lists of rational terms. }~\cite{Alwall2014} at $\sqrt{s} = 13$ TeV for runs I and II and $\sqrt{s} = 14$ TeV for HL-LHC.  More specifically, the package {\tt SMEFT@NLO} \footnote{This corresponds to one-loop diagrams for the selected dimension-6 operators. The only tree-level contribution case is for the bottom channel, making it the only case where NLO-QCD is included.} \cite{Degrande2021smeftNLO} is used to obtain the effects of the four-heavy-quark operators discussed in the previous section. This package is an automation of one-loop QCD computations in the SMEFT and, not only involves all the CP-even operators discussed above, but also relevant operators for the Higgs and gauge boson sectors. Generated by using the {\tt NLOCT} package~\cite{Degrande2015NLOCT}, the UV and rational counter-terms are included. A correct definition of schemes for the handling of gauge anomalies and evanescent operators is required to provide consistent rational terms. Thus, throughout this paper, we use the same evanescent operators  definitions of the {\tt SMEFT@NLO} package. It is noteworthy that the {\tt SMEFT@NLO} works with a slightly different basis from the one defined in Eq. \eqref{eq:4heavy}, and hence, the effects of evanescent operators must be taken into account when we compare  at the amplitude level the analytical results reported below and the computations obtained via \mg. 

Our results obtained from this model at NLO are renormalized in a fixed-scale renormalization scheme, which introduces a new 
scale $\mu_{\mathrm{EFT}}$ in the counterterms  of the Wilson coefficients.
Therefore, the full NLO accuracy can be reached without the implementation of the running of the Wilson coefficients in \mg \footnote{The implementation of RGE effects in Monte Carlo generators had not been possible until the recent developments in \mg~\cite{Aoude2023}.}. This scheme is similar to the on-shell renormalization of the top quark and, therefore, has similar properties. Namely, large logarithms only appear when the scales probed in the processes are far from $\mu_{\mathrm{EFT}}$ which is not the case for the processes we are considering here. In practice, the scheme is achieved by putting not only the pole but also a logarithm in the UV-counterterms. To go from the $\overline{\mathrm{MS}}$ to our fixed scale scheme, the pole of the EFT operators related to the renormalization of their coefficients is replaced by 
\begin{equation}
    \frac{1}{\bar{\epsilon}}\to\frac{1}{\bar{\epsilon}}-\log\frac{\mu_{\mathrm{EFT}}^2}{\mu_r^2},
\end{equation}
in the UV counterterms. As a result, the $\overline{\mathrm{MS}}$ predictions are recovered when $\mu_{\mathrm{EFT}}=\mu_r$, but the errors are not necessarily the same as they are obtained by varying the renormalization scale $\mu_r$ and not $\mu_{\mathrm{EFT}}$, since this would correspond to changing the renormalization scheme. Hence, we set $\mu_{\mathrm{EFT}}=m_t$ whenever we present results from simulations. 

Unless specified, renormalization and factorization  scales are set to the half of the sum of the masses in the final state. Scale uncertainties are obtained by varying the renormalization scale by a factor of two above and below the central value. The NLO sets of NNPDF3.0~\cite{NNPDF2015} for the PDFs are used with $\alpha_s(M_Z)=0.118$  (tagged as {\tt NNPDF30\_nlo\_as\_0118}). We consider five massless flavours in the proton, including the bottom quark. In addition, the masses of the heavy SM particles are set to
\begin{align}
    m_t=172.5 \; \mathrm{GeV},\quad m_h=125\; \mathrm{GeV},\quad m_Z=91.1876 \; \mathrm{GeV},\quad 
    m_W = 80.41 \; \mathrm{GeV},  
\end{align}
while all other masses are set to zero. Finally, we set the Fermi constant to  $G_F= 1.16637\cdot 10^{-5}$ GeV$^{-2}$.

\subsection{Top-pair production}
\label{sec:ttproduction}

The leading top quark production mechanism at hadron colliders is the $t\bar{t}$ process. In this subsection we study the impact of the heavy-quark four-fermion operators on this process, first through analytical partonic results, then followed by a detailed simulation at the LHC.

We start by presenting  the analytical results of the differential cross-sections of the $t\bar{t}$ process at the parton level for the interference between SM and SMEFT contributions arising 
from the four-heavy-quark operators in Eq. \eqref{eq:4heavy}. 
These analytical results serve as checks for the implementation of the four-fermion operators in the Monte-Carlo-generated predictions used in our analysis of section \ref{sec:pheno}.
They have also been checked using {\tt FeynArts}\cite{feynarts} - {\tt FormCalc}\footnote{We use the FormCalc v8.4, version known to treat correctly four-fermion interactions.}\cite{formcalc} supplemented with  {\tt LoopTools}\cite{formcalc}
with the naive-dimensional regularization scheme (NDR) treatment of the $\gamma_5$ matrix in $d$-dimensions.

At tree-level in the SM, the partonic production of a top-pair has contributions from quark-antiquark and gluon-gluon initial states given by the differential cross-sections
\begin{align}
    \frac{d\hat{\sigma}_{qq\rightarrow t\bar{t}}}{d\Omega} = & \, \frac{\alpha_s^2 \beta_t}{9\s^3} \left(2m_t^{2}\s+\left(\T-m_t^{2}\right)^{2}+\left(\U-m_t^{2}\right)^{2}\right), \label{eq:treeSM} \\
    \frac{d\hat{\sigma}_{gg\rightarrow t\bar{t}}}{d\Omega} = & \,  \frac{1}{64}  \frac{\alpha_s^2\beta_t}{6\s^3}  \frac{7\s^2 + 9 (\T-\U)^2}{\left( \T- m_t^2 \right)^2\left( \U- m_t^2 \right)^2 } \nonumber \\
    & \left( m_t^4 (3 \T^2 +14 \T \U +3 \U^2) - m_t^2 (\T+\U) (\T^2 + 6 \T \U + \U^2) + \T \U (\T^2 + \U^2) - 6 m_t^8 \right),
\end{align}
with $\beta_t=\sqrt{1-4m_t^2/\s}$. The gluon channel gives around 90\% of the total production rate at the LHC, while at the Tevatron the quark channels were the dominant sub-processes.

We show below the relevant one-loop contributions of the SMEFT operators to the parton-level cross-section arising from the interference term with the SM for both the quark and gluon channels. They are expected to be the dominant effect of the four-heavy-quark operators in the SMEFT expansion when the scale of new physics is large enough for the EFT approximation to be valid.

\subsubsection{Parton-level analytical results for the quark channel }
\label{sec:quarkChannel}

In what follows, we omit any color factors at first for simplicity. We present the results below in terms of the Passarino-Veltman integrals written as $A_0=A_0(m_t^2)$, $B_0=B_0(\s;m_t^2,m_t^2)$ and $C_0=C_0(0,\s,0;m_t^2,m_t^2,m_t^2)$.

\begin{figure}
\centering
\unitlength = 0.6mm
\begin{subfigure}{0.3\textwidth}
\centering
\includegraphics{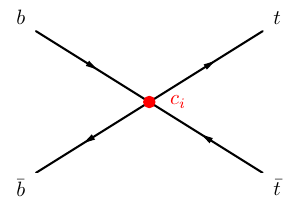}
\caption{}
\label{fig:tree_level_bottom}
\end{subfigure}
\quad
\begin{subfigure}{0.3\textwidth}
\includegraphics{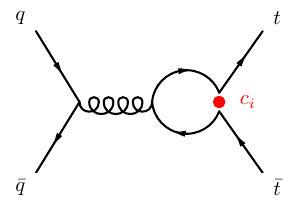}
\caption{}
\label{fig:Loop1qq}
\end{subfigure}
\quad
\begin{subfigure}{0.3\textwidth}
\includegraphics{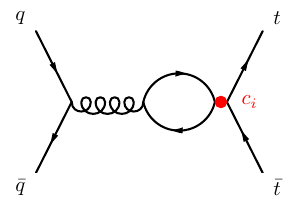}
\caption{}
\label{fig:Loop2qq}
\end{subfigure}
\caption{EFT diagrams for the relevant dimension-6 operators at (a) tree-level and (b-c) one-loop for the quark-initiated process $q \bar{q} \to t\bar{t}$. }
\label{fig:LoopEFTqq}
\end{figure}

The Feynman diagrams in Fig. \ref{fig:LoopEFTqq} present the possible insertions of the four-heavy-quark operators, which, with the exception of the operator $\mathcal{O}_{tt}^{(1)}$, at tree-level enter only in the bottom quark-initiated process (Fig. \ref{fig:tree_level_bottom}). At loop level, the four-heavy-quark operators induce NLO corrections to the s-channel of the SM and enter through all the quark channels  (Fig. \ref{fig:Loop1qq}-\ref{fig:Loop2qq}). Thus, we refer to our results as of NLO order, even though the results in section \ref{sec:pheno} also include the loop-induced gluon-initiated sub-processes (See diagrams in Fig. \ref{fig:LoopEFTgg}). The results presented in this paper do not consider squares of one-loop diagrams, which technically are next-to-NLO corrections. Hence, whenever we refer to  the quadratic contributions of effective operators we mean the square of diagrams of the type shown in Fig. \ref{fig:tree_level_bottom}, their possible interference, and the interference between diagrams having insertions at tree-level with diagrams having insertions at one-loop. From this, it follows that the quadratic contributions come exclusively from bottom-initiated sub-processes.  

In general, there are two types of loop structures arising from four-fermion operators.  We can have 
\begin{itemize}
\item Structure 1: Diagrams in which fermion-flow goes from one of the external spinors through the internal fermion lines all the way to the other external spinor, \textit{i.e.} the two fermion currents from the effective operator are involved in the loop. This corresponds to the diagram in Fig. \ref{fig:Loop1qq}.
\item Structure 2: Diagrams in which the spinor indices contract in such a way that a trace over the Lorentz structures emerges in the numerator, \textit{ i.e.} only one of the fermion currents from the effective operator is involved in the loop. This corresponds to the diagram in Fig. \ref{fig:Loop2qq}.
\end{itemize}

When we consider the operators listed in Eq. \eqref{eq:4heavy}, there are amplitudes containing insertions of operators with chirality structures $\bar{L}L\bar{L}L$, $\bar{L}L\bar{R}R$ and $\bar{R}R\bar{R}R$ for each of the structures above. The last chirality structure can be obtained from the first one by parity transformations. Thus, in total, there are four cases to be computed, two for each structure, which we proceed to discuss.

The amplitude for the structure 1 with $\bar{L}L\bar{L}L$ can be written as
\begin{align}
\mathcal{M}_{\mathrm{NP}}^{(1)}= & \,\mathcal{C}_i  \frac{g^{2}}{\s}\frac{c_{i}}{\Lambda^{2}}\bar{u}\left(p_{4}\right)\Gamma_{\mu}^{\mathrm{LL}}v\left(p_{3}\right)\cdot\bar{v}\left(p_{2}\right)\gamma^{\mu}u\left(p_{1}\right), \label{eq:amp_struct1}
\end{align}
with $\mathcal{C}_i$ standing for the color structure of the amplitude with an insertion of the effective operator $\mathcal{O}_i$. The vertex function $\Gamma_{\mu}^{\mathrm{LL}}$ is simplified to have the form 

\begin{align}
\Gamma_{\mu}^{\mathrm{LL}} & =-\frac{i}{4\pi^2}\left(\s\gamma_{\mu}-q_{\mu}q\!\!\!/\right)P_R F_1\left(\s,m \right), \label{eq:GammaLL}
\end{align}
with $q=p_1+p_2$, $m$ the mass of the fermion in the loop and $F_1$ defined as 
\begin{align}
F_1\left(\s,m \right) & \equiv  -\frac{1}{6\s}\mathrm{Re}\left[\left(2m^{2}+\s\right)B_{0}\left(\s,m^{2}\right)-2A_{0}\left(m^{2}\right)+2m^{2}-\frac{4}{3}\s\right],\nonumber \\
& \xrightarrow{\mathrm{Finite}} \frac{1}{6\s} \mathrm{Re}\left[ \left( 2 m^2+\s\right)\left(\beta_t\log\frac{\beta_t+1}{\beta_t-1} -2\right)-\s\log\frac{\mu^2}{m^2}+\frac{4}{3}\s\right]. \label{eq:F1def}
\end{align}
We expanded around $D=4-2\epsilon$ to obtain the second line of Eq. \eqref{eq:F1def} and, during this and the next subsection, the divergence is subtracted in the $\overline{\mathrm{MS}}$ scheme with the counter-term provided by the $gt\bar{t}$-vertex. Similarly, the amplitude for the structure 2 with $\bar{L}L\bar{L}L$ can be written as
\begin{align}
\mathcal{M}_{\mathrm{NP}}^{(2)} = & \,\mathcal{C}_i\frac{g^{2}}{\s}\frac{c_{i}}{\Lambda^{2}}\bar{u}\left(p_{4}\right)\bar{\gamma}^{\rho}P_{L}v\left(p_{3}\right)\cdot \bar{v}\left(p_{2}\right)\bar{\gamma}^{\mu}u\left(p_{1}\right)\cdot\mathcal{I}_{\rho\mu}, \label{eq:amp_struct2}
\end{align}
where the tensor carrying the information about the loop effects is simplified to 
\begin{align}
\mathcal{I}_{\rho\mu}= & -\frac{i}{4\pi^2} \left(\s g_{\mu\rho}-q_{\mu}q_{\rho}\right)F_2 \left(\s,m\right). \label{eq:int_trace_1}
\end{align}
The factor $F_2$ is defined as
\begin{align}
F_2\left( \s, m \right) & \equiv -\frac{1}{6\s}\mathrm{Re}\left[\left(2m^{2}+\s\right)B_{0}\left(\s,m^{2}\right)-2 A_{0}\left(m^{2}\right)+2m^2 - \frac{\s}{3}\right], \nonumber \\
& \xrightarrow{\mathrm{Finite}} \frac{1}{6\s} \mathrm{Re}\left[ \left( 2 m^2+\s\right)\left(\beta_t\log\frac{\beta_t+1}{\beta_t-1} -2\right)-\s\log\frac{\mu^2}{m^2}+\frac{\s}{3}\right]. \label{eq:F2def}
\end{align}
We notice that the axial parts of the amplitudes in Eq. \eqref{eq:amp_struct1}-\eqref{eq:amp_struct2}  do not contribute when the interference with the SM tree-level amplitudes is performed. Hence, the results above stand also for the operators with chirality $\bar{R}R\bar{R}R$. The result for the structure 2 with $\bar{L}L\bar{R}R$ is given by the same quantity $\mathcal{I}_{\rho\mu}$ in Eq. \eqref{eq:int_trace_1}. This is a consequence of the fact that the amplitudes for the $\bar{L}L\bar{L}L$ and $\bar{L}L\bar{R}R$  cases only differ in the sign of the terms with Levi-Civita tensors, but such terms vanish in the final result. 

Finally, we consider the structure 1 with $\bar{L}L\bar{R}R$, where the right-handed fermions in the effective vertex are taken to be the top quarks in the final state. For this case, the amplitude is given by Eq. \eqref{eq:amp_struct1} with the replacement $\Gamma^{\mathrm{LL}}_{\mu} \rightarrow \Gamma^{\mathrm{LR}}_{\mu} $, where the vertex factor now has the form 
\begin{align}
\Gamma^{\mathrm{LR}}_{\mu} & =-\frac{i}{16\pi^2}\left( D-4\right) m\left(\gamma_{\mu}q\!\!\!/-q_{\mu}\right)B_{0}\left(\s,m^{2}\right)P_R ,\label{eq:vertex_llrr0}  \\
& \xrightarrow{\mathrm{Finite}} -\frac{i}{4\pi^2}\cdot 2m\left(\gamma_{\mu}q\!\!\!/-q_{\mu}\right)P_R.\label{eq:vertex_llrr}
\end{align}
The factor $(D-4)$ in Eq. \eqref{eq:vertex_llrr0} implies that the finite amplitude for this case is given purely by rational terms\footnote{Rational terms appear in the implementation of the Passarino-Veltman reduction of one-loop amplitudes as the finite product of poles of order $\mathcal{O}(\epsilon^{-1})$ and terms in the numerator of order $\mathcal{O}(\epsilon)$. The rational part of a one-loop amplitude can be identified as the terms that do not involve logarithms or dilogarithms at order $\mathcal{O}(\epsilon^0)$.}. 

Including the appropriate color factors in the results above, we find the partonic differential cross-section for the interference between tree-level SM and the SMEFT at one-loop in the quark channel for the five heavy four-quark operators:
\begin{align}
\allowdisplaybreaks
\frac{d\hat{\sigma}}{d\Omega}\Bigr|_{\mathcal{O}_{tt}^{(1)}}^{\mathrm{Int}} & = \frac{\alpha_{s}^{2}}{18\pi^2}\frac{c_{tt}^1}{\Lambda^{2}}\frac{\beta_t}{\s^2}\left(2m_t^{2}\s+\left(\T-m_t^{2}\right)^{2}+\left(\U-m_t^{2}\right)^{2}\right) F_1\left( \s, m_t \right), \label{eq:quark_ctt1} \\
\frac{d\hat{\sigma}}{d\Omega}\Bigr|_{\mathcal{O}_{QQ}^{(1)}}^{\mathrm{Int}} & = \frac{\alpha_{s}^{2}}{36\pi^2}\frac{c_{QQ}^1}{\Lambda^{2}}\frac{\beta_t}{\s^2}\left(2m_t^{2}\s+\left(\T-m_t^{2}\right)^{2}+\left(\U-m_t^{2}\right)^{2}\right) F_1\left( \s, m_t \right),  \\
\frac{d\hat{\sigma}}{d\Omega}\Bigr|_{\mathcal{O}_{QQ}^{(8)}}^{\mathrm{Int}} & = \frac{\alpha_{s}^{2}}{216\pi^2}\frac{c_{QQ}^8}{\Lambda^{2}}\frac{\beta_t}{\s^2}  \nonumber \\ 
& \: \:  \left(2m_t^{2}\s+\left(\T-m_t^{2}\right)^{2}+\left(\U-m_t^{2}\right)^{2}\right) \left(3 \left( F_2\left( \s, m_b \right)+ F_2\left( \s, m_t \right) \right)  - F_1\left( \s, m_t \right) \right), \label{eq:quark_cqq8}  \\
\frac{d\hat{\sigma}}{d\Omega}\Bigr|_{\mathcal{O}_{Qt}^{(1)}}^{\mathrm{Int}} & = -\frac{\alpha_{s}^{2}}{18\pi^2} \, \frac{c_{Qt}^1}{\Lambda^{2}}\frac{m_t^2\beta_t}{\s},  \\
\frac{d\hat{\sigma}}{d\Omega}\Bigr|_{\mathcal{O}_{Qt}^{(8)}}^{\mathrm{Int}} & = \frac{\alpha_{s}^{2}}{216\pi^2}\frac{c_{Qt}^8}{\Lambda^{2}}\frac{\beta_t}{\s^2} \nonumber \\ 
&   \qquad\left(3\left(2m_t^{2}\s+\left(\T-m_t^{2}\right)^{2}+\left(\U-m_t^{2}\right)^{2}\right)  \left( F_2\left( \s, m_b \right) + 2 F_2\left( \s, m_t \right) \right) + 2 m_t^2\s \right). \label{eq:quark_cqt8}
\end{align}
In the results above, we keep the explicit dependence on the bottom mass, although the results in section \ref{sec:pheno} are obtained in the five-flavour scheme. These differential rates can be compared to the SM differential cross-section in Eq. \eqref{eq:treeSM}, from which we notice that, with the exception of the result for $\mathcal{O}_{Qt}^{(1)}$, the results above can be written as corrections in the form of overall factors multiplying the SM result. We also note that the interference terms change signs at different center-of-mass energies, and this effect will manifest itself in the full analysis performed in section  \ref{sec:pheno}.

As a further check of our results, the formulas in Eq. \eqref{eq:quark_ctt1}-\eqref{eq:quark_cqt8} have been compared to a toy UV model with vector bosons  ($X_\mu$) as mediators that could generate the four-heavy-quark operators once the heavy states are integrated out. In the case of the color-octet operators, the mediator must bear the corresponding color structure. As expected, we found a good agreement between our SMEFT predictions and the toy model for energies below the mass of the heavy states, $\sqrt{\s}\ll m_X$.

For later use, we will need to study the result for a slightly different operator basis (used by the {\tt SMEFT@NLO}), where the operator $\mathcal{O}_{QQ}^{(8)}$ is written in terms of a color-singlet operator by means of Fierz transformations as
\begin{align}
    c_{QQ}^8 = 8 [C_{qq}^{(3)}]^{3333}, & \qquad \mathcal{O}_{qq}^{(3)} = \left( \bar{Q}_L \gamma_{\mu} \tau^i Q_L \right) \left( \bar{Q}_L \gamma^{\mu} \tau^i Q_L \right). \label{eq:cqq8def2}
\end{align}
The differential cross-section starting from the definition in Eq. \eqref{eq:cqq8def2} can be computed to yield as a result the formula in Eq. \eqref{eq:quark_cqq8} with the substitution $F_2\left( \s, m \right) \rightarrow F_1\left( \s, m \right)$. By inspection of Eq. \eqref{eq:F1def} and Eq.   \eqref{eq:F2def}, we notice that the difference of the two results arises from the rational part. As it is well known, even though the two definitions of the operator $\mathcal{O}_{QQ}^{(8)}$ are equivalent at tree-level, the amplitudes at one-loop are not. The inclusion of evanescent operators \cite{aebischer2022} is required to find an equivalence between these results in different bases.

\subsubsection{Parton-level analytical results for the gluon channel}

\begin{figure}
\centering
\unitlength = 0.6mm
\begin{subfigure}{0.3\textwidth}
\includegraphics{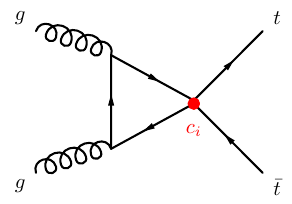}
\caption{ }
\label{fig:Loop1EFT}
\end{subfigure}
\hfill
\begin{subfigure}{0.3\textwidth}
\includegraphics{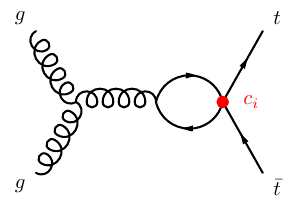}
\caption{ }
\label{fig:Loop2EFT}
\end{subfigure}
\hfill
\begin{subfigure}{0.3\textwidth}
\includegraphics{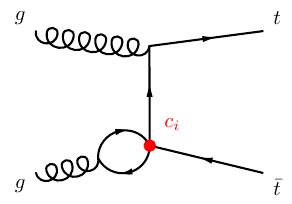}
\caption{}
\label{fig:Loop3EFT}
\end{subfigure}
\caption{EFT one-loop diagrams for the partonic process $g g \to t\bar{t}$.}
\label{fig:LoopEFTgg}
\end{figure}

The LO contribution of the four-heavy-quark operators to the gluon-initiated process is at one-loop through the topologies shown in Fig. \ref{fig:LoopEFTgg}. Each of the diagrams of the type shown in Fig. \ref{fig:Loop3EFT} leads to vanishing amplitudes for the operators $\mathcal{O}_{tt}^{(1)}$, $\mathcal{O}_{QQ}^{(1)}$ and $\mathcal{O}_{QQ}^{(8)}$, while in the cases of $\mathcal{O}_{Qt}^{(1)}$ and $\mathcal{O}_{Qt}^{(8)}$ they contribute. This can be understood through the definitions of the loop corrections $\Gamma_{\mu}^{\mathrm{LL},\mathrm{RR}}$, $\Gamma_{\mu}^{\mathrm{LR}}$ and $\mathcal{I}_{\rho\mu}$ in Eq. \eqref{eq:GammaLL}, \eqref{eq:int_trace_1} and \eqref{eq:vertex_llrr}. These quantities enter the gluon-initiated process by substituting $q^\mu \rightarrow p^\mu_i$, with $p_i$ the momentum of the external gluons. Hence, only $\Gamma_{\mu}^{\mathrm{LR}}$ leads to non-vanishing amplitudes  once these quantities get contracted with the polarization vector of the gluons. Accordingly, diagrams of the type of Fig. \ref{fig:Loop3EFT} only contribute in the cases of the operators $\mathcal{O}_{Qt}^{(1)}$ and $\mathcal{O}_{Qt}^{(8)}$, and do so with purely rational terms.

In addition to the diagrams in Fig. \ref{fig:LoopEFTgg}, there are diagrams that contribute to the self-energy of the top quark in the form of tadpole contributions, which come solely from four-fermion operators mixing helicities: $\mathcal{O}_{Qt}^{(1)}$ and $\mathcal{O}_{Qt}^{(8)}$. These tadpoles turn out to be non-physical as they can be absorbed by the mass counter-term of the top quark. Finally, divergences appear in diagrams of the type shown in Fig. \ref{fig:Loop1EFT}  and cancel those corresponding to the diagrams in Fig. \ref{fig:Loop2EFT} for the operators $\mathcal{O}_{tt}^{(1)}$, $\mathcal{O}_{QQ}^{(1)}$, $\mathcal{O}_{QQ}^{(8)}$ and $\mathcal{O}_{Qt}^{(8)}$. The respective diagrams with insertions of the $\mathcal{O}_{Qt}^{(1)}$ operator are finite. 

The partonic differential cross-sections for the interference between SM at tree-level and SMEFT at one-loop in the gluon channel (where sub-leading contributions proportional to the bottom mass are neglected) are given by the expressions 

{\small
\allowdisplaybreaks
\begin{align}
\frac{d\hat{\sigma}}{d\Omega}\Bigr|_{\mathcal{O}_{tt}^{(1)}}^{\mathrm{Int}} = & \frac{1}{64 \pi^2} \frac{c_{tt}^1}{\Lambda^2} \frac{\alpha_s^2 m_t^2 \beta_t}{24 \s^2 }     \frac{1}{(\T - m_t^2 )(\U - m_t^2 )}  \nonumber \\
    & \: \left( 2 m_t^2 \s (9\left(\T-\U\right)^2 -13 \s^2) C_0 + 36(\T-\U)^2( m_t^2 B_0 -  A_0)   + 3 ( \s + 12 m_t^2)\left(\T-\U\right)^2 - 13 \s^3 \right),\label{eq:gluonFormula1}\\
\frac{d\hat{\sigma}}{d\Omega}\Bigr|_{\mathcal{O}_{QQ}^{(1)}}^{\mathrm{Int}} = & \frac{1}{64 \pi^2} \frac{c_{QQ}^1}{\Lambda^2} \frac{\alpha_s^2 m_t^2 \beta_t}{24 \s^2 }     \frac{1}{(\T - m_t^2 )(\U - m_t^2 )}  \nonumber \\
    & \: \left( 2 m_t^2 \s (9\left(\T-\U\right)^2 -13 \s^2) C_0 + 36(\T-\U)^2( m_t^2 B_0 -  A_0)   + 3 ( \s + 12 m_t^2)\left(\T-\U\right)^2 - 19 \s^3 \right),  \\
\frac{d\hat{\sigma}}{d\Omega}\Bigr|_{\mathcal{O}_{QQ}^{(8)}}^{\mathrm{Int}} = & \frac{1}{64 \pi^2} \frac{c_{QQ}^8}{\Lambda^2} \frac{\alpha_s^2 m_t^2 \beta_t}{144 \s^2 }     \frac{1}{(\T - m_t^2 )(\U - m_t^2 )}  \nonumber \\
    & \: \left( 4 m_t^2 \s (9\left(\T-\U\right)^2 -13 \s^2) C_0 + 72(\T-\U)^2( m_t^2 B_0 -  A_0)   + 3 ( 5\s + 24 m_t^2)\left(\T-\U\right)^2 - 41 \s^3 \right),  \\
\frac{d\hat{\sigma}}{d\Omega}\Bigr|_{\mathcal{O}_{Qt}^{(1)}}^{\mathrm{Int}} = & \frac{1}{64 \pi^2} \frac{c_{Qt}^1}{\Lambda^2} \frac{\alpha_s^2 m_t^2 \beta_t}{24 \s }     \frac{1}{(\T - m_t^2 )(\U - m_t^2 )}  \nonumber \\
    & \qquad \qquad \qquad \qquad  \left( 2 \s (7 \s^2 -22 m_t^2 \s +56 m_t^4) C_0 +  \s ( 56 m_t^2 - 19 \s) - 18 \left(\T-\U\right)^2 \right),\\
\frac{d\hat{\sigma}}{d\Omega}\Bigr|_{\mathcal{O}_{Qt}^{(8)}}^{\mathrm{Int}} = & \frac{1}{64 \pi^2} \frac{c_{Qt}^8}{\Lambda^2}\frac{\alpha_s^2 m_t^2 \beta_t}{288 \s^2}\frac{1}{(\T - m_t^2 )(\U - m_t^2 )}  \nonumber \\ &   \left(4 \s \left( 11 \s^3 + 88 m_t^4 \s  - m_t^2 ( 29 \s^2 - 27 \left( \T - \U\right)^2 ) \right) C_0 + 216 (\T-\U)^2 \left( m_t^2 B_0 -A_0  \right)  \right. \nonumber \\ 
    & \qquad \qquad \qquad \qquad \qquad \qquad \; \; \; \left. + \s (29 \s^2 + 63 (\T-\U)^2)  + 8 m_t^2 \left( 22 \s^2 + 27 (\T-\U)^2 \right) \right). \label{eq:gluonFormula2}
\end{align}}
The differential cross-sections corresponding to the operators $\mathcal{O}_{QQ}^{(1)}$ and $\mathcal{O}_{tt}^{(1)}$ are different only by the rational term. This is due to the contribution of a diagram of the type \ref{fig:Loop1EFT} with the bottom quark running in the loop. Such diagram is given purely as a rational contribution. To see this, we first realize that out of the two structures discussed in section \ref{sec:quarkChannel}, only the second structure contributes when the bottom runs in the loop. Such triangle diagrams lead to the loop factor $(2 m^2 C_0 + \mathcal{R})$, where $m$ is the mass of the particle in the loop and $\mathcal{R}$ is the rational term. When the mass of the bottom quark is negligible, the contribution of the diagram comes from the rational term, which turns out to be $\mathcal{R} = 1$. 

The analytical expressions for the gluon channel computed above agree with the results obtained in  \cite{Muller2021} except for the cases of $\mathcal{O}_{Qt}^{(8)}$ and $\mathcal{O}_{QQ}^{(8)}$ \footnote{The comparison between Eq. \eqref{eq:gluonFormula1}-\eqref{eq:gluonFormula2} and the results in Ref. \cite{Muller2021} is readily performed after noticing the equivalence 
\begin{equation*}    S_1 (s,m)=-m^2 \mathrm{Re} ( C_0 ), \qquad \qquad  S_2 (s,m) = \frac{1}{s} \mathrm{Re} (m^2(B_0+1)-A_0) \end{equation*}}.

\subsubsection{Full computation of differential distributions}

We now present in detail the deformations in the differential distributions of $pp \rightarrow t\bar{t}$ due to the presence of heavy quark four-fermion operators obtained using \mg. The comparison between these results and the analytical expressions presented above allowed the identification of a coding \textit{logic error} in \mg~consisting of the wrong selection of rational terms corresponding to a given one-loop amplitude. The differential cross-section of the top-pair process with insertions of dimension-6 operators can be schematically organized as 
\begin{equation}
    \frac{d\sigma}{d m_{t\bar{t}}}\Big(\frac{c_i}{\Lambda^{2}}\Big) = \frac{d\sigma^{\mathrm{SM}}}{d m_{t\bar{t}}} + \frac{d\sigma^{\mathrm{Int.}}}{d m_{t\bar{t}}} \frac{c_i }{\Lambda^{2}} + \frac{d\sigma^{\mathrm{Quad.}}}{d m_{t\bar{t}}}  \frac{c_i c_j}{\Lambda^{4}},
\end{equation}
with 
\begin{align}
    \frac{d\sigma^{\mathrm{Int.}}}{d m_{t\bar{t}}} & =  \frac{d\sigma_{\mathrm{Int.}}^{(0)}}{d m_{t\bar{t}}}+\alpha_s \frac{d\sigma_{\mathrm{Int.}}^{(1)}}{d m_{t\bar{t}}} + \alpha_s^2 \frac{d\sigma_{\mathrm{Int.}}^{(2)}}{d m_{t\bar{t}}} , \\
    \frac{d\sigma^{\mathrm{Quad.}}}{d m_{t\bar{t}}}  & =  \frac{d\sigma_{\mathrm{Quad.}}^{(0)}}{d m_{t\bar{t}}} +   \alpha_s \frac{d\sigma_{\mathrm{Quad.}}^{(1)}}{d m_{t\bar{t}}} +   \alpha_s^2 \frac{d\sigma_{\mathrm{Quad.}}^{(2)}}{d m_{t\bar{t}}}.
\end{align}
The superscript in round brackets stands for the extra order in $\alpha_s$ compared to the first non-zero contribution. We emphasize that dimension-8 operators are not considered. The simulation of events involves the full interference of the four-heavy-quark operators with leading QCD contributions in the SM, $\sigma_{\mathrm{Int}}$,
while the quadratic contribution solely comes from the square of the dimension-six operators, $\sigma_{\mathrm{Quad}}$, at the LHC. The quadratic contribution requires the computation of the square of bottom-pair initiated tree-level diagrams ($d\sigma_{\mathrm{Quad.}}^{(0)}/d m_{t\bar{t}}$) and their NLO QCD corrections involving both the virtual and real corrections ($d\sigma_{\mathrm{Quad.}}^{(1)}/d m_{t\bar{t}}$). The quadratic contribution also contains the square of one-loop diagrams with one insertion of effective operators ($d\sigma_{\mathrm{Quad.}}^{(2)}/d m_{t\bar{t}}$). This term can be computed in \MG as a loop-induced contribution only if the bottom quark is not in the protons, since this order in $\alpha_s$ corresponds to NNLO for the bottom induced process with all its extra difficulties including the computation of two-loop diagrams. Although $\sigma_{\mathrm{Quad.}}^{(2)}$ is further suppressed in $\alpha_s$, $\sigma_{\mathrm{Quad.}}^{(0)}$ is bottom PDF suppressed. As a matter of fact, we have computed the loop-induced contribution of $\sigma_{\mathrm{Quad.}}^{(2)}$ and found that it is of the same order of magnitude as $\sigma_{\mathrm{Quad.}}^{(0)}$. Since $\sigma_{\mathrm{Quad.}}^{(2)}$ cannot be fully computed at this stage, and is similar in shape and magnitude to $\sigma_{\mathrm{Quad.}}^{(0)}$,  we only include the latter in our analysis given that they are regarded as an estimate of the higher order EFT contributions. 
In other contexts, recent works explored double insertions of effective vertices~\cite{ElFaham2022,Asteriadis2023} showing that, in the four-top production and in the gluon-fusion Higgs production, the differences between single and double insertions do not bear phenomenological importance. Finally, the interference between the SM and dimension-8 operators has been shown to be as important as dimension-6 squared contribution in diboson productions~\cite{Degrande2023}, which are processes where the terms of order $\mathcal{O}(\Lambda^{-2})$ are suppressed.

\begin{figure}
\centering
\begin{subfigure}{0.49\textwidth}
   \resizebox{\columnwidth}{!}{\includegraphics{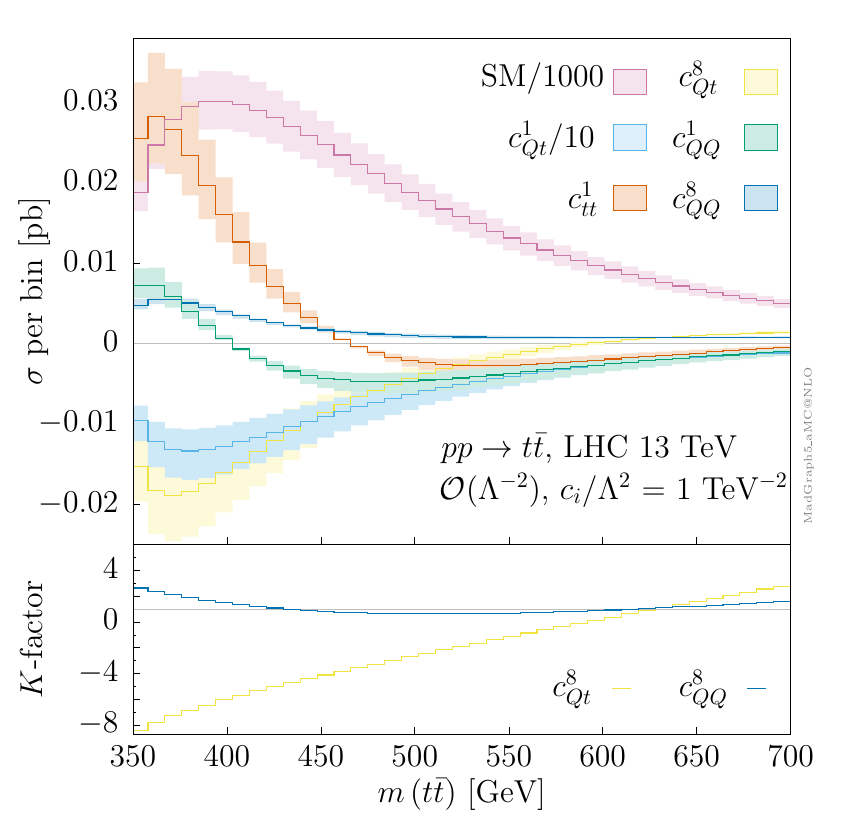}}
    \caption{ }
    \label{fig:4heavy_int_Minv}
\end{subfigure}
\,
\begin{subfigure}{0.49\textwidth}
   \resizebox{\columnwidth}{!}{\includegraphics{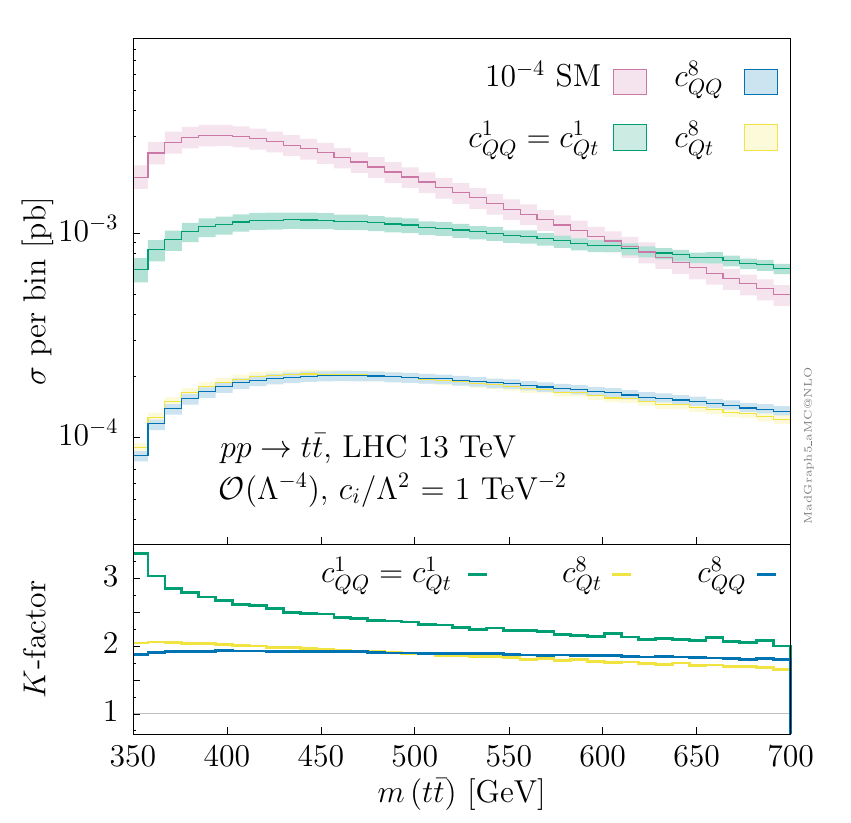}}
    \caption{ }
    \label{fig:4heavy_quad_Minv}
\end{subfigure}
\caption[Caption for LOF]{Invariant-mass distribution of the top-pair production at NLO of the (a) interference between four-heavy quark operators and the SM and (b) new physics square (without the interference terms shown in (a)). The Wilson coefficients are set to $c_i/\Lambda^2 = 1\,\mathrm{TeV}^{-2}$. In (b), since the curves corresponding to the $c_{QQ}^1$ and $c_{Qt}^1$ coefficients are the same, only one is shown in the plot. }
\label{fig:4heavy_Minv}
\end{figure}

In Fig. \ref{fig:4heavy_Minv} we present the invariant-mass distribution for the interference and quadratic contributions (without the interference) of these operators, including the renormalization scale uncertainties as discussed in the beginning of this section, which are shown as bands. In the inset at the bottom, the $K$-factors\footnote{The $K$-factors in this paper are defined as the ratio $\sigma_{\mathrm{NLO}}/\sigma_{\mathrm{LO}}$. When considering $K$-factors of new physics, we take into consideration purely NLO and LO SMEFT  contributions. } are displayed for the effective operators that contribute at tree-level. The operators $\mathcal{O}_{QQ}^{(8)}$ and  $\mathcal{O}_{Qt}^{(8)}$ can interfere with the SM at tree-level through the bottom quark-initiated channel due to their color structure. The $K$-factor corresponding to the coefficient $c_{QQ}^{8}$ shows that the loop corrections are comparable to the tree-level contributions. This means that, even though it is suppressed by PDFs, the bottom channel presents a sizable cross-section except for the near-threshold region, where the loop induced gluon channel contributions can be almost three times larger. A similar behaviour occurs with the $\mathcal{O}_{Qt}^{(8)}$ operator, with a larger $K$-factor in the near threshold region favoring the gluon channel. 

The total one-loop interference with SM amplitudes of the four-heavy-quark operators suffers from phase-space cancellations, but when we consider the differential distributions, as presented in Fig. \ref{fig:4heavy_int_Minv}, we notice that there are portions of the phase-space that are favored. Fortunately, differential distributions have been measured for the top-pair production, and we can exploit this to get a better sensitivity to the four-heavy-quark operators than just considering total rates. In particular, the distributions in Fig. \ref{fig:4heavy_int_Minv} show that the coefficient $c_{Qt}^1$ leads to contributions one order of magnitude larger than the other coefficients, and distinctively, although out of the range in the plot, presents a change in sign at high energy, \textit{i.e.} in between 1 - 1.5 TeV (see Fig. \ref{fig:HLinvMass}). On the contrary, for the coefficients $c_{QQ}^1$ and $c_{tt}^1$ such flip of sign happens at around 400 GeV and 460 GeV, respectively. Finally, the operator $c_{Qt}^8$ presents a change in sign at an invariant-mass of roughly $ 600$ GeV. As discussed previously in our analytical computations, these changes in sign are present at the partonic level in each of the channels, although at different energies. When the results are weighted with PDFs, we find cancellations in some regions of the phase-space with the negative interference being traced back to the partonic amplitudes.

The quadratic contributions shown in Fig. \ref{fig:4heavy_quad_Minv} correspond to operators that enter the bottom channel at tree-level, \textit{i.e.} the operators $\mathcal{O}_{QQ}^{(1)}$, $\mathcal{O}_{QQ}^{(8)}$, $\mathcal{O}_{Qt}^{(1)}$  and $\mathcal{O}_{Qt}^{(8)}$. The quadratic contributions from the square of one-loop diagrams are suppressed by loop factors, thus the gluon-initiated diagrams are sub-leading at order $\mathcal{O}(\Lambda^{-4})$.  As an additional consequence, the quadratic contributions of the operator $\mathcal{O}_{tt}^{(1)}$ only appear at two-loop. With $c_i/\Lambda^2=1$ TeV$^{-2}$, we can see that the interference terms dominate over the quadratic terms in the top-pair production at the LHC for most of the experimental phase-space for the operators studied here.
{ 
\renewcommand{\arraystretch}{1.7}
\begin{table}
\begin{center}
{\footnotesize
\begin{tabular}{|c|c|c|}
\cline{2-3}
\multicolumn{1}{c|}{} & \multicolumn{1}{c|}{$q\bar{q}\rightarrow t\bar{t}$} & \multirow{1}{*}{$gg\rightarrow t\bar{t}$}\tabularnewline
\hline 
\hline 
\multirow{1}{*}{SM} & $\frac{32}{9} \pi^2 \alpha_s^2 (1+\cos^2 \theta) $ & $\frac{1}{6} \pi^2 \alpha_s^2  \frac{(1 + \cos^2 \theta ) (7 + 9 \cos^2 \theta )}{\sin^2 \theta}$     \tabularnewline

\multirow{1}{*}{$c_{tt}^1$} & $\frac{8}{81} \frac{\alpha_s^2}{\Lambda^2} \s (1+\cos^2 \theta) (3 \log \frac{\s}{\mu^2}-2) $ & $\frac{1}{6} \frac{\alpha_s^2}{\Lambda^2} m_t^2 \frac{(3 \cos^2 \theta - 13)}{\sin^2 \theta}$     \tabularnewline

\multirow{1}{*}{$c_{QQ}^1$} &  $\frac{4}{81} \frac{\alpha_s^2}{\Lambda^2} \s (1+\cos^2 \theta) (3 \log \frac{\s}{\mu^2}-2) $ &     $\frac{1}{6} \frac{\alpha_s^2}{\Lambda^2} m_t^2 \frac{(3 \cos^2 \theta - 19)}{\sin^2 \theta}$  \tabularnewline

\multirow{1}{*}{$c_{QQ}^8$}& $\frac{2}{243} \frac{\alpha_s^2}{\Lambda^2} \s (1+\cos^2 \theta) (15\log \frac{\s}{\mu^2}-28) $  &    $\frac{1}{36} \frac{\alpha_s^2}{\Lambda^2} m_t^2 \frac{(15 \cos^2 \theta - 41)}{\sin^2 \theta}$      \tabularnewline

\multirow{1}{*}{$c_{Qt}^1$}& $ \frac{32}{9} \frac{\alpha_s^2}{\Lambda^2} m_t^2 $ &     $\frac{1}{6} \frac{\alpha_s^2}{\Lambda^2} m_t^2  \frac{1}{\sin^2 \theta} (7 (\log^2 \frac{\s}{m_t^2} - \pi^2 ) - 18 \cos^2 \theta  -19) $        \tabularnewline

\multirow{1}{*}{$c_{Qt}^8$}& $\frac{2}{27} \frac{\alpha_s^2}{\Lambda^2} \s (1+\cos^2 \theta) (3 \log \frac{\s}{\mu^2} - 5 ) $  &   
$\frac{1}{72} \frac{\alpha_s^2}{\Lambda^2} m_t^2  \frac{1}{\sin^2 \theta} (22 (\log^2 \frac{\s}{m_t^2} - \pi^2 ) + 63 \cos^2 \theta  + 29) $      \tabularnewline
\hline 
\end{tabular}
}
\caption{ Ultra-high-energy ($\s \gg m_t^2$) behaviour of the unpolarized amplitude square of the interference between SMEFT diagrams and the SM in the quark-initiated and gluon-initiated sub-processes. The dependence on the Wilson coefficients is kept implicit.  } \label{tab:high_energy}
\end{center}
\end{table}
}

As a last comment, we discuss the growth with energy of the amplitudes. In Table \ref{tab:high_energy} we list the growth with energy of the unpolarized squared amplitude produced by the interference between SMEFT effects and the SM. Such expressions have been obtained by keeping the leading term after taking the limit $\s \gg m_t^2$ in the squared amplitudes that led to Eq. \eqref{eq:quark_ctt1}-\eqref{eq:quark_cqt8}  and Eq. \eqref{eq:gluonFormula1}-\eqref{eq:gluonFormula2}. Most of the considered operators display the $s$ factor enhancement for the quark-initiated process also present for the tree-level contribution of 2L2H operators. The main difference with those operators is the extra logarithmic growth, which could be used to distinguish them. We also notice that the $\theta$-dependence is the same as in the SM for high-energies, from which we expect that forward-backward asymmetry observables will not enhance the sensitivity to four-heavy-quark operators, instead spin correlations is recommended.  It is observed that the  $\mathcal{O}_{Qt}^{(1)}$ operator leads to a constant expression in the quark-channel and to a logarithmic growth in the gluon channel. Due to this, of all the five operators, $\mathcal{O}_{Qt}^{(1)}$ presents the weakest growth with energy.  This behaviour will have a strong impact in the fits obtained at HL-LHC (See Fig. \ref{fig:HLinvMass}).

\subsection{Four-top production}
\label{sec:4topproduction}

\begin{figure}[t]
\centering
\unitlength = 0.6mm
\begin{subfigure}{0.3\textwidth}
\centering
\includegraphics{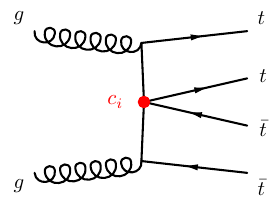}
\label{fig:tttt1EFT}
\caption{ }
\end{subfigure}
\;
\begin{subfigure}{0.3\textwidth}
\centering
\includegraphics{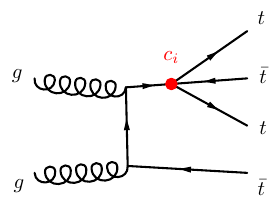}
\label{fig:tttt2EFT}
\caption{}
\end{subfigure}
\;
\begin{subfigure}{0.3\textwidth}
\centering
\includegraphics{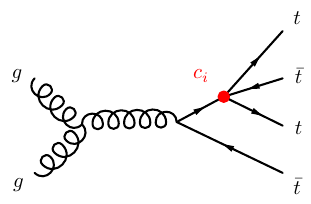}
\label{fig:tttt3EFT}
\caption{}
\end{subfigure}
\caption{EFT tree-level diagrams for the process $g g \to t\bar{t}t\bar{t}$.}
\label{fig:ttttSMEFTdiagrams}
\end{figure}

We proceed to describe the main features of the $pp\rightarrow t\bar{t}t\bar{t}$ process and the contributions of the four-heavy-quark operators to it in comparison to the $pp\rightarrow t\bar{t}$ case discussed previously. Just like in the case of the top-pair production, a large amount of work has been done to understand the four-top production at hadron colliders. In the SM, the $t\bar{t}t\bar{t}$ production cross-section is dominated by the gluon channel. The Born amplitudes receive contributions of the order $\mathcal{O}(\alpha_s^2)$ and $\mathcal{O}(\alpha_s\alpha)$, with $\alpha$ indicating couplings of electroweak origin. The theoretical prediction for the production rate at $\sqrt{s}=13$ TeV is $\sigma\left( p p \rightarrow t\bar{t}t\bar{t}\right) = 11.97_{-21\%}^{+18\%} $ fb at NLO considering QCD+EW  corrections \cite{Pagani2018}, where the errors come from scale uncertainties as specified in the beginning of this section. Recently, theoretical predictions have been obtained at NLO-QCD in the four-leptons decay channel of the four top quarks with a center-of-mass energy of $\sqrt{s} = 13.6 $ TeV~\cite{dimitrakopoulos2024}.

All the five operators in Eq.~\eqref{eq:4heavy} contribute at tree-level to both gluon- and quark-induced four-top production. 
The Feynman diagrams of the gluon-initiated process with insertions of four-fermion operators are shown in the Fig. \ref{fig:ttttSMEFTdiagrams} and they are the dominant contribution. We also include the sub-dominant quark-initiated processes in our analysis.

Since the typical cross-section of the four-top production is small,  naively its constraining power is expected to be limited. In reality, this is compensated by the high sensitivity of the four-top production to the four-heavy-quark operators. Furthermore, such sensitivity is enhanced by the behaviour of the partonic cross-sections at high-energy in the quadratic contribution due to the energy scaling of the four-fermion operators.  The increasing cross-section at high energy hints, however, that there might be EFT validity issues. In Ref. \cite{Zhang2018} a complete discussion about the validity of the SMEFT implementation for this process is presented, where it is shown that for $c_i/\Lambda^{2}<1$ TeV$^{-2}$ the EFT expansion is under control at LHC energies. When considering the contributions of the order $\mathcal{O} (\Lambda^{-4})$ that arise from the square of single insertion of effective operators, in principle, the contributions from the interference of dimension-8 operators with the SM should also be included. The effect of dimension-8 operators is left for future work. In addition, contributions with double insertions of operators should also be included. However, since double insertions of the four-heavy-quark operators are only possible in the bottom induced sub-process, those contributions are suppressed by the parton distribution function of the bottom. Thus, we will only consider single insertions of dimension-6 operators. 

\begin{figure}
\centering
\begin{subfigure}{0.496\textwidth}
   \resizebox{\columnwidth}{!}{\includegraphics{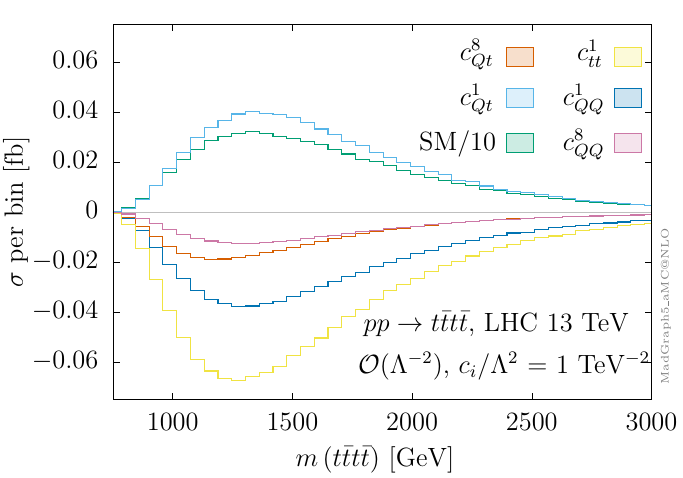}}
    \caption{ }
    \label{fig:4t_int_Minv}
\end{subfigure}
\begin{subfigure}{0.496\textwidth}
   \resizebox{\columnwidth}{!}{\includegraphics{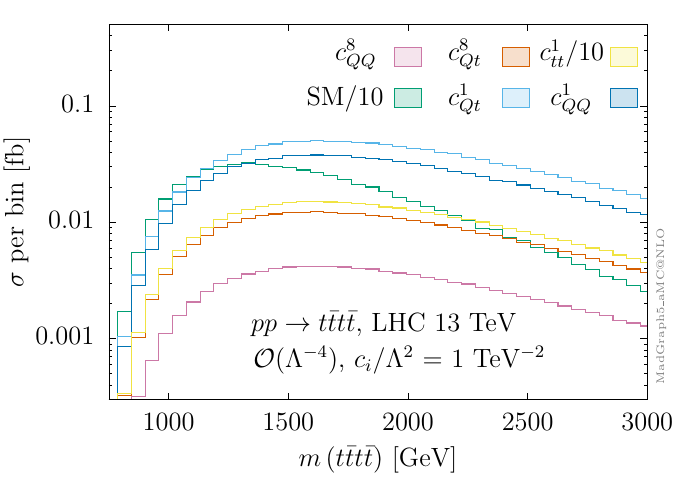}}
    \caption{ }
    \label{fig:4t_sq_Minv}
\end{subfigure}
\caption{Four-top invariant-mass distribution at tree-level of the (a) interference between four-heavy quark operators and the QCD+EW SM and (b) new physics square (without the interference). The Wilson coefficients are set to $c_i/\Lambda^2 = 1\,\mathrm{TeV}^{-2}$. Renormalization and factorization scales are set to $\mu_r=\mu_f=2 m_t =345$ GeV.}
\label{fig:4t_Minv}
\end{figure}

The inclusive cross-section of the four-top process is computed with single insertions of dimension-6 operators as 
 \begin{equation}
    \sigma_{t\bar{t}t\bar{t}}\, \Big(\frac{c_i}{\Lambda^{2}}\Big) = \sigma_{t\bar{t}t\bar{t}}^{\mathrm{SM}} + \sigma_{t\bar{t}t\bar{t}}^{\mathrm{Int.}} \frac{c_i}{\Lambda^2}   + \sigma_{t\bar{t}t\bar{t}}^{\mathrm{Quad.}} \frac{c_i c_j}{\Lambda^4}, \label{eq:4topxsec}
\end{equation}
with
\begin{equation}
    \sigma^{\mathrm{Int.}}_{t\bar{t}t\bar{t}} = \alpha_s^3 \sigma^{\mathrm{Int. 1}}_{t\bar{t}t\bar{t}} + \alpha_s^2 \alpha \, \sigma^{\mathrm{Int. 2}}_{t\bar{t}t\bar{t}} .
\end{equation}
The linear terms arise as new physics interfering with tree-level SM amplitudes at orders $\alpha_s^3$ and $\alpha_s^2 \alpha$.

In Table \ref{tab:tttt} we present the contributions at linear and quadratic order in the effective theory expansion following the conventions of Eq. \eqref{eq:4topxsec} (the corresponding QCD-NLO corrections for the four-top production, including the relevant four-heavy-quark operators using {\tt SMEFT@NLO}, have been computed in \cite{Degrande2021smeftNLO}). 
Since the main goal of this paper is to probe the constraining reach at leading order on the four-heavy-quark operators of the top-pair and four-top production, we will not consider QCD corrections on the latter when including effective operators. At order $\mathcal{O}(\Lambda^{-4})$ only diagonal contributions proportional to the square of each operator coefficient are listed. 

It is noteworthy the drastic change in the interference pattern due to the inclusion of tree-level electroweak contributions \cite{ElFaham2022,Darme2021}. First, the electroweak contribution to the interference term is larger than the corresponding QCD one.
This can be understood from phase-space cancellations, similar to the di-boson production case found in Ref. \cite{MMaltoni2021}.
Furthermore, the QCD and EW interference also have opposite signs, which leads to further suppression of the interference terms. 
The scale uncertainties are also large (50–70\%) such that only leading effects can meaningfully be constrained, as we  will see later. 

We notice that when we restrict our studies to the pure four-top component of the operators $\mathcal{O}_{QQ}^{(8)}$ and $\mathcal{O}_{QQ}^{(1)}$, a degeneracy arises from the tree-level relation
\begin{equation}
    \mathcal{O}_{QQ}^{(8)} \stackrel{tttt-\mathrm{only}}{=} \frac{1}{3} \mathcal{O}_{QQ}^{(1)}, 
\end{equation}
obtained by means of Fierz identities. The relation above is reflected in Table \ref{tab:tttt}, where the rows corresponding to the Wilson coefficients $c_{QQ}^8$ and $c_{QQ}^1$ are related by roughly a factor of three in the interference terms and of nine in the quadratic terms. This degeneracy can be lifted when combining these results with the top-pair production.

{ 
\renewcommand{\arraystretch}{1.5}
\begin{table}
\begin{center}
{\footnotesize
\begin{tabular}{|c|c|c|c|c|}
\hline
\multirow{2}{*}{$c_i$} & \multicolumn{3}{c|}{$\mathcal{O} (\Lambda^{-2})$} & \multirow{2}{*}{$\mathcal{O} (\Lambda^{-4})$}\tabularnewline
\cline{2-4}
 & $\mathcal{O}(\alpha_s^3 \Lambda^{-2})$ & $\mathcal{O}(\alpha_s^2 \alpha \, \Lambda^{-2})$ & Total $\mathcal{O}(\Lambda^{-2})$ &  \tabularnewline
\hline 
\hline 
\multirow{1}{*}{$c_{tt}^1$} & \multirow{1}{*}{ $0.552^{+71\%}_{-39\%}$ } & \multirow{1}{*}{ $-1.74_{-27\%}^{+42\%}$ } & \multirow{1}{*}{$-1.24_{-25\%}^{+36\%} $} &\multirow{1}{*}{ $4.25^{+73\%}_{-39\%}$ }    \tabularnewline

\multirow{1}{*}{$c_{QQ}^1$} & \multirow{1}{*}{  $0.272^{+71\%}_{-39\%}$ } &  \multirow{1}{*}{ $-0.991_{-27\%}^{+42\%}$ } & \multirow{1}{*}{$-0.737_{-25\%}^{+38\%}$} & \multirow{1}{*}{$1.06^{+73\%}_{-39\%}$}     \tabularnewline

\multirow{1}{*}{$c_{QQ}^8$}& \multirow{1}{*}{ $0.0889^{+71\%}_{-39\%}$ } & \multirow{1}{*}{ $-0.329_{-27\%}^{+43\%}$ }  & \multirow{1}{*}{$-0.245_{-25\%}^{+38\%}$} & \multirow{1}{*}{ $0.118^{+73\%}_{-39\%} $}       \tabularnewline

\multirow{1}{*}{$c_{Qt}^1$}& \multirow{1}{*}{ $-0.0392^{+71\%}_{-39\%}$ } & \multirow{1}{*}{ $0.747_{-26\%}^{+42\%}$ }  & \multirow{1}{*}{$0.745_{-27\%}^{+42\%}$} & \multirow{1}{*}{$1.44^{+73\%}_{-39\%} $}        \tabularnewline

\multirow{1}{*}{$c_{Qt}^8$}& \multirow{1}{*}{ $0.282^{+70\%}_{-39\%}$ } &  \multirow{1}{*}{$-0.605_{-27\%}^{+42\%}$}   & \multirow{1}{*}{$-0.322_{-22\%}^{+30\%}$} & \multirow{1}{*}{$0.349^{+73\%}_{-39\%} $}       \tabularnewline
\hline 
\end{tabular}
}
\caption{  Tree-level contributions (in fb.) of the four-heavy-quark operators  to the four-top production at the LHC $\sqrt{s}=13$ TeV and with $c_i/\Lambda^2=1$ TeV$^{-2}$, organized according to Eq. \eqref{eq:4topxsec}. The $\mathcal{O} (\Lambda^{-4})$ indicates quadratic contributions purely, without the interference effects at $\mathcal{O} (\Lambda^{-2})$. The SM at NLO considering QCD+EW corrections is $\sigma_{ t\bar{t}t\bar{t}}^{\mathrm{SM}} = 11.97_{-21\%}^{+18\%} $ fb \cite{Pagani2018}. Uncertainties are obtained by varying the renormalization scale by a factor of 2 above and below the central value.}
\label{tab:tttt}
\end{center}
\end{table}
}

In Fig. \ref{fig:4t_Minv}, we present the tree-level contributions from the SM and from the new physics to the invariant-mass distribution of the four-top final state. The dominant terms of the SM amplitude are of the order $\mathcal{O}(\alpha_s^2)$ and $\mathcal{O}(\alpha_s \alpha)$. They interfere with the new physics amplitudes to generate the linear contributions presented in Fig.  \ref{fig:4t_int_Minv}. For the quadratic contributions, only diagonal terms of  $c_ic_j$ (with $i=j$) are plotted in Fig. \ref{fig:4t_sq_Minv}. Following the conventions described at the beginning of section \ref{sec:ttproduction}, the renormalization and factorization scales are set as $\mu_r=\mu_f=2m_t=345$ GeV. The invariant-mass distributions linear in the $c_i$ peak around  1.3 TeV. On the other hand, the square contributions tend to dominate in the high-energy regime. As presented in Fig. \ref{fig:4t_sq_Minv}, the peaks tend to be at around 1.7 TeV and fall off gradually, slower than the corresponding linear counterparts. Remarkably, the square of the new physics effects is comparable to the SM and even larger at high enough energies for $c_i/\Lambda^2\sim 1$ TeV${}^{-2}$. This is worrisome when we consider the SMEFT expansion validity, indicating that either we impose severe cuts in the phase-space or we require only stringent bounds to be valid.

By comparison of the results for the top-pair invariant-mass distribution in Fig. \ref{fig:4heavy_Minv} with the corresponding invariant-mass distributions for the four-top in Fig. \ref{fig:4t_Minv}, we notice that the interference terms dominate in the case of the top-pair, whereas the quadratic terms are more important in the four-top process. However, the quadratic terms in top pair production are either PDF suppressed or $\alpha_s$ suppressed (2-loop for gluon annihilation, for example), with the latter not included here. In addition, the peaks of the invariant-mass distributions are different, being close to threshold for the top-pair and above 1 TeV for the four-top production. These facts will have an impact on our sensitivity analysis and validity discussion presented in the next section.

\section{Analysis and Results}
\label{sec:pheno}

{ 
\setlength{\tabcolsep}{4.7pt}
\renewcommand{\arraystretch}{1.48}
\begin{table}[t]
\begin{center}
{\small
\begin{tabular}{|c|c|c|c|c|c|c|}

\hline
Proc. & Tag & $\sqrt{s}$, $\mathcal{L}$ & Final state & Observable & $n_{\mathrm{dat}}$ & Ref. \tabularnewline
\hline
\hline
\multirow{6}{*}{$t\bar{t}$} & CMS$_{tt}$-1 & 13 TeV, 2.3 fb$^{-1}$  & lepton+jets     & $d\sigma/dm_{t\bar{t}}$ & 8  &\cite{CMSttdiff2017}  \tabularnewline
& CMS$_{tt}$-2 & 13 TeV, 35.8 fb$^{-1}$ & lepton+jets & $d\sigma/dm_{t\bar{t}}$ & 10  & \cite{CMSttdiff2018a} \tabularnewline
& CMS$_{tt}$-3 & 13 TeV, 2.1 fb$^{-1}$  & dilepton     & $d\sigma/dm_{t\bar{t}}$ & 6   &\cite{CMSttdiff2018b}  \tabularnewline
& CMS$_{tt}$-4 & 13 TeV, 35.9 fb$^{-1}$ & dilepton & $d\sigma/dm_{t\bar{t}}$ & 7  & \cite{CMSttdiff2019} \tabularnewline
& ATLAS$_{tt}$ & 13 TeV, 36.1 fb$^{-1}$ & lepton+jets & $d\sigma/dm_{t\bar{t}}$ & 9   & \cite{ATLASttdiff2019}\tabularnewline
& HL-LHC &  14 TeV, 3 ab$^{-1}$ & Total & $d\sigma/dm_{t\bar{t}}$ &  24 &  \tabularnewline
\hline
\hline
\multirow{6}{*}{$t\bar{t}t\bar{t}$} & CMS$_{4t}$-1 & 13 TeV, 35.9 fb$^{-1}$  & Two same-sign or multi-leptons & $\sigma_{\mathrm{Tot}} (t\bar{t}t\bar{t})$ & 1 & \cite{CMStttt2018} \tabularnewline
& CMS$_{4t}$-2 & 13 TeV, 137 fb$^{-1}$  & Two same-sign or multi-leptons  & $\sigma_{\mathrm{Tot}} (t\bar{t}t\bar{t})$ & 1 & \cite{CMStttt2020} \tabularnewline
& CMS$_{4t}$-3 & 13 TeV, 138 fb$^{-1}$  & Two same-sign or multi-leptons  & $\sigma_{\mathrm{Tot}} (t\bar{t}t\bar{t})$ & 1 & \cite{CMStttt2023} \tabularnewline
& ATLAS$_{4t}$-1 & 13 TeV, 139 fb$^{-1}$  & Two same-sign or multi-leptons & $\sigma_{\mathrm{Tot}} (t\bar{t}t\bar{t})$ & 1 &  \cite{ATLAStttt2020} \tabularnewline
& ATLAS$_{4t}$-2 & 13 TeV, 140 fb$^{-1}$  & Two same-sign or multi-leptons & $\sigma_{\mathrm{Tot}} (t\bar{t}t\bar{t})$ & 1 &  \cite{ATLAStttt2023} \tabularnewline
& HL-LHC &  14 TeV, 3 ab$^{-1}$ & Total & $d\sigma/dm_{t\bar{t}t\bar{t}}$ &  11 &  \tabularnewline
\hline

\end{tabular}
}
\caption{Experimental measurements of top-pair (Top block of the table) and four-top production (Bottom block of the table) at the LHC considered in the analysis of section \ref{sec:pheno}. The second column shows the label used to present the results obtained from the corresponding dataset. Additionally, we present the setup used in our simulations regarding the HL-LHC. }
\label{tab:datasets}
\end{center}
\end{table}
}

In this section, we present the analysis of the constraining reach of the $pp \rightarrow t\bar{t}$ and $pp\rightarrow t\bar{t} t\bar{t}$ processes on the four-heavy-quark operators. The theoretical predictions are computed with the setup presented at the beginning of section \ref{sec:calculation_Frame}.

Our sensitivity study is based on the fit of the $\chi^2$-distribution. The 95\% confidence level (CL) bounds on the effective operators couplings are obtained by using the datasets listed in Table \ref{tab:datasets}. This analysis has been done at the level of the top quark, thus we do not consider experimental reports given only at the level of the decay products of the top. We also note that the recent first measurement of the total cross-section for the top-pair production at $\sqrt{s}=13.6$ TeV~\cite{CMSttInclusive2023} provides a new different data point. However this measurement is not considered in this analysis as inclusive cross-sections are less sensitive to the four-heavy-quark operators due to the cancellations among different kinematical regions. In fact, we checked that the addition of this measurement does not lead to substantial modifications to the bounds here presented. We construct the $\chi^2$-distribution depending on the set of Wilson coefficients $c_i=\{c_{Qt}^{1},c_{tt}^1,c_{Qt}^{8},c_{QQ}^{1},c_{QQ}^{8}\}$ as
\begin{align}
    \chi^2_i \Bigl( \frac{c_i}{\Lambda^2} \Bigr) & = \sum_{\mathrm{Bins}} \frac{\left(\mathrm{O}_{\mathrm{SMEFT}}\bigl( \frac{c_i}{\Lambda^2} \bigr)  -\mathrm{O}_{\mathrm{Exp}}\right)^2}{(\delta \mathrm{O})^2}, \label{eq:chi2}
\end{align}
where the observable O can be, for instance, the invariant-mass differential distribution or the total cross-section of the $t\bar{t}$ or the $t\bar{t}t\bar{t}$ processes.  The errors of the theoretical predictions considered in this analysis originating from PDFs and scale uncertainties are not considered in the total uncertainties entering the $\chi^2$-distribution
since they are much smaller than the reported uncertainties from the measurements  ($\delta \mathrm{O} = \delta \mathrm{O}_{\mathrm{Exp}}$) in Table \ref{tab:datasets}. 
We define the Best Fit Point (BFP) as the values of the coefficients that minimize the total $\chi^2$-distribution, where:
\begin{equation}
\chi^2 = \sum_i  \chi^2_i \Bigl( \frac{c_i}{\Lambda^2} \Bigr).
\end{equation}
For top-pair production, CMS reports results with Gaussian errors, which we consider as they are published, while in the case of ATLAS we keep only the Gaussian errors, neglecting the small non-Gaussian uncertainties.
The measured total cross-sections of the four-top production are reported with non-Gaussian uncertainties, and in this case, we shift the cross-section in such a way that the error bands are symmetric, which is sufficient for the goals of our analysis. Finally, we assume that all uncertainties are not correlated.  Experimental collaborations have put efforts in the recent years on how to handle correlations between different processes, however such correlations are not yet available in the experimental results and thus estimates on their effects are beyond the scope of this work.

\begin{figure}[t]
    \centering
    \resizebox{0.7\columnwidth}{!}{\includegraphics{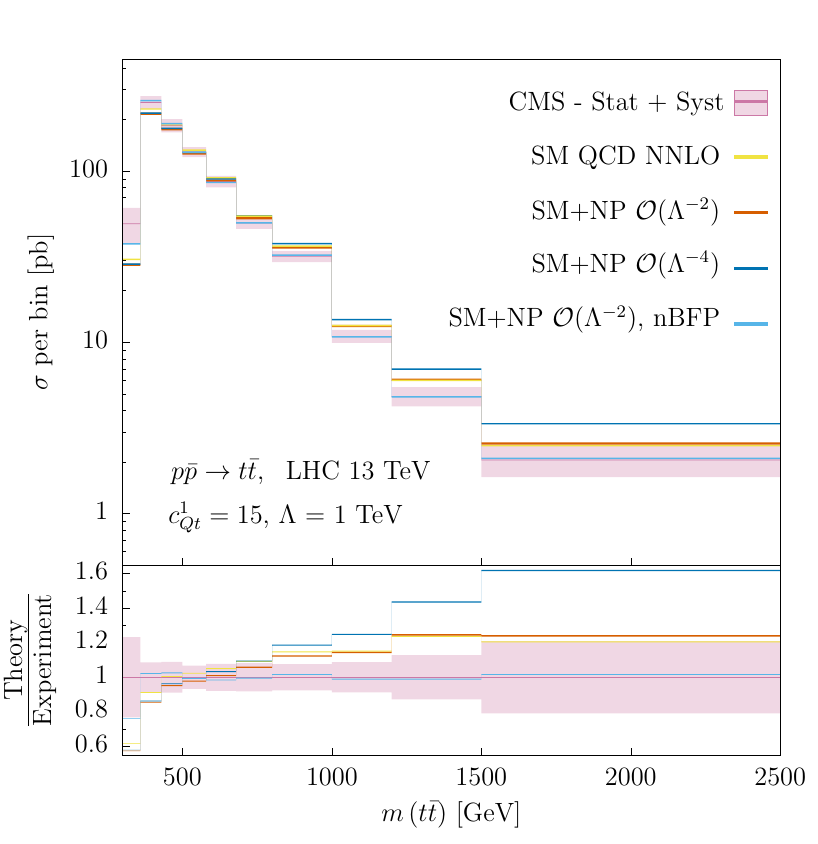}}
    \caption{Invariant-mass distribution of the top-pair production for new physics parametrized by the $\mathcal{O}_{Qt}^{(1)}$ operator at the linear and quadratic orders in the effective expansion, in red and in blue respectively. The effects of the coefficients set near to the Best Fit Point (nBFP) are also presented, in light blue. The SM prediction at NNLO-QCD order is shown in yellow.  The experimental data corresponds to measurements with 35.8 fb$^{-1}$ of integrated luminosity at CMS (CMS$_{tt}$-2 in Table \ref{tab:datasets}) and is shown in pink. The ratio between the different theoretical predictions and the CMS value is displayed in the inset at the bottom.  }
    \label{fig:expCMS2}
\end{figure}

The theoretical computation of the observable $\mathrm{O}_{\mathrm{SMEFT}}$  in Eq. \eqref{eq:chi2} is organized as follows
\begin{align}
   \mathrm{O}_{\mathrm{SMEFT}}\Bigl( \frac{c_i}{\Lambda^2} \Bigr) & = \mathrm{O}_{\mathrm{SM}} + \sum_i a_i \frac{c_i}{\Lambda^2} + \sum_{ij} b_{ij} \frac{c_i c_j}{\Lambda^4}, \label{eq:observExpansion}
\end{align}
so that the bounds at the interference order ($\mathcal{O}(\Lambda^{-2})$) in the tables below refer to numbers obtained from a truncation up to the second term in the right-hand side of the Eq. \eqref{eq:observExpansion}, while bounds at the quadratic order ($\mathcal{O}(\Lambda^{-4})$) consider interference plus quadratic terms, including the quadratic off-diagonal elements $b_{ij}$. The latter terms arise from multiplying two diagrams with insertions at tree-level of effective operators $\mathcal{O}_i$, and from multiplying diagrams having insertions at tree-level with diagrams having insertions at one-loop. Contributions coming from the square of one-loop diagrams with one insertion are not included, since they are of order next-to-NLO (see section \ref{sec:quarkChannel}). Similarly, contributions of diagrams with double insertions are not taken into account as they only arise in diagrams that are bottom-initiated and so are expected to be suppressed by the corresponding PDFs. The bounds at the quadratic order are obtained to estimate the uncertainties and thus assess the stability of the EFT expansion for the bounds at the interference order.

\subsection{Fits to the measurements of the top-pair production}

To obtain the theoretical prediction of the SM to the top-pair production, $d\sigma_{\mathrm{SM}}/dm_{t\bar{t}}$, we perform the computations in \mg~at  QCD-NLO and use  $K$-factors at a differential level extracted from Ref. \cite{Mitov2017, Czakon2016} to account for NNLO effects. The SM prediction obtained by this procedure is in agreement (at the order of 3-4\%) within the error bands of the results from Ref. \cite{Tsinikos2017}, which contain the invariant-mass distributions with the same bin size as the experimental results of the dataset CMS$_{tt}$-4.  Since the analysis of Ref. \cite{Tsinikos2017} includes EW-NLO corrections in the SM, we use their predictions in the fit of the dataset CMS$_{tt}$-4.

To illustrate the sizable NLO contributions to the top-pair process, in Fig. \ref{fig:expCMS2} we show the invariant-mass distribution in the LHC 13 TeV. We consider the experimental results of the dataset CMS$_{tt}$-2, which has the largest number of bins, and compare them to the effects of the effective operators at linear and quadratic orders. We present the case of the Wilson coefficient $c_{Qt}^1=15$ with the others set to zero, for which the $t\bar{t}$  process is the most sensitive.  In the region between 1-1.5 TeV the interference and the SM bins are on top of each other, which is a consequence of the flip in the sign for the $c_{Qt}^1$ contributions in this phase-space region. The differential distribution obtained from near the best fit point (nBFP) with only interference terms is also shown in Fig. \ref{fig:expCMS2}, where the BFP is found at 
\begin{align}
& c_{tt}^1=116, \quad \quad \quad c_{Qt}^1=-64.9, \quad \quad \quad c_{QQ}^1=484\,(150), \nn \\
&\quad \quad   c_{Qt}^8=164\,(150), \quad \quad c_{QQ}^8=-1113\,(-150), \label{eq:5.2}
\end{align}
with $\Lambda=1$ TeV. By near the BFP, we mean that when the coefficients take large values regarding the effective expansion, we set them to $c_i=\pm 150$, presented in round brackets in Eq \eqref{eq:5.2}. This $c_i$ value, although arbitrary, is chosen such that the absurdly large values from the BFP are more sensible, but large enough to illustrate the capability of the effective operators to fit data. Finally, the SM prediction at QCD-NNLO order is also included, which seems to present a different shape from the one measured by CMS. In particular, strong deviations are observed in the first bin. The issue of the first bin has been addressed in Ref. \cite{Ellis2021}, where the effects of the $\mathcal{O}_G$ operator are discussed, which can bring the theoretical predictions closer to the measured value without spoiling the tail behaviour. However, more stringent bounds on $c_G$ are found from multijet data \cite{Hirschi2018,Krauss2017,Goldouzian2020} (at the order $\mathcal{O}(\Lambda^{-4})$), suggesting that  this operator alone cannot fully parametrize this apparent deviation near threshold. Despite the diverse shapes of the four-top operators contribution to top pair invariant mass, we can only partially improve the fit, suggesting that, even with all the dimension-six operators included, this first bin deviation might be hard to accommodate in the SMEFT. However, a global fit is needed to get a definitive answer. 

A final note regarding the datasets: the reported values from the measurement CMS$_{tt}$-4 do not agree at 95\% CL with our best prediction of the SM. For the latter, we use the results from Ref.~\cite{Tsinikos2017}, where predictions at  NNLO-QCD and NLO-EW are provided with the same bin size as those used in the CMS analysis. We notice that the tension resides in the first bin. A first possible explanation is that the theoretical predictions do not include resummation of threshold logarithms and small-mass logarithms. However, this option seems to be discarded as resummation effects are not large enough \cite{Zaro2020}.

{ 
\setlength{\tabcolsep}{2pt}
\renewcommand{\arraystretch}{1.48}
\begin{table}[t]
\begin{center}
{\tiny
\begin{tabular}{|c|c|c|c|c|c|c|c|c|}
\cline{4-9}
    \multicolumn{3}{c|}{} & \multicolumn{1}{c}{CMS$_{tt}$-1 } & \multicolumn{1}{c}{CMS$_{tt}$-2}& \multicolumn{1}{c}{CMS$_{tt}$-3} & \multicolumn{1}{c}{CMS$_{tt}$-4} & \multicolumn{1}{c|}{ATLAS$_{tt}$} & \multicolumn{1}{c|}{Combined} \tabularnewline
\hline 
\hline 
        \multirow{4}{*}{$c_{tt}^1$} & \multirow{2}{*}{Ind.} &  $\mathcal{O}(\Lambda^{-2})$ &  \multicolumn{1}{c}{\multirow{1}{*}{ $[-148 , 64.4 ]$ }} & \multicolumn{1}{c}{\multirow{1}{*}{ $ [-58.9 , 0.99] $ }}  & \multicolumn{1}{c}{\multirow{1}{*}{ $ [-129 ,  332] $ }} & \multicolumn{1}{c}{ $ [ -56.4, -0.81 ] $ } & \multicolumn{1}{c|}{ $ [-26.4 , 52.2] $ } & \multicolumn{1}{c|}{ $ [-28.1 , 7.16] $ } \tabularnewline
        
        &  & $\mathcal{O}(\Lambda^{-4})$ & \multicolumn{1}{c}{\multirow{1}{*}{ $[-148 , 64.4 ]$ }} & \multicolumn{1}{c}{\multirow{1}{*}{ $ [-58.9 , 0.99] $ }}  & \multicolumn{1}{c}{\multirow{1}{*}{ $ [-129 ,  332] $ }} & \multicolumn{1}{c}{ $ [ -56.4, -0.81 ] $ } & \multicolumn{1}{c|}{ $ [-26.4 , 52.2] $ }  & \multicolumn{1}{c|}{ $ [- 28.1 , 7.16] $ }  \tabularnewline

        &  \multirow{2}{*}{Marg.} & \multirow{2}{*}{$\mathcal{O}(\Lambda^{-4})$} & \multicolumn{1}{c}{\multirow{2}{*}{ $[- 122, 3.22]$ }} & \multicolumn{1}{c}{\multirow{1}{*}{ $ [- 50.8, -10.8 ] $ }}  & \multicolumn{1}{c}{\multirow{2}{*}{ - }} &        \multicolumn{1}{c}{\multirow{2}{*}{ - }} & \multicolumn{1}{c|}{\multirow{2}{*}{ $ [- 232, 129 ] $ }}  & \multicolumn{1}{c|}{\multirow{2}{*}{ $ [- 48.0, 2.83] $ }}  \tabularnewline
        
        & &  & \multicolumn{1}{c}{} & \multicolumn{1}{c}{\multirow{1}{*}{ $ \cup \: [4.55, 255] $}}   & \multicolumn{1}{c}{}  & \multicolumn{1}{c}{}  & \multicolumn{1}{c|}{}  & \tabularnewline
\hline
\hline
        \multirow{3}{*}{$c_{QQ}^1$} & \multirow{2}{*}{Ind.} & $\mathcal{O}(\Lambda^{-2})$ & \multicolumn{1}{c}{\multirow{1}{*}{  $[-292,139]$ }} & \multicolumn{1}{c}{\multirow{1}{*}{ $ [-107 , 2.17 ] $ }}  & \multicolumn{1}{c}{\multirow{1}{*}{ $ [-335 , 462] $ }} & \multicolumn{1}{c}{ $ [-109 , -1.66 ] $ } & \multicolumn{1}{c|}{ $ [94.3 , -51.3] $ }  &  \multicolumn{1}{c|}{ $ [-51.7 , 14.9] $ } \tabularnewline

        &  & $\mathcal{O}(\Lambda^{-4})$ & \multicolumn{1}{c}{\multirow{1}{*}{ $[-18.2,16.2]$ }} & \multicolumn{1}{c}{\multirow{1}{*}{ $ [-3.04 , 1.27] $ }}  & \multicolumn{1}{c}{\multirow{1}{*}{ $ [-21.4 , 21.1] $ }} & \multicolumn{1}{c}{ - } & \multicolumn{1}{c|}{ $ [-19.7 , 18.1] $ }  & \multicolumn{1}{c|}{ $ [-5.72 , 4.29] $ }  \tabularnewline

        &  \multirow{1}{*}{Marg.} & \multirow{1}{*}{$\mathcal{O}(\Lambda^{-4})$ } & \multicolumn{1}{c}{\multirow{1}{*}{ $[- 12.7, 13.1]$ }} & \multicolumn{1}{c}{\multirow{1}{*}{ $ [- 15.3, 12.1 ]   $}}   & \multicolumn{1}{c}{\multirow{1}{*}{ - }} & \multicolumn{1}{c}{\multirow{1}{*}{ - }} & \multicolumn{1}{c|}{\multirow{1}{*}{ $ [- 26.5, 24.0  ] $ }}  & \multicolumn{1}{c|}{\multirow{1}{*}{ $ [- 8.05, 4.95] $ }}  \tabularnewline
\hline
\hline
         \multirow{3}{*}{$c_{QQ}^8$} &  \multirow{2}{*}{Ind.} &  $\mathcal{O}(\Lambda^{-2})$ & \multicolumn{1}{c}{\multirow{1}{*}{ $[-323, 126]$ }} & \multicolumn{1}{c}{\multirow{1}{*}{ $ [-157 , 1.74] $ }}  & \multicolumn{1}{c}{\multirow{1}{*}{ $ [-575 , 334] $ }} & \multicolumn{1}{c}{ $ [-119, -2.53 ] $ } & \multicolumn{1}{c|}{ $ [-60.1 , 105] $ }  & \multicolumn{1}{c|}{ $ [-66.9 , 15.0] $ } \tabularnewline
 
        & &  $\mathcal{O}(\Lambda^{-4})$  & \multicolumn{1}{c}{\multirow{1}{*}{ $[-43.0,32.1]$ }} & \multicolumn{1}{c}{\multirow{1}{*}{ $ [-11.9 , 1.52] $ }}  & \multicolumn{1}{c}{\multirow{1}{*}{ $ [-48.9 , 43.1] $ }} & \multicolumn{1}{c}{ - } & \multicolumn{1}{c|}{ $ [-40.2 , 29.2] $ }  & \multicolumn{1}{c|}{ $ [-16.1 , 7.90] $ } \tabularnewline
 
        & \multirow{1}{*}{Marg.} & \multirow{1}{*}{$\mathcal{O}(\Lambda^{-4})$} & \multicolumn{1}{c}{\multirow{1}{*}{ $[- 31.5, 26.7]$ }} & \multicolumn{1}{c}{\multirow{1}{*}{ $ [- 316, 163] $ }}  & \multicolumn{1}{c}{\multirow{1}{*}{ - }} & \multicolumn{1}{c}{\multirow{1}{*}{ - }} & \multicolumn{1}{c|}{\multirow{1}{*}{ $ [- 75.2, 68.8 ] $ }}  & \multicolumn{1}{c|}{\multirow{1}{*}{ $ [- 18.7 , 14.8] $ } } \tabularnewline
\hline
\hline
        \multirow{3}{*}{$c_{Qt}^1$} & \multirow{2}{*}{Ind.} &   $\mathcal{O}(\Lambda^{-2})$ & \multicolumn{1}{c}{\multirow{1}{*}{ $[-53.7,78.8]$ }} & \multicolumn{1}{c}{\multirow{1}{*}{ $ [-3.23 , 11.4] $ }}  & \multicolumn{1}{c}{\multirow{1}{*}{ $ [-451 , 28.0] $ }} & \multicolumn{1}{c}{ - } & \multicolumn{1}{c|}{ $ [-33.2 , 29.0] $ }  & \multicolumn{1}{c|}{ $ [-11.4 , 12.7] $ } \tabularnewline

        &  &  $\mathcal{O}(\Lambda^{-4})$ & \multicolumn{1}{c}{\multirow{1}{*}{ $[-15.9,17.7]$ }} & \multicolumn{1}{c}{\multirow{1}{*}{ $ [ -1.52, 2.32] $ }}  & \multicolumn{1}{c}{\multirow{1}{*}{ $ [-30.4 , 14.8] $ }} & \multicolumn{1}{c}{ - } & \multicolumn{1}{c|}{ $ [-20.7 , 12.3] $ }  & \multicolumn{1}{c|}{ $ [-4.94 , 4.80] $ } \tabularnewline

        & \multirow{1}{*}{Marg.} & \multirow{1}{*}{$\mathcal{O}(\Lambda^{-4})$} & \multicolumn{1}{c}{\multirow{1}{*}{ $[- 6.79, 18.2]$ }} & \multicolumn{1}{c}{\multirow{1}{*}{ $ [-  50.3, 30.2]  $ }}  & \multicolumn{1}{c}{\multirow{1}{*}{ - }} & \multicolumn{1}{c}{\multirow{1}{*}{ - }} & \multicolumn{1}{c|}{\multirow{1}{*}{ $ [-  43.8 , 24.7 ] $ }}  & \multicolumn{1}{c|}{\multirow{1}{*}{ $ [- 6.33, 7.24] $ }} \tabularnewline
\hline
\hline
        \multirow{4}{*}{$c_{Qt}^8$} & \multirow{2}{*}{Ind.}   & $\mathcal{O}(\Lambda^{-2})$ & \multicolumn{1}{c}{\multirow{1}{*}{ $[-177,69.5]$  }} & \multicolumn{1}{c}{\multirow{1}{*}{ $ [-100 , 0.88] $ }}  & \multicolumn{1}{c}{\multirow{1}{*}{ $ [-322 , 64.3] $ }} & \multicolumn{1}{c}{ $ [ -95.8, -0.77 ] $ } & \multicolumn{1}{c|}{ $ [-32.3 , 44.9] $ }  & \multicolumn{1}{c|}{ $ [-44.6 , 5.92] $ } \tabularnewline
 
        &  &  $\mathcal{O}(\Lambda^{-4})$   & \multicolumn{1}{c}{\multirow{1}{*}{ $[-55.5, 31.1]$ }} & \multicolumn{1}{c}{\multirow{1}{*}{ $ [-26.0 , 0.85] $ }}  & \multicolumn{1}{c}{\multirow{1}{*}{ $ [-72.8 , 34.2] $ }} & \multicolumn{1}{c}{ $ [ -27.3, -0.79] $ } & \multicolumn{1}{c|}{ $ [-59.7 , 25.7] $ }  &  \multicolumn{1}{c|}{ $ [-31.4 , 5.02] $ } \tabularnewline

        & \multirow{2}{*}{Marg.} & \multirow{2}{*}{$\mathcal{O}(\Lambda^{-4})$} & \multicolumn{1}{c}{\multirow{2}{*}{ $[- 35.6, 25.2]$ }} & \multicolumn{1}{c}{\multirow{1}{*}{ $ [- 142, -6.50 ] $ }}  & \multicolumn{1}{c}{\multirow{2}{*}{ - }} & \multicolumn{1}{c}{\multirow{2}{*}{ - }} & \multicolumn{1}{c|}{\multirow{2}{*}{ $ [- 100, 58.2] $ }}  & \multicolumn{1}{c|}{\multirow{2}{*}{ $ [- 23.7, 1.77 ] $ }}  \tabularnewline

        & &  & \multicolumn{1}{c}{} & \multicolumn{1}{c}{\multirow{1}{*}{ $ \cup \: [2.21, 82.5] $}}   & \multicolumn{1}{c}{}  & \multicolumn{1}{c}{}  & \multicolumn{1}{c|}{}  & \tabularnewline
\hline 
\end{tabular}
}
\caption{ The 95\% CL bounds (assuming $\Lambda = 1$ TeV) for the coefficients of the four-heavy-quark operators in the process $pp \rightarrow t\bar{t}$ individual and marginalized. The intervals are presented for the different datasets introduced in Table \ref{tab:datasets}. The missing entries correspond to cases where the SM does not provide a good fit to the data. Notice that $\mathcal{O}(\Lambda^{-4})$ includes terms of the order $\mathcal{O}(\Lambda^{-2})$ (see Eq. \eqref{eq:observExpansion})}
\label{tab:boundstt02}
\end{center}
\end{table}
}

{ 
\setlength{\tabcolsep}{3pt}
\renewcommand{\arraystretch}{1.48}
\begin{table}[b]
\begin{center}
{\tiny
\begin{tabular}{|c|c|c|c|c|c|c|}
\cline{2-7}
    \multicolumn{1}{c|}{} & \multicolumn{1}{c}{CMS$_{tt}$-1 } & \multicolumn{1}{c}{CMS$_{tt}$-2}& \multicolumn{1}{c}{CMS$_{tt}$-3} & \multicolumn{1}{c}{CMS$_{tt}$-4} & \multicolumn{1}{c|}{ATLAS$_{tt}$} & \multicolumn{1}{c|}{Combined} \tabularnewline
\hline 
\hline 
    \multirow{1}{*}{$c_1$}   & \multicolumn{1}{c}{\multirow{1}{*}{ $ [- 48.1, 60.7] $ }}& \multicolumn{1}{c}{\multirow{1}{*}{ $ [- 8.96, 49.9] $ }}  & \multicolumn{1}{c}{\multirow{1}{*}{ $ [-40.7 , 1.92] $ }} & \multicolumn{1}{c}{ $ [-0.27 , 44.1] $ } & \multicolumn{1}{c|}{ $ [ 8.39, 30.5] $ } & \multicolumn{1}{c|}{ $ [ -28.4, -4.90] $ }  \tabularnewline

    \multirow{1}{*}{$c_2$}   & \multicolumn{1}{c}{\multirow{1}{*}{ $ [-24.1, 90.4] $ }} & \multicolumn{1}{c}{\multirow{1}{*}{ $ [-47.8, 21.1 ] $ }}  & \multicolumn{1}{c}{\multirow{1}{*}{ $ [-86.7 , 160] $ }} & \multicolumn{1}{c}{ $ [-26.6 , 33.3 ] $ } & \multicolumn{1}{c|}{ $ [- 44.6 , 1.53 ] $ } & \multicolumn{1}{c|}{ $ [ -17.2, 19.1] $ }  \tabularnewline

    \multirow{1}{*}{$c_3$}   & \multicolumn{1}{c}{\multirow{1}{*}{ $ [-282, 526] $ }} & \multicolumn{1}{c}{\multirow{1}{*}{ $ [- 8.97, 425] $ }}  & \multicolumn{1}{c}{\multirow{1}{*}{ $ [-4.72, 460] $ }} & \multicolumn{1}{c}{ $ [-321, -1.91 ] $ } & \multicolumn{1}{c|}{ $ [- 144, 227] $ } & \multicolumn{1}{c|}{ $ [ -261, 5.60] $ }  \tabularnewline

    \multirow{1}{*}{$c_4(\cdot 10^3)$}   & \multicolumn{1}{c}{\multirow{1}{*}{ $ [- 17.1 ,  17.8 ] $ }}  & \multicolumn{1}{c}{\multirow{1}{*}{ $ [- 4.58 , 4.09 ] $ }}  & \multicolumn{1}{c}{\multirow{1}{*}{ $ [- 12.5 , 19.3 ] $ }} & \multicolumn{1}{c}{ $ [-0.35 , 0.078] $ } & \multicolumn{1}{c|}{ $ [- 35.0 , 53.3] $ } & \multicolumn{1}{c|}{ $ [ -0.14, 0.38] $ }  \tabularnewline
    
    \multirow{1}{*}{$c_5(\cdot 10^4)$}   & \multicolumn{1}{c}{\multirow{1}{*}{ $ [- 31.0,   23.1 ] $ }}  & \multicolumn{1}{c}{\multirow{1}{*}{ $ [- 32.6,  32.3 ] $ }}  & \multicolumn{1}{c}{\multirow{1}{*}{ $ [-21.7 , 15.4] $ }} & \multicolumn{1}{c}{ $ [-23.5 , 23.5 ] $ } & \multicolumn{1}{c|}{ $ [- 30.1 ,  29.8] $ }  & \multicolumn{1}{c|}{ $ [ -0.048, 0.46] $ }  \tabularnewline
\hline
\end{tabular}
}
\caption{ The 95\% CL marginalized bounds (assuming $\Lambda = 1$ TeV) for the coefficients of the four-heavy operators at linear order in the diagonal basis in the process $pp \rightarrow t\bar{t}$. The marginalized intervals are presented for the different datasets introduced in Table \ref{tab:datasets}}
\label{tab:boundsttDiag}
\end{center}
\end{table}
}

The individual 95\% CL bounds on the Wilson coefficients of the four-heavy-quark operators obtained from the $t\bar{t}$ datasets of Table \ref{tab:datasets} are given in Table \ref{tab:boundstt02}. In order to avoid possible correlations between datasets, the last column stands for the bounds obtained by considering only the datasets CMS$_{tt}$-2 (which is an update of CMS$_{tt}$-1) , CMS$_{tt}$-4 (which is an update of CMS$_{tt}$-3 )  and ATLAS$_{tt}$. The bounds from these differential measurements are expected to be in general more stringent than the bounds obtained from the inclusive measurements because of the increasing sensitivity in the high $m_{t\bar{t}}$ region. For the EFT assumption to be valid, the new physics scale is bound to $\Lambda \gtrsim 500$ GeV. Assuming coefficients $c_i$ of order one, the bounds are, in general, of order a few hundreds of GeV. Only the coefficient $c_{Qt}^1$ presents the tightest bounds at the interference level. The other coefficients are poorly constrained at the interference level. The constraints are much tighter  when quadratic contributions are included for $c^1_{QQ}$ and $c^8_{QQ}$, which raises again the validity question. It should be kept in mind also that the dominant $\Lambda^{-4}$ could come from terms not included here, such as the square of one-loop diagrams with insertions of  four-heavy-quark operators. Such diagrams lead to contributions that  are smaller or, at the most, of the same order of the square of the corresponding bottom-induced sub-process for $c_i/\Lambda^{2}= 1$ TeV$^{-2}$. Moreover, the energy-growth of those contributions to the cross-section is at the most quadratic in the center-of-mass energy. We have checked that for $c_i/\Lambda^{2} = 1$ TeV$^{-2}$ it is safe to disregard such terms within the cuts considered in our analysis. However, the resulting bounds in Table \ref{tab:boundstt02} are in most of the cases larger than one, signalling that there might be regions where it is not safe to disregard the diagrams in question. In addition, the effects of squares of one-loop diagrams could have a larger impact when considering HL-LHC measurements for which the reach could be above 5 TeV in the invariant-mass distribution. Finally, in the five-flavour scheme, a complete consideration of the quadratic terms requires the inclusion of diagrams with real emissions, which we also disregard as they cannot be computed yet in \mg. 

The marginalized bounds at quadratic order on the Wilson coefficients are also presented in Table \ref{tab:boundstt02}, for which we allow all the $c_i$ to vary at the same time. The allowed volumes in the parameter space of the Wilson coefficients are found by acceptance and rejection methods. In general, the results from the marginalized fit do not change drastically the individual bounds at the quadratic level, just widening slightly the allowed intervals. Finally, the missing entries marked with a dashed line are configurations for which we did not find a solution at $95\%$ CL due to the apparent discrepancy between the measurements and the SM prediction in the first bin. 
  
The intervals for a marginalized fit at the linear expansion can also be obtained. This fit presents strong flat directions, but nevertheless stringent bounds can be obtained in some  directions (linear combinations of parameters). In Table \ref{tab:boundsttDiag} the 95\% CL bounds are listed for the combinations $c_i$ with $i=1,...,5$ given by the change of basis
\begin{align}
\mathbf{c}' & =\mathcal{R}\cdot\mathbf{c}
\end{align}
with 
\begin{align}
\mathbf{c}^T  =  \left[c_{Qt}^{1},c_{tt}^1,c_{Qt}^{8},c_{QQ}^{1},c_{QQ}^{8}\right] \quad \mathrm{and} \quad \mathbf{c}'^T = \left[c_1,c_2,c_3,c_4,c_5 \right].
\end{align}
Given the fact that the $\chi^2$-distribution is a quadratic polynomial in the $c_i$ at the interference level, the rotation matrix is obtained by finding the eigenvectors of the matrix of coefficients of the quadratic terms. Hence, the $\mathcal{R}$ matrix is different for each dataset.  In particular, for the combination of datasets the rotation matrix has the form
\begin{align}
\mathcal{R}^{t\bar{t}}_{\mathrm{Combined}} &=  \left[\begin{array}{ccccc}
-0.99   & -0.012      & -0.12     & -0.062      & 0.0024          \\
0.097   & -0.70       & -0.55     & -0.33       & -0.31           \\
-0.072  & -0.51       & 0.79      & -0.32       & 0.073           \\
0.039   & -0.19       & -0.24     & -0.068      & 0.95            \\
-0.057  & -0.46       & 0.053     & 0.88        & -0.016          \\
\end{array}\right].
\end{align}
From this, we observe that the most constrained direction $c_1$ is close to the $c_{Qt}^{1}$ axis. In the diagonal basis, the BFP can be found at
\begin{equation}
c_1=-16.6, \quad   c_2=0.944, \quad    c_3=-128, \quad     c_4=123 \quad c_5 = 2038. 
\end{equation}
The rotation matrices for each of the datasets (listed in the appendix \ref{app:rotation_mat}) show that in most of the cases the two best constrained directions are close to the $c_{Qt}^{1}$ and $c_{tt}^{1}$ axis. 
Special care must be taken for the marginalized analysis for the datasets CMS$_{tt}$-3 and ATLAS$_{tt}$ since the values reported by the experimental collaborations are given as normalized distributions. The $\chi^2$-distribution for normalized distributions has an involved dependence on the Wilson coefficients, which can appear in the denominator. In these situations, the diagonalization approach described above is no longer valid, since such transformations on the $\chi^2$-distribution would reshape the allowed space of Wilson coefficients. We solve this issue by multiplying the normalized differential distributions by the total cross-section reported in Ref. \cite{Tsinikos2017}.

\subsection{Fits to the measurements of the four-top production}

As indicated in Table \ref{tab:datasets}, measurements of the $pp\rightarrow t\bar{t}t\bar{t}$ only consider inclusive cross-sections. Following Eq. \eqref{eq:observExpansion}, we take the SM and SMEFT predicitions as presented in Table \ref{tab:tttt}.

The individual 95\% CL bounds on the Wilson coefficients of the four-heavy-quark operators obtained from the $t\bar{t}t\bar{t}$ datasets are given in Table \ref{tab:boundstttt}. The last column stands for the bounds obtained from considering only datasets from a different final state and collaboration, \textit{i.e.} the results quoted in CMS$_{4t}$-2 (which is an update of CMS$_{4t}$-1 and ) and ATLAS$_{4t}$. The last column stands for the bounds obtained from considering only datasets CMS$_{4t}$-2 and ATLAS$_{4t}$-1, which are datasets from different collaborations and thus expected to be uncorrelated. The bounds are in a similar range as those from the $t\bar{t}$ process. In the best case, a bound of the order $\Lambda \gtrsim 500$ GeV for the new physics scale is found, although the bounds can get as weak as to constraint a few hundreds of GeV. Additionally, the bounds in Table \ref{tab:boundstttt} are much more stringent when quadratic contributions are included compared to bounds coming from only interference contributions. We also notice that the most recent measurements reported in \cite{CMStttt2023,ATLAStttt2023}, corresponding to CMS$_{4t}$-3 and ATLAS$_{4t}$-2, do not improve substantially the constraints when compared to previous measurements, as expected since the additional luminosity is only incremental.

{ 
\setlength{\tabcolsep}{4.7pt}
\renewcommand{\arraystretch}{1.4}
\begin{table}[t]
\begin{center}
{\scriptsize
\begin{tabular}{|c|c|c|c|c|c|c|c|}
\cline{3-8}
\multicolumn{2}{c}{} &  \multicolumn{1}{|c}{CMS$_{4t}$-1 } & \multicolumn{1}{c}{CMS$_{4t}$-2} & \multicolumn{1}{c}{CMS$_{4t}$-3} & \multicolumn{1}{c}{ATLAS$_{4t}$-1} & \multicolumn{1}{c}{ATLAS$_{4t}$-2} & \multicolumn{1}{c|}{Combined} \tabularnewline
\hline 
\hline 
        \multirow{2}{*}{$c_{tt}^1$} &  $\mathcal{O}(\Lambda^{-2})$   & \multicolumn{1}{c}{\multirow{1}{*}{ $ [-24.9 , 15.0] $ }} & \multicolumn{1}{c}{ $ [-9.44, 7.94 ] $ }  &  \multicolumn{1}{c}{ $ [-13.9, 4.15 ] $ } &  \multicolumn{1}{c}{ $ [-20.4 , 0.17] $ } &  \multicolumn{1}{c}{ $ [ -18.5, 0.63 ] $ } &   \multicolumn{1}{c|}{ $ [-9.42 , 0.12] $ } \tabularnewline

        & $\mathcal{O}(\Lambda^{-4})$   & \multicolumn{1}{c}{\multirow{1}{*}{ $ [-2.55 , 2.84] $ }} & \multicolumn{1}{c}{ $ [-1.52, 1.81 ] $ }  & \multicolumn{1}{c}{ $ [ -1.87, 2.16 ] $ } &  \multicolumn{1}{c}{ $ [-2.30 , 2.59] $ } &  \multicolumn{1}{c}{ $ [ -2.18, 2.47 ] $ } &  \multicolumn{1}{c|}{ $ [-1.52 , 1.81] $ } \tabularnewline
\hline
\hline
        \multirow{2}{*}{$c_{QQ}^1$}  &  $\mathcal{O}(\Lambda^{-2})$  & \multicolumn{1}{c}{\multirow{1}{*}{ $ [-41.8 , 25.2] $ }} & \multicolumn{1}{c}{ $ [-15.9 , 13.4 ] $ }   &  \multicolumn{1}{c}{ $ [ -23.3, 6.98 ] $ } &  \multicolumn{1}{c}{ $ [-34.3 , 0.28] $ } &  \multicolumn{1}{c}{ $ [ -31.1, 1.06 ] $ } &  \multicolumn{1}{c|}{ $ [-15.9 , 0.20] $ } \tabularnewline

        & $\mathcal{O}(\Lambda^{-4})$  & \multicolumn{1}{c}{\multirow{1}{*}{ $ [-5.06 , 5.75] $ }} & \multicolumn{1}{c}{ $ [ -2.99, 3.69 ] $ }   &  \multicolumn{1}{c}{ $ [ -3.7, 4.39 ] $ } & \multicolumn{1}{c}{ $ [-4.55, 5.24] $ } &  \multicolumn{1}{c}{ $ [ -4.32, 5.01 ] $ } &  \multicolumn{1}{c|}{ $ [-2.99 , 3.67] $ } \tabularnewline
\hline
\hline
        \multirow{2}{*}{$c_{QQ}^8$} &  $\mathcal{O}(\Lambda^{-2})$ & \multicolumn{1}{c}{\multirow{1}{*}{ $[-126, 75.8]$ }} & \multicolumn{1}{c}{\multirow{1}{*}{ $ [-47.8 , 40.2] $ }}  & \multicolumn{1}{c}{ $ [ -70.2, 21.0 ] $ } &  \multicolumn{1}{c}{ $ [-103 , 0.86] $ }  &  \multicolumn{1}{c}{ $ [ -93.6, 3.17 ] $ } &  \multicolumn{1}{c|}{\multirow{1}{*}{ $ [-47.7 , 0.60] $ }} \tabularnewline
 
        & $\mathcal{O}(\Lambda^{-4})$ & \multicolumn{1}{c}{\multirow{1}{*}{ $[-15.2,17.2]$ }} & \multicolumn{1}{c}{\multirow{1}{*}{ $ [-8.98 , 11.1] $ }}  & \multicolumn{1}{c}{ $ [ -11.1, 13.2 ] $ } &  \multicolumn{1}{c}{ $ [ -13.6, 15.7 ] $ }  &  \multicolumn{1}{c}{ $ [ -12.9, 15.0 ] $ } &  \multicolumn{1}{c|}{\multirow{1}{*}{ $ [-8.97 , 11.0] $ }} \tabularnewline
\hline
\hline
        \multirow{2}{*}{$c_{Qt}^1$}  & $\mathcal{O}(\Lambda^{-2})$ & \multicolumn{1}{c}{\multirow{1}{*}{ $[-24.9,41.4]$ }} & \multicolumn{1}{c}{\multirow{1}{*}{ $ [-13.2 , 15.7] $ }}  &  \multicolumn{1}{c}{ $ [  -6.9, 23.1 ] $ } & \multicolumn{1}{c}{ $ [-0.28 , 33.9 ] $ }  &  \multicolumn{1}{c}{ $ [  -1.04, 30.8 ] $ } &  \multicolumn{1}{c|}{\multirow{1}{*}{ $ [-0.20 , 15.7] $ }} \tabularnewline

        & $\mathcal{O}(\Lambda^{-4})$ & \multicolumn{1}{c}{\multirow{1}{*}{ $[-4.89,4.38]$ }} & \multicolumn{1}{c}{\multirow{1}{*}{ $ [ -3.12, 2.60] $ }}  &  \multicolumn{1}{c}{ $ [ -3.72, 3.21 ] $ } & \multicolumn{1}{c}{ $ [ -4.46 , 3.94 ] $ }  &  \multicolumn{1}{c}{ $ [ -4.26, 3.74 ] $ } &  \multicolumn{1}{c|}{\multirow{1}{*}{ $ [-3.12 , 2.60] $ }} \tabularnewline
\hline
\hline
        \multirow{2}{*}{$c_{Qt}^8$} & $\mathcal{O}(\Lambda^{-2})$ & \multicolumn{1}{c}{\multirow{1}{*}{ $[-95.7, 57.7]$  }} & \multicolumn{1}{c}{\multirow{1}{*}{ $ [-36.4 , 30.6] $ }}   &  \multicolumn{1}{c}{ $ [ -53.4, 16.0 ] $ } & \multicolumn{1}{c}{ $ [ -78.5 , 0.65 ] $ } &  \multicolumn{1}{c}{ $ [ -71.2, 2.42  ] $ } &  \multicolumn{1}{c|}{\multirow{1}{*}{ $ [-36.3 , 0.45] $ }} \tabularnewline
 
        & $\mathcal{O}(\Lambda^{-4})$ & \multicolumn{1}{c}{\multirow{1}{*}{ $[-8.95, 9.87]$ }} & \multicolumn{1}{c}{\multirow{1}{*}{ $ [-5.35 , 6.27] $ }}   & \multicolumn{1}{c}{ $ [ -6.57, 7.50 ] $ } &  \multicolumn{1}{c}{ $ [ -8.06, 8.98] $ } &  \multicolumn{1}{c}{ $ [ -7.66, 8.58 ] $ } &  \multicolumn{1}{c|}{\multirow{1}{*}{ $ [-5.34 , 6.27] $ }} \tabularnewline
\hline 
\end{tabular}
}
\caption{Same as Table \ref{tab:boundstt02}, now in the $pp \rightarrow t\bar{t}t\bar{t}$ process. Only individual bounds are presented in this case. }
\label{tab:boundstttt}
\end{center}
\end{table}
}

A marginalized analysis, allowing the five Wilson coefficients to be non-zero at the same time, is not possible in the four-top production. Given the uncorrelated measurements from ATLAS and CMS combined, we are fitting two data points $N_{\mathrm{data}}=2$ from the same observable (total cross-section) with five parameters. From this, the $\chi^2$-distribution can only yield meaningful bounds along one direction in the parameter space. The marginalized bounds with one degree of freedom, $N_{\mathrm{dof}}=1$, are
\begin{equation}
    c_1 \in [-6.98, 0.087] 
\end{equation}
with
\begin{equation}
c_1 =  0.74\, c_{tt}^1 + 0.44\, c_{QQ}^1 + 0.15\, c_{QQ}^8 - 0.44 \, c_{Qt}^1 + 0.19 \,  c_{Qt}^8. \label{eq:5.3}
\end{equation}
We can observe a correspondence in the dominant coefficients in Eq. \eqref{eq:5.3} and the quadratic contributions shown in the last column of Table \ref{tab:tttt} for the total cross-section of the four-top production.

\subsection{2D comparison between top-pair and four-top processes}

\begin{figure}
\setlength{\tabcolsep}{-4pt}
\renewcommand{\arraystretch}{0}
 \resizebox{\columnwidth}{!} {\includegraphics{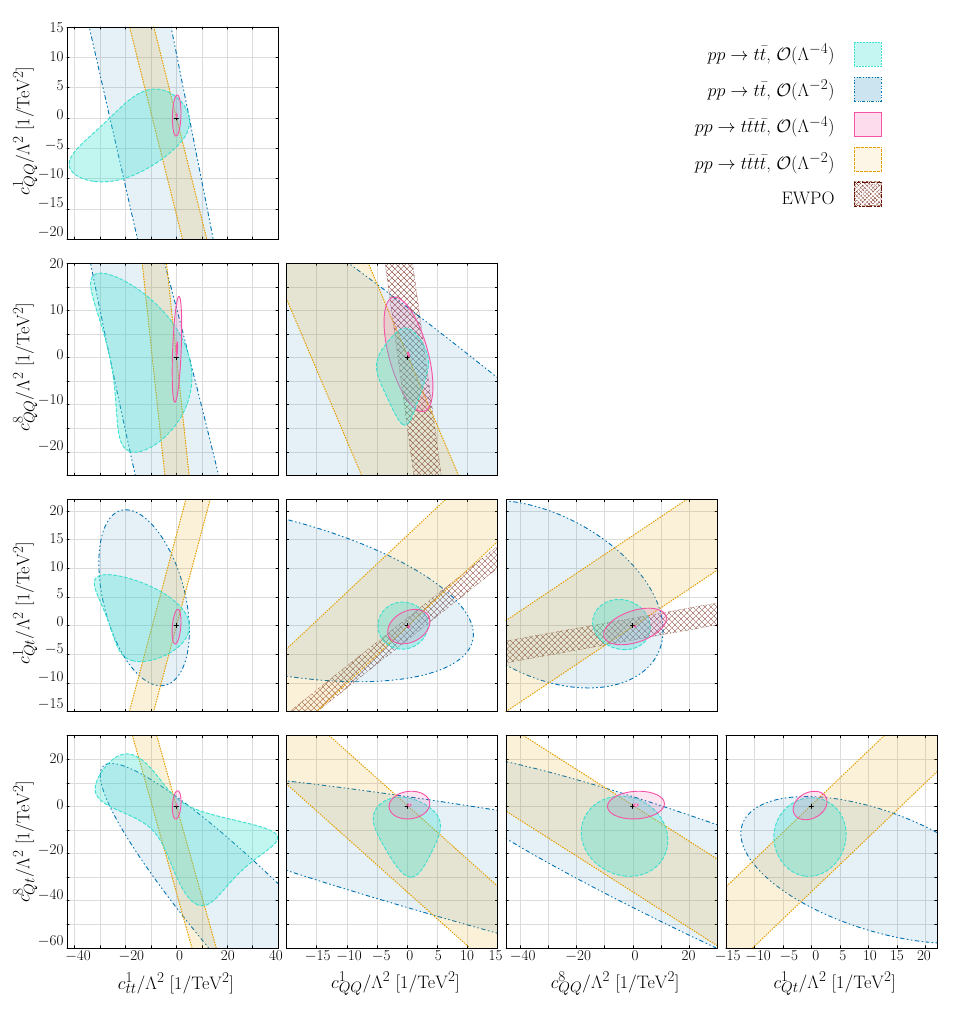}}

\caption{Exclusion regions in the $(c_i,c_j)$-plane obtained from measurements at the LHC with datasets listed in Table \ref{tab:datasets} for top-pair and four-top production. Bounds obtained from observables with a linear dependence on the Wilson coefficient are shown in orange and blue, while bounds obtained from quadratic dependencies are shown in magenta and cyan. Bounds coming from EWPO are displayed in a plaid pattern.  The black cross stands for SM values. Points that lie outside the ellipses are excluded at 95\% CL.}
\label{fig:bounds2d}
\end{figure}

In Fig. \ref{fig:bounds2d} the results of considering two effective operators at a time are shown (the remaining operators are set to zero). The exclusion regions are obtained at 95\% CL, so that points outside the coloured boundaries are excluded. At the interference level, the regions corresponding to the top-pair production are represented by ellipses.  For the four-top production, those regions are represented by planes bounded only along one axis as a consequence of only having two data points in the fit of the two corresponding Wilson coefficients.  The inclusion of the quadratic contributions drastically reduces the allowed region of the Wilson coefficients. 

In general, the results in Fig. \ref{fig:bounds2d} indicate that the top-pair production can render limits on the Wilson coefficients comparable to those extracted from four-top production. Specifically, the regions from considering the quadratic contributions in the plane $c_{QQ}^1$-$c_{Qt}^1$ are about the same size. In the interest of performing a global fit in the top sector, the results at the interference level suggest that the two processes are complementary in most of the cases, \textit{i.e.} each of these processes constrains the Wilson coefficients along different directions. This will be even more clear in the next subsection, with the projected bounds for the HL-LHC.

By inspection of the plots involving the $c_{tt}^1$ coefficient in Fig. \ref{fig:bounds2d}, we can infer that at the interference level the two processes are complementary, while at the quadratic level the best sensitivity is clearly provided by the four-top production. Notoriously from Fig. \ref{fig:bounds2d}, strong bounds along the $c_{tt}^1$ coefficient are found, which in the end is expected as the contributions of the corresponding operator are the largest at the linear and quadratic orders (see Table \ref{tab:tttt}). 

Finally, our results can be compared to the sensitivity of EWPO to four-heavy-quark operators~\cite{Dawson2022}. The operators $\mathcal{O}_{QQ}^{(1)}$, $\mathcal{O}_{QQ}^{(8)}$ and $\mathcal{O}_{Qt}^{(1)}$ enter through loop corrections in the observables 
\begin{equation}  \Gamma_Z,\,\sigma_{\mathrm{h}},\,R_l,\,R_b,\,R_c,\,A_b,A_{b,\mathrm{FB}},  \label{eq:EWPO}
\end{equation}
where $\sigma_h$ stands for the cross-section of the process $e^+e^- \rightarrow$ hadrons. The corresponding experimental measurements of these quantities are found in \cite{Dawson2020}. In Fig. \ref{fig:bounds2d} the 95\% CL exclusion regions are presented in the planes of the three operators aforementioned at the linear order in $c_i/\Lambda^2$. Only experimental uncertainties were considered to get these regions. Theoretical errors do not change substantially the bounds presented here, just widening slightly the region bands. The individual bounds from EWPO are
\begin{align}
   & c^{1}_{QQ} \in  [-1.61, 2.68], \\
   & c^{8}_{QQ} \in  [-15.23, 25.41], \\
   & c^{1}_{Qt} \in  [-2.24, 1.35],
\end{align}
which seem to be competitive when compared to those obtained  from the $t\bar{t}$ process up to interference contributions. The $c_{tt}^1$ and $c_{Qt}^8$ coefficients do not receive constraints from EWPO. In the former case, there are no one-loop possible insertions of the corresponding effective operator into any of the observables in Eq. \eqref{eq:EWPO}. For the latter, the only possible insertion is of the type that corrects the $b\bar{b}Z$-vertex with the top quark running in the loop, but such correction presents a color-octet structure that does not interfere with the color-singlet structure of the SM.

The production and decays of the Higgs boson also show sensitivity to the four-heavy-quark operators~\cite{Alasfar2022}. However, only the associated production of a Higgs boson with top quarks leads to competitive constraints for operators with mixed chiralities. In particular, the bounds on the Wilson coefficients $c^1_{Qt}$  and $c^8_{Qt}$ are found to be $[-1.1,1.2]$ and $[-4.6,4.9]$ at order $\mathcal{O}(\Lambda^{-2})$, respectively. The results in Ref. \cite{Alasfar2022} present better stability in the SMEFT expansion when the terms at order $\mathcal{O}(\Lambda^{-4})$ are included.

Finally, let us notice that, when considering the four-heavy-quark operators, the EWPO and Higgs processes impose constraints on the new-physics scale of the same order of magnitude as those constraints obtained from the four-top process. We also observe that, in general terms, if the four-top process is included in constraining those operators, there is no apparent reason for not considering the top-pair production.

\section{Sensitivity projection at HL-LHC}
\label{sec:HLLHC}

In this section, we focus on the projections of the top quark processes at the HL-LHC with $\sqrt{s}=14$ TeV. A higher sensitivity to new physics effects is expected due to the larger luminosity of the HL-LHC, but in the case of the four-top process, a higher sensitivity also arises from larger contributions of the four-heavy-quark operators when the energy is increased to $\sqrt{s}=14$ TeV. Remarkably, the increment of the quadratic contribution is around 50\% for the five operators. In the case of the interference with the QCD terms, the contributions at the energy of the HL-LHC are around 25\% larger, except for the $\mathcal{O}_{Qt}^{(1)}$ operator, due to phase-space cancellations, and $\mathcal{O}_{tt}^{(1)}$ for which the increment is of 35\%. Such increment of the interference is analogous to the pure SM case, for which we find that the HL-LHC results are 25\% larger. 

\begin{figure}
\centering
\begin{subfigure}{0.49\textwidth}
   \resizebox{\columnwidth}{!}{\includegraphics{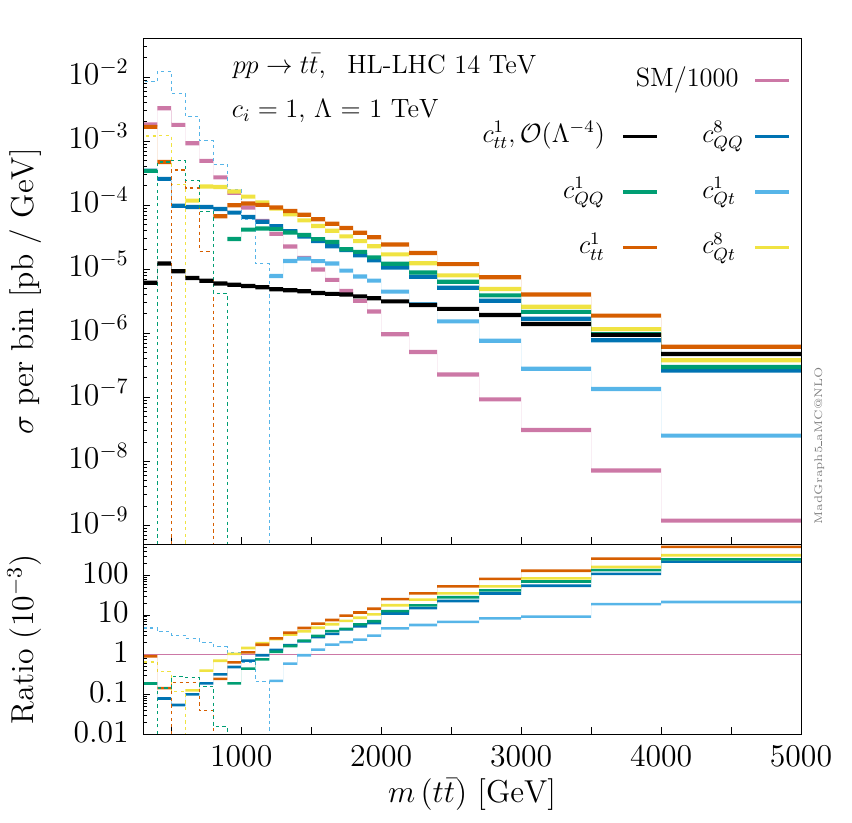}}
    \caption{ }
    \label{fig:HLinvMass}
\end{subfigure}
\,
\begin{subfigure}{0.49\textwidth}
    \resizebox{\columnwidth}{!}{\includegraphics{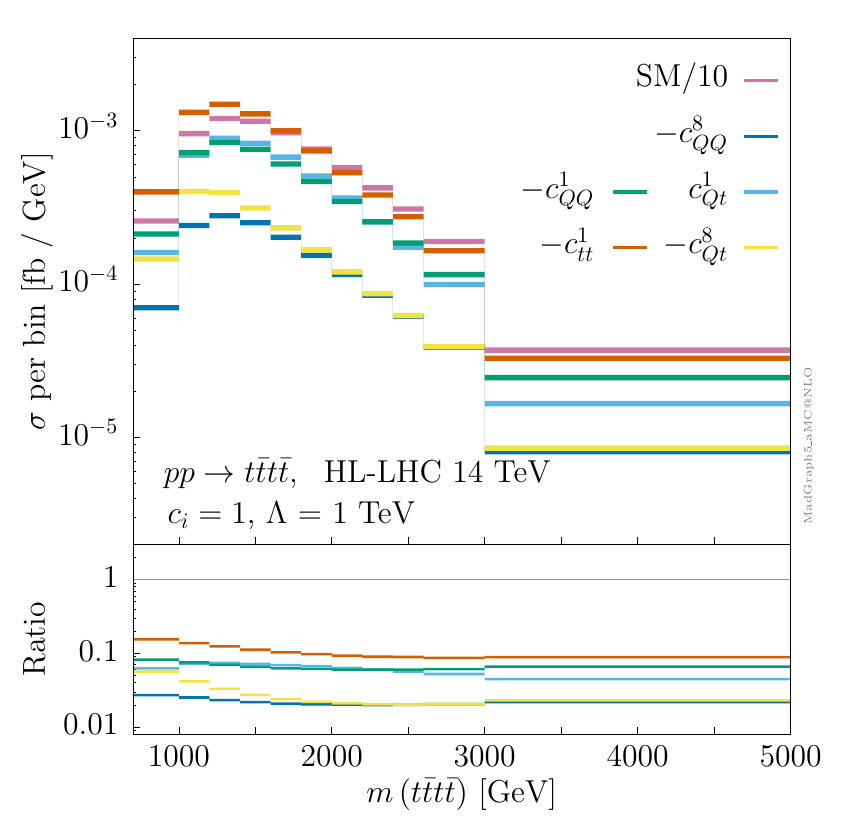}}
    \caption{ }
    \label{fig:HLinvMass4t}
\end{subfigure}
\caption[Caption for LOF]{Invariant-mass distribution of (a) the top-pair production at one-loop and (b) the four-top production at tree-level for the interference between four-heavy-quark operators and the SM at the HL-LHC. The Wilson coefficients are set to $c_i/\Lambda^2 = 1\,\mathrm{TeV}^{-2}$. The bins of the interference distributions with negative weights are represented by dashed lines. The insets at the bottom of the plots present the ratio between the new-physic effects and the SM. The black line in (a) shows the contribution from the square of the one-loop diagrams with a four-top vertex.}
\end{figure}

We study the constraining power of the top-pair and four-top processes at the HL-LHC, considering the measurement of the invariant-mass distribution of the final state in each production process. In Fig. \ref{fig:HLinvMass} we present the invariant-mass distribution of the top-pair process used in our projections, while in Fig. \ref{fig:HLinvMass4t} the four-top process case is presented. Comparing those two figures, a better behaviour is observed in the tail of the four-top production for the interference as its ratio with the SM tends to smaller values in the high-energy bins, while for the top-pair production the ratio approaches one. Finally, we also notice that in the top-pair production, the histograms in  Fig. \ref{fig:HLinvMass} arising from the effective operators are almost on top of each other in the high energy regime, except for the $c_{Qt}^1$ case. Hence, distinguishing the contributions of this effective operators at high energies might be difficult. The largest difference between the histograms is observed at energies below 1.5 TeV. Finally, the particular behaviour at high energies of the $\mathcal{O}_{Qt}^{(1)}$ operator is expected from the energy growth of the interference amplitudes presented in Table \ref{tab:high_energy}.

The projected sensitivities are obtained by assuming that the measured observables $\mathrm{O}_n$ (In this case, the invariant-mass distributions) coincide with the SM predictions. Valid for a counting observable, the uncertainties are constructed as
\begin{equation}
    \delta \mathrm{O}_n  = \sqrt{(\delta \mathrm{O}_n)^2_{\mathrm{stat}} + (\delta \mathrm{O}_n)^2_{\mathrm{syst}}} = \sqrt{\frac{ \sigma_n^{\mathrm{SM}}}{\mathcal{L}} + \alpha^{2} (\sigma_n^{\mathrm{SM}})^2 },
\end{equation}
so that the statistical uncertainty is taken to be $(\delta \mathrm{O}_n)_{\mathrm{stat}}=\sqrt{\sigma_n^{\mathrm{SM}}/\mathcal{L}}$, where $\mathcal{L}$ is the integrated luminosity and $\sigma_n^{\mathrm{SM}}$ is the cross-section in the $n$-bin of the invariant-mass distribution. The systematic uncertainty has been parametrized by $\delta (\mathrm{O}_n)_{\mathrm{syst}}= \alpha\, \sigma_n^{\mathrm{SM}} $, following the study performed in Ref. \cite{Durieux:2017rsg}, where $\alpha$ is a dimensionless coefficient that represents the magnitude of the systematic error in relation to the SM cross-section. For the top-pair process, we choose a value of $\alpha_{tt} = 0.05$ corresponding to a 5\% of systematic errors, while for the four-top process, we expect systematic errors to be   around 20\%,  $\alpha_{4t} = 0.2$, given the large scale uncertainties and the fact that computations at NNLO for the four-top final state seem to be out of reach in the near future.

{ 
\setlength{\tabcolsep}{3pt}
\renewcommand{\arraystretch}{1.62}
\begin{table}[t]
\begin{center}
{\tiny
\begin{tabular}{|c|c|c|c|c|c|c|c|c|}
\cline{3-9}
    \multicolumn{2}{c|}{} & \multicolumn{3}{c|}{ $pp\rightarrow t\bar{t}$} & \multicolumn{3}{c|}{ $pp\rightarrow t\bar{t} t\bar{t}$} & \multicolumn{1}{c|}{$t\bar{t}$ + $t\bar{t}t\bar{t}$} \tabularnewline
\hline
\hline
    \multirow{2}{*}{$c_i$} &\multirow{2}{*}{Cut} & \multicolumn{2}{c}{ Individual } & \multicolumn{1}{c|}{ Marginalized } & \multicolumn{2}{c}{ Individual } & \multicolumn{1}{c}{ Marginalized } & \multicolumn{1}{|c|}{ Marginalized }  \tabularnewline

    &  & \multicolumn{1}{c}{ $\mathcal{O}(\Lambda^{-2})$ } & \multicolumn{1}{c}{ $\mathcal{O}(\Lambda^{-4})$ }& \multicolumn{1}{c|}{ $\mathcal{O}(\Lambda^{-4})$} & \multicolumn{1}{c}{ $\mathcal{O}(\Lambda^{-2})$} & \multicolumn{1}{c}{ $\mathcal{O}(\Lambda^{-4})$ }& \multicolumn{1}{c}{ $\mathcal{O}(\Lambda^{-4})$} & \multicolumn{1}{|c|}{ $\mathcal{O}(\Lambda^{-4})$} \tabularnewline
\hline 
\hline 
    \multirow{2}{*}{$c_{tt}^1$} & $m_{\mathrm{Tot.}}<5$ TeV & \multicolumn{1}{c}{\multirow{1}{*}{ $ [-0.51, 0.51] $ }}& \multicolumn{1}{c}{\multirow{1}{*}{ $ [-0.51, 0.51]  $ }}   & \multicolumn{1}{c|}{ $[-11.3, 10.6]  $ } & \multicolumn{1}{c}{ $[ -2.37 , 2.37 ]  $ } & \multicolumn{1}{c}{ $[ -0.55, 0.66 ]  $ } & \multicolumn{1}{c}{ $[ - 0.26 , 0.33 ]  $ } & \multicolumn{1}{|c|}{ $[ -  0.71, 0.80 ]  $ }  \tabularnewline
   
    & $m_{\mathrm{Tot.}}<3$ TeV & \multicolumn{1}{c}{\multirow{1}{*}{ $ [-2.58, 2.58] $ }}& \multicolumn{1}{c}{\multirow{1}{*}{ $ [-2.58, 2.58] $ }}   & \multicolumn{1}{c|}{ $ [ -38.1 , 13.2 ] $ } & \multicolumn{1}{c}{ $[ -2.35, 2.35 ]  $ } & \multicolumn{1}{c}{ $[  -0.62 , 0.78 ]  $ } & \multicolumn{1}{c}{ $[ - 0.30, 0.40 ]  $ }  & \multicolumn{1}{|c|}{ $[ -0.82, 0.94 ]  $ }  \tabularnewline
\hline 
\hline 
    \multirow{2}{*}{$c_{QQ}^1$} & $m_{\mathrm{Tot.}}<5$ TeV  & \multicolumn{1}{c}{\multirow{1}{*}{ $ [-1.02, 1.02] $ }} & \multicolumn{1}{c}{\multirow{1}{*}{ $ [-1.11, 0.96] $ }}   & \multicolumn{1}{c|}{ $ [-5.82, 5.38] $ } & \multicolumn{1}{c}{ $[ -3.91, 3.91 ]  $ }  & \multicolumn{1}{c}{ $[ -1.07, 1.35 ]  $ } & \multicolumn{1}{c}{ $[ - 2.30 , 2.35 ]  $ }  & \multicolumn{1}{|c|}{ $[ - 2.50, 3.94 ]  $ } \tabularnewline
    
    & $m_{\mathrm{Tot.}}<3$ TeV & \multicolumn{1}{c}{\multirow{1}{*}{ $ [-5.0, 5.0] $ }} & \multicolumn{1}{c}{\multirow{1}{*}{ $ [-7.71, 3.07]  $ }}   & \multicolumn{1}{c|}{ $ [-10.3, 11.4] $ }   & \multicolumn{1}{c}{ $[ -3.95, 3.95 ]  $ } & \multicolumn{1}{c}{ $[ -1.21 , 1.61 ]  $ } & \multicolumn{1}{c}{ $[ - 2.37 , 2.44 ]  $ }  & \multicolumn{1}{|c|}{ $[ -3.17 , 5.08 ]  $ } \tabularnewline
\hline 
\hline 
    \multirow{2}{*}{$c_{QQ}^8$} & $m_{\mathrm{Tot.}}<5$ TeV  & \multicolumn{1}{c}{\multirow{1}{*}{ $ [-1.21, 1.21] $ }} & \multicolumn{1}{c}{\multirow{1}{*}{ $ [-1.24, 1.18] $ }}   & \multicolumn{1}{c|}{ $ [-13.1, 12.7] $ } & \multicolumn{1}{c}{ $[ -11.8 , 11.8 ]  $ } & \multicolumn{1}{c}{ $[ -3.22, 4.07 ]  $ } & \multicolumn{1}{c}{ $[ - 6.88 , 7.14 ]  $ }   & \multicolumn{1}{|c|}{ $[ - 9.87 , 5.47 ]  $ } \tabularnewline
    
    & $m_{\mathrm{Tot.}}<3$ TeV & \multicolumn{1}{c}{\multirow{1}{*}{ $ [-6.01, 6.01] $ }} & \multicolumn{1}{c}{\multirow{1}{*}{ $ [-21.1, 4.74] $ }}   & \multicolumn{1}{c|}{ $ [-26.3, 28.7] $ }  & \multicolumn{1}{c}{ $[ -11.9 , 11.9 ]  $ } & \multicolumn{1}{c}{ $[ -3.62 , 4.82 ]  $ } & \multicolumn{1}{c}{ $[ - 7.05 , 7.35 ]  $ }  & \multicolumn{1}{|c|}{ $[ -15.2 , 7.73 ]  $ } \tabularnewline
\hline 
\hline 
    \multirow{2}{*}{$c_{Qt}^1$} & $m_{\mathrm{Tot.}}<5$ TeV  & \multicolumn{1}{c}{\multirow{1}{*}{ $ [-9.03, 9.03] $ }}  & \multicolumn{1}{c}{\multirow{1}{*}{ $ [-4.24, 2.92] $ }}  & \multicolumn{1}{c|}{ $ [-6.45, 5.39] $ } & \multicolumn{1}{c}{ $[ -4.07 , 4.07]  $ } & \multicolumn{1}{c}{ $[ -1.12, 0.94 ]  $ } & \multicolumn{1}{c}{ $[ - 0.55, 0.44 ]  $ }   & \multicolumn{1}{|c|}{ $[ -1.36 , 1.21 ]  $ } \tabularnewline
    
    & $m_{\mathrm{Tot.}}<3$ TeV & \multicolumn{1}{c}{\multirow{1}{*}{ $ [-17.7, 17.7] $ }}  & \multicolumn{1}{c}{\multirow{1}{*}{ $ [-5.44, 4.31] $ }}   & \multicolumn{1}{c|}{ $ [-10.8, 10.2] $ } & \multicolumn{1}{c}{ $[ -4.0 , 4.0 ]  $ } & \multicolumn{1}{c}{ $[ -1.35 , 1.06 ]  $ } & \multicolumn{1}{c}{ $[ - 0.70 , 0.51 ]  $ }  & \multicolumn{1}{|c|}{ $[ -1.63 , 1.41 ]  $ } \tabularnewline
\hline 
\hline     
    \multirow{2}{*}{$c_{Qt}^8$} & $m_{\mathrm{Tot.}}<5$ TeV  & \multicolumn{1}{c}{\multirow{1}{*}{ $ [-0.82, 0.82] $ }}  & \multicolumn{1}{c}{\multirow{1}{*}{ $ [-0.82,-0.82] $ }}  & \multicolumn{1}{c|}{ $ [-16.4, 12.0] $ } & \multicolumn{1}{c}{ $[ -8.58, 8.58 ]  $ } & \multicolumn{1}{c}{ $[ -1.96, 2.29 ]  $ } & \multicolumn{1}{c}{ $[ - 0.91 , 1.12 ]  $ }  & \multicolumn{1}{|c|}{ $[ -2.50 , 2.56 ]  $ } \tabularnewline
    
    & $m_{\mathrm{Tot.}}<3$ TeV  & \multicolumn{1}{c}{\multirow{1}{*}{ $ [-3.86, 3.86]  $ }}  & \multicolumn{1}{c}{\multirow{1}{*}{ $ [-4.21, 3.61] $ }}  & \multicolumn{1}{c|}{ $ [-27.7, 20.8] $ } & \multicolumn{1}{c}{ $[ -8.47 , 8.47 ]  $ } & \multicolumn{1}{c}{ $[ -2.23 , 2.71 ]  $ } & \multicolumn{1}{c}{ $[ - 1.06, 1.32 ]  $ }  & \multicolumn{1}{|c|}{ $[ -2.91, 3.04 ]  $ } \tabularnewline
\hline
\end{tabular}
}
\caption{ The 95\% confidence level bounds (assuming $\Lambda = 1$ TeV) for the coefficients of the four-heavy-quark operators in the processes $pp \rightarrow t\bar{t}$ and $pp \rightarrow t\bar{t}t\bar{t}$ at the HL-LHC with $\sqrt{s}=14$ TeV. The intervals are presented for two different cuts in the invariant-mass distribution. }
\label{tab:boundsHL}
\end{center}
\end{table}
}

The luminosity for the HL-LHC is expected to be around $\mathcal{L}=3$ ab$^{-1}$. Thus, given the typical values of the cross-sections for both top production processes, the total uncertainties tend to be systematics dominated. Because of this, the binning of our projections for the invariant-mass distribution is chosen in such a way that the systematics are comparable to the statistical uncertainties. As shown in Table~\ref{tab:datasets}, we chose 24 bins for the invariant mass in the top-pair channel and 11 bins for the invariant mass in the four-top channel (details on the binning can be found in the appendix \ref{app:binning}). We present the results for two different cuts in the invariant-mass distribution at $m_{\mathrm{Tot.}}<3$ TeV and $m_{\mathrm{Tot.}}<5$ TeV. These two different cuts provide information about the sensitivity of the bounds to the tail of the distributions. 

The individual 95\% CL bounds on the Wilson coefficients of
the four-heavy-quark operators obtained from the top-pair and four-top production processes are given in Table \ref{tab:boundsHL}. The last column stands for the bounds obtained from combining the theoretical predictions from both processes. Marginalized limits are also tabulated for predictions including quadratic terms. Considering the individual bounds, we observe that both processes tend to be more sensitive to the $c_{tt}^1$ coefficient due to an enhanced sensitivity from the tails. In some particular entries, such as those for $c_{Qt}^1$, the four-top is more sensitive than the top-pair process, but in others, like those for $c_{QQ}^8$ and $c_{Qt}^8$, the situation is inverted. This suggests that both processes are sensitive to different directions in the parameter space and, consequently, are complementary. We also notice through the change in the bounds when terms at order $\mathcal{O}(\Lambda^{-4})$ are included that  the top-pair production appears to be slightly more stable in comparison to the four-top production regarding the convergence of the SMEFT expansion. As already mentioned, those fits only contain the $\mathcal{O}(\Lambda^{-4})$ term due to the bottom PDF suppressed contributions of orders $\alpha_s^0$ and  $\alpha_s^1$ for the NLO QCD correction. Fig.~\ref{fig:HLinvMass} also shows the $\alpha_s^2\Lambda^{-4}$ contribution for the operator $\mathcal{O}_{tt}^{(1)}$, \textit{i.e.} the one loop amplitude squared. Since for this operator there is no tree-level bottom-initiated contribution, that contribution is the full $\alpha_s^2\Lambda^{-4}$ contribution, and can be computed in \mg~as a loop-induced process. It shows that the high energy slope is higher than the interference and consistent with a $s^2$ enhancement compared to the SM. For $c_i/\Lambda^2\sim 1$ TeV, this contribution has a similar magnitude as the interference when the center of mass energy is about 4 TeV. Our constraints are in this range, and thus are just around the limit of the validity region. The other four-top operators $\alpha_s^2\Lambda^{-4}$ contributions can be computed in the same way if we assume the four flavour scheme. By doing this, we found that they are of the same order of magnitude with similar high-energy behaviour. Those contributions are also similar to the $(\alpha_s^0+\alpha_s^1)\Lambda^{-4}$ due to the bottom-induced processes, showing that they can be used as a reliable estimate of the $\Lambda^{-4}$ corrections. Finally, by comparing the results for both cuts, we can infer that high-energy effects from the tails of the distribution are important. Especially, the difference is drastic for the $c_{QQ}^8$ operator, raising questions about the validity of including those high-energy bins.  Moreover, the more stringent cut affects marginally the bounds at linear order of the four-top process because the shape of the differential invariant mass distributions is similar to the SM, unlike the case of top-pair production, where the largest deviations are in the high-energy bins. Thus, there is no large improvement in the sensitivity of the four-top process to new physics by going to the very high-energy region unless quadratic contributions are included. In fact, the four-top process at the interference level constrains basically the same direction, since the shapes of differential distributions for all the four-heavy-quark operators are very similar. The reduction of the interval in four-top while going from individual to marginalised bound comes from the fact that the larger number of parameters  does not really increase the number of degrees of freedom in this case due to the similar and overlapping shapes in Fig. \ref{fig:HLinvMass4t}.

{ 
\setlength{\tabcolsep}{2pt}
\renewcommand{\arraystretch}{1.48}
\begin{table}[t]
\begin{center}
{\scriptsize
\begin{tabular}{|c|c|c|c|c|}
\cline{1-5}
    \multicolumn{1}{|c|}{$c_i$} & \multicolumn{1}{c|}{Cut} & \multicolumn{1}{c}{ $pp\rightarrow t\bar{t}$} & \multicolumn{1}{c}{ $pp\rightarrow t\bar{t} t\bar{t}$} & \multicolumn{1}{c|}{$t\bar{t}$ + $t\bar{t}t\bar{t}$} \tabularnewline
\hline 
\hline 
    \multirow{2}{*}{$c_1$} & $m_{\mathrm{Tot.}}<5$ TeV  & \multicolumn{1}{c}{\multirow{1}{*}{ $ [- 0.35, 0.35] $ }}& \multicolumn{1}{c}{\multirow{1}{*}{ $ [- 1.46, 1.46] $ }}  & \multicolumn{1}{c|}{\multirow{1}{*}{ $ [- 0.42 , 0.42] $ }}   \tabularnewline

    & $m_{\mathrm{Tot.}}<3$ TeV  & \multicolumn{1}{c}{\multirow{1}{*}{ $ [- 1.71, 1.71] $ }}& \multicolumn{1}{c}{\multirow{1}{*}{ $ [- 1.42, 1.42] $ }}  & \multicolumn{1}{c|}{\multirow{1}{*}{ $ [-1.71 , 1.71] $ }}   \tabularnewline
\hline 
\hline 
    \multirow{2}{*}{$c_2$} & $m_{\mathrm{Tot.}}<5$ TeV  & \multicolumn{1}{c}{\multirow{1}{*}{ $ [- 17.6, 17.6] $ }} & \multicolumn{1}{c}{\multirow{1}{*}{ $ [- 18.6, 18.6] $ }}  & \multicolumn{1}{c|}{\multirow{1}{*}{ $ [-4.95, 4.95] $ }}  \tabularnewline

    & $m_{\mathrm{Tot.}}<3$ TeV  & \multicolumn{1}{c}{\multirow{1}{*}{ $ [- 29.8, 29.8] $ }}& \multicolumn{1}{c}{\multirow{1}{*}{ $ [- 17.5, 17.5] $ }}  & \multicolumn{1}{c|}{\multirow{1}{*}{ $ [-5.36 , 5.36] $ }}  \tabularnewline
\hline 
\hline     
    \multirow{2}{*}{$c_3$} & $m_{\mathrm{Tot.}}<5$ TeV  & \multicolumn{1}{c}{\multirow{1}{*}{ $ [- 39.6, 39.6] $ }} & \multicolumn{1}{c}{\multirow{1}{*}{ $ [- 37.5, 37.5] $ }}  & \multicolumn{1}{c|}{\multirow{1}{*}{ $ [-26.3, 26.3] $ }}  \tabularnewline

    & $m_{\mathrm{Tot.}}<3$ TeV  & \multicolumn{1}{c}{\multirow{1}{*}{ $ [- 85.5, 85.5] $ }}& \multicolumn{1}{c}{\multirow{1}{*}{ $ [- 55.5, 55.5] $ }}  & \multicolumn{1}{c|}{\multirow{1}{*}{ $ [-61.6 , 61.6] $ }}   \tabularnewline
\hline 
\hline 
    \multirow{2}{*}{$c_4$} & $m_{\mathrm{Tot.}}<5$ TeV  & \multicolumn{1}{c}{\multirow{1}{*}{ $ [- 62.1,  62.1] $ }}  & \multicolumn{1}{c}{\multirow{1}{*}{ $ [- 477, 477] $ }}  & \multicolumn{1}{c|}{\multirow{1}{*}{ $ [- 63.3 , 63.3 ] $ }}  \tabularnewline
    
    & $m_{\mathrm{Tot.}}<3$ TeV  & \multicolumn{1}{c}{\multirow{1}{*}{ $ [- 289, 289] $ }}& \multicolumn{1}{c}{\multirow{1}{*}{ $ [- 509, 509] $ }}  & \multicolumn{1}{c|}{\multirow{1}{*}{ $ [-68.9 , 68.9] $ }}   \tabularnewline
\hline 
\hline 
    \multirow{2}{*}{$c_5$} & $m_{\mathrm{Tot.}}<5$ TeV  & \multicolumn{1}{c}{\multirow{1}{*}{ $ [- 403, 403] $ }}  & \multicolumn{1}{c}{\multirow{1}{*}{ $ [- 1785, 1785] $ }}  & \multicolumn{1}{c|}{\multirow{1}{*}{ $ [- 74.9 , 74.9] $ }}  \tabularnewline
    
    & $m_{\mathrm{Tot.}}<3$ TeV  & \multicolumn{1}{c}{\multirow{1}{*}{ $ [- 727, 727] $ }}& \multicolumn{1}{c}{\multirow{1}{*}{ $ [ - 2213 , 2213 ] $ }}  & \multicolumn{1}{c|}{\multirow{1}{*}{ $ [-217, 217] $ }}   \tabularnewline
\hline
\end{tabular}
}
\caption{Marginalized 95\% CL bounds ($\Lambda = 1$ TeV) for the interference given by the coefficients of the four-heavy-quark operators in the diagonal basis of the processes  $pp \rightarrow t\bar{t}$ and  $pp \rightarrow t\bar{t}t\bar{t}$. The intervals are presented for the two different cuts in the invariant-mass distribution.  Notice that there is an abuse of notation, since the $c_i$ are not the same for both processes, they are defined through the matrices in Appendix \ref{app:rotation_mat}. For example, by comparing the $c_1$ of both processes, we are comparing the most constrained directions of the processes but not the same quantity. }
\label{tab:boundsDiagHL}
\end{center}
\end{table}
}
Marginalized bounds for the predictions truncated at the interference order are presented in Table \ref{tab:boundsDiagHL} for the diagonal directions. The rotation procedure is done as in the section \ref{sec:pheno}, with the rotation matrix for the cut $m_{\mathrm{Tot.}}<5$ TeV in the combined case given by
\begin{align}
\mathcal{R}^{\mathrm{Comb.}}_{\mathrm{HL}} &=  \left[\begin{array}{ccccc}
-0.029  & -0.75       & -0.46     & -0.37       & -0.31           \\
-0.86   & 0.21        & -0.35     & 0.24        & -0.20           \\
0.34    & -0.17       & -0.16     & 0.83        & -0.38           \\
-0.16    & -0.43        & -0.10     & 0.33       & 0.82           \\
0.34  & 0.43       & -0.80     & -0.10        & 0.23            \\
\end{array}\right].  \label{eq:HL_rotation}
\end{align}
The respective rotation matrices for the top-pair and four-top production cases are given in the Appendix \ref{app:rotation_mat}. It is observed that the matrix for the combined results presents several entries similar to the entries of the matrix for the top-pair production. From Eq. \eqref{eq:HL_rotation} we also notice that the most constrained direction is along the second row corresponding to $c_{tt}^1$, unlike the situation at $\sqrt{s}=13$ TeV  for which the most constrained direction is along $c_{Qt}^{1}$. Such change arises from the constant behaviour at high-energy of the square amplitude from the $\mathcal{O}_{Qt}^{(1)}$ operator (See Table \ref{tab:high_energy}).

The last column of Table \ref{tab:boundsDiagHL} presents the situation for which the complementary character of both processes is the strongest, this being more notorious in the directions of the $c_2$ and $c_5$ coefficients. We also observe the significance of the top-pair compared to the four-top production from the bounds along the directions $c_1$. Finally, the cut removing high-energy bins does not have a large impact on most of the limits presented. As we saw it also earlier, the intervals of those diagonal coefficients increase rapidly, and question the validity of the looser bounds when all the uncertainties are added. The PDF uncertainties for both quark-antiquark and gluon-gluon luminosity increase significantly around 3 TeV and should be added to our predictions unless they are strongly reduced using new data from other processes.

The effects of choosing a different magnitude of the systematic errors for the four-top process can be readily extracted from the bounds presented above. For a value of $\alpha_{4t} = 0.05$, \textit{i.e.} a value four times smaller than the used to obtain Table \ref{tab:boundsHL}, to parametrize the systematic errors of the four-top process leads to more stringent constraint bands by a factor of roughly four on each extreme at the order $\mathcal{O}(\Lambda^{-2})$. This is a consequence of the uncertainties being systematics dominated. At order $\mathcal{O}(\Lambda^{-4})$, the bounds also get more stringent by a factor of roughly four on each extreme. A scenario with such small uncertainties seems rather optimistic, considering the large scale uncertainties in the four-top process.

We also explore the consequences of the four-top process being measured only through some of its decaying channels. It is expected that the differential distributions for this process will be measured through the decay channels with two leptons with same-sign, and channels with three isolated leptons, corresponding in total to a branching ratio of
\begin{equation}
    \mathrm{BR}_{t\bar{t}t\bar{t}}(\mathrm{semi-leptonic}+\mathrm{multi-leptons})\approx0.12. \label{eq:BRtttt}
\end{equation}
By considering this branching ratio in the analysis leading to the individual bounds obtained from the differential distributions with 11 bins, we obtained wider limits of around 10\% of the ones displayed in Table \ref{tab:boundsHL}. Considering the size of the branching ratio, the change in the limits is not large. This is due to the fact that the uncertainties are systematic dominated, leading to the effect of the branching ratio canceling out between the numerator and the denominator of the $\chi^2$-distribution. Let us notice that the high-luminosity would allow the measurement of invariant-mass distributions with the same 11 bins in Fig. \ref{fig:HLinvMass4t} even when only the decay products in Eq. \eqref{eq:BRtttt} are considered. It is important to point out that the invariant-mass distribution might not be the best observable to look for deviations in the four-top process, as there are strong phase-space cancellations from the QCD interference with new physics effects. Although other differential distributions could be more suitable, we used the invariant-mass distribution as a first estimate.

\section{Conclusions}
\label{sec:conclusions}

Top-pair production at hadron colliders computed at NLO offers the possibility to probe dimension-6 effective operators involving only the third generation of quarks. These enter at tree-level with bottom-quark-initiated processes and at one-loop with contributions from four-top effective operators. The SMEFT expansion is considered including ${\cal O} (\Lambda^{-4})$ terms involving a product of bottom-quark-initiated and one-loop processes in a five-flavour scheme, since the product of two one-loop contributions is suppressed. We compare the sensitivity to these operators arising from top-pair production with the tree-level four-top production at the current LHC run and at the future HL-LHC upgrade.

The results are obtained using \mg. We also present compact analytical results computed for the first time for the top-pair production partonic cross-sections  at $\mathcal{O}(\Lambda^{-2})$ from the interference of four-heavy-quark operators and the SM in the quark-initiated production channels. Moreover, our results for the gluon-initiated production channels complete other existing results. The insertion in loop diagrams of four-fermion operators composed by chiral currents requires delicate computations, involving an adequate scheme definition to treat possible anomalous contributions and a clear definition of the evanescent operators. These subtleties were addressed in our computations, leading to the analytical formulas in Eq. \eqref{eq:quark_ctt1}-\eqref{eq:quark_cqt8} and Eq. \eqref{eq:gluonFormula1}-\eqref{eq:gluonFormula2}. These results were used to obtain, for the first time, a full validation of one-loop computations in the SMEFT performed by \mg.  

Using a $\chi^2$ analysis, we find that not only are the constraints competitive when compared to the sensitivity from other processes, such as four-top production, but they are complementary, yielding limits over different directions in the space of Wilson coefficients. This is promising in pursuit of a global statistical analysis. However, even though the top-pair production process seems to behave better in the SMEFT expansion, the loop suppression leads to loose constraints resulting in small values of the scale $\Lambda$, thus arising concerns about the validity of such bounds. The four-top process gets the sensitivity from the 1.5 TeV region, which also seems questionable in terms of SMEFT validity since the bounds cover $\Lambda \lsim  1$ TeV for $c_i = {\cal O}(1)$, thus also leading to unreliable bounds. Hence, we infer from our analysis that both processes should be used with care when obtaining bounds on the four-heavy-quark operators. In these cases, measurements with much larger precision that could push the bounds presented here by at least one order of magnitude are required. They could be obtained through more optimal observables, such as the high energy slope of the distribution and the inclusion of spin information, for example.

The sensitivity projections for the HL-LHC were also explored. We observed an enhancement in sensitivity in all the four-heavy-quark operators with respect to the sensitivity found from the current measurements at the LHC. Hence, we obtained that in the best scenario, the operators can probe scales $c_i/\Lambda^2\approx 0.5 \,\,\mathrm{TeV}^{-2} \approx 1/(1.5 \,\,\mathrm{TeV})^2$, an improvement with respect to current bounds. 

Our analysis implemented only invariant-mass differential distributions, but we expect that the definitions of new observables could enhance the sensitivity of both processes. In the four-top production, this is motivated by the strong cancellations in the contributions of order $\mathcal{O}(\alpha_s^3\Lambda^{-2})$. In the top-pair production, cancellations from different phase-space regions are also present for some of the operators. We have observed that differential distributions on the rapidity are less sensitive to such cancellations, but the shapes of the four-heavy-quark distributions are similar to those of the SM, which makes the discrimination of new physics effects challenging. Therefore, more optimized observables are required to improve the constraints. Observables  more sensitive to the chiral structure of the operators seem promising. In this vein, spin correlations offer an opportunity to be explored in the future. 

To conclude, we have shown that the top-pair production has the potential to probe new physic effects coming from four-heavy-quark operators, despite their LO being at one-loop. This work is intended as a starting point for a more comprehensive analysis of the sensitivity of the top-pair production to those operators. Global analyses that consider the four-top production to bound the four-heavy-quark operators will benefit from the inclusion of the one-loop SMEFT contributions in the top-pair production studied here.

\acknowledgments

The work of AV is supported by the DFG project 499573813 “EFTools”. The work of AV and CD is supported by the F.R.S.-FNRS through the IISN convention "Theory of Fundamental Interactions" (N : 4.4517.08)  and the MISU convention F.6001.19, and a CNPq Ph.D. fellowship. Computational resources have been provided by the supercomputing facilities of the Universit\'e Catholique de Louvain (CISM/UCL) and the Consortium des \'Equipements de Calcul Intensif en F\'ed\'eration Wallonie Bruxelles (C\'ECI) funded by the Fond de la Recherche Scientifique de Belgique (F.R.S.-FNRS) under convention 2.5020.11 and by the Walloon Region. The work of RR is supported in part by the São Paulo Research Foundation (FAPESP) through grant \# 2021/10290-2 and by the Brazilian Agency CNPq through a productivity grant \# 311627/2021-8.


\appendix

\section{Appendix: Binning at HL-LHC}
\label{app:binning}

The chosen binning of the invariant-mass distribution of the top-pair production for the HL-LHC is such that the bin size is 100 GeV in the range of energies $(300,2000)$ GeV, thus providing 17 bins in this range. Then, the intervals of the remaining 7 bins in the differential distribution are 
\begin{align}
   & \{(2000,2200),(2200,2400),(2400,2700),(2700,3000),\nn \\
    & \quad\quad \qquad (3000,3500),(3500,4000),(4000,5000)\},
\end{align}
with units in GeV. Analogously, for the four-top process we have the 11 bins
\begin{align}
    & \{  (700,1000), (1000,1200), (1200,1400), (1400,1600), (1600,1800), (1800,2000), \nn \\
    & \qquad \quad (2000,2200), (2200,2400), (2400,2600), (2600,3000), (3000,5000)     \}.
\end{align}

\section{Appendix: Rotation Matrices for linear fits. }
\label{app:rotation_mat}

The rotation matrices for each of the datasets in the results of the Table \ref{tab:boundsttDiag} are

{
\allowdisplaybreaks
\begin{align}
\mathcal{R}_{\mathrm{CMS}_{tt}-1} &=  \left[\begin{array}{ccccc}
0.82    & 0.37        & 0.35      & 0.22        & 0.15            \\
0.57    & -0.56       & -0.46     & -0.23       & -0.29           \\
-0.012  & -0.56       & 0.76      & -0.31       & 0.098           \\
0.014   & 0.46        & 0.15      & -0.63       & -0.60           \\
0.068   & 0.16        & -0.24     & -0.63       & 0.72            \\
\end{array}\right], \\
\mathcal{R}_{\mathrm{CMS}_{tt}-2} &=  \left[\begin{array}{ccccc}
-0.30   & -0.65       & -0.53     & -0.34       & -0.29           \\
0.95    & -0.22       & -0.15     & -0.052      & -0.14           \\
0.0032  & -0.52       & 0.78      & -0.32       & 0.11            \\
0.049   & 0.49        & -0.015    & -0.86       & -0.11           \\
0.049   & -0.12       & -0.28     & -0.18       & 0.93            \\
\end{array}\right], \\
\mathcal{R}_{\mathrm{CMS}_{tt}-3} &=  \left[\begin{array}{ccccc}
0.99    & -0.066      & 0.077     & 0.025       & -0.034          \\
0.016   & -0.54       & -0.73     & -0.24       & -0.34           \\
-0.083  & -0.69       & 0.61      & -0.37       & 0.035           \\
-0.0090 & -0.44       & -0.17     & 0.59        & 0.65            \\
-0.069  & -0.19       & 0.23      & 0.67        & -0.68           \\
\end{array}\right], \\
\mathcal{R}_{\mathrm{CMS}_{tt}-4} &=  \left[\begin{array}{ccccc}
-0.20   & -0.69       & -0.54     & -0.35       & -0.24           \\
-0.98   & 0.18        & 0.071     & 0.030       & 0.083           \\
0.050   & 0.22        & -0.71     & 0.18        & 0.64            \\
-0.0034 & -0.48       & 0.44      & -0.23       & 0.72            \\
-0.057  & -0.45       & 0.043     & 0.89        & -0.045          \\
\end{array}\right],\\
\mathcal{R}_{\mathrm{ATLAS}_{tt}} &=  \left[\begin{array}{ccccc}
-0.99   & 0.022       & -0.11     & -0.045      & 0.013           \\
0.058   & -0.68       & -0.56     & -0.32       & -0.33           \\
-0.081  & -0.54       & 0.77      & -0.32       & 0.10            \\
-0.055  & -0.46       & 0.042     & 0.88        & 0.010           \\
0.043   & -0.18       & -0.28     & -0.086      & 0.94            \\
\end{array}\right].
\end{align}
The rotation matrix corresponding to the combined bounds in Table \ref{tab:boundsttDiag} is
\begin{align}
\mathcal{R}^{t\bar{t}}_{\mathrm{Combined}} &=  \left[\begin{array}{ccccc}
-0.99   & -0.012      & -0.12     & -0.062      & 0.0024          \\
0.097   & -0.70       & -0.55     & -0.33       & -0.31           \\
-0.072  & -0.51       & 0.79      & -0.32       & 0.073           \\
0.039   & -0.19       & -0.24     & -0.068      & 0.95            \\
-0.057  & -0.46       & 0.053     & 0.88        & -0.016          \\
\end{array}\right].
\end{align}
The rotation matrices corresponding to the results in Table \ref{tab:boundsDiagHL} for the top-pair and four-top production processes at the HL-LHC are
\begin{align}
\mathcal{R}^{t\bar{t}}_{\mathrm{HL}}&=  \left[\begin{array}{ccccc}
-0.038  & -0.74       & -0.46     & -0.37       & -0.31           \\
0.86    & -0.24       & 0.16      & 0.37        & -0.19           \\
-0.36   & -0.018      & -0.36     & 0.80        & -0.32           \\
-0.36   & -0.52       & 0.76      & 0.16        & -0.015          \\
0.032   & -0.34       & -0.25     & 0.25        & 0.87            \\
\end{array}\right], \\
\mathcal{R}^{4t}_{\mathrm{HL}} &=  \left[\begin{array}{ccccc}
0.43    & -0.74       & -0.20     & -0.45       & -0.15           \\
0.70    & 0.37      & 0.58     & -0.17        & -0.060            \\
0.47   & -0.22       & -0.16     & 0.80        & 0.26            \\
0.32   & 0.51       & -0.78      & -0.19       & -0.024            \\
-0.0056   & -0.019    & 0.028     & -0.30       & 0.95           \\
\end{array}\right]. 
\end{align}

\section{Appendix: \texorpdfstring{$\chi^2$}{Lg}-distributions.}

The $\chi^2$-distribution for the interference at HL-LHC for the top-pair production used to obtain the bounds for $m_{\mathrm{Tot.}}<5$ TeV in Table \ref{tab:boundsHL} is

\begin{align}
   \chi^2_{tt}(c_i) & = 135.35 (c_{tt}^{1})^2 + 0.43167 (c_{Qt}^{1})^2 + 51.740 (c_{Qt}^{8})^2 + 32.815 (c_{QQ}^{1})^2+ 24.103 (c_{QQ}^{8})^2  \nonumber \\
   & \qquad + 56.195 c_{QQ}^{1} c_{QQ}^{8} + 6.8853c_{QQ}^{1} c_{Qt}^{1} + 82.372 c_{QQ}^{1} c_{Qt}^{8} + 133.21 c_{QQ}^{1} c_{tt}^{1} \nonumber \\
   & \qquad + 5.8323 c_{QQ}^{8} c_{Qt}^{1} + 70.612 c_{QQ}^{8} c_{Qt}^{8}   + 114.22 c_{QQ}^{8} c_{tt}^{1}  + 8.6116 c_{Qt}^{1} c_{Qt}^{8} \nonumber \\
   & \qquad + 13.848 c_{Qt}^{1} c_{tt}^{1}  +  167.33 c_{Qt}^{8} c_{tt}^{1},
\end{align}
where we set $\Lambda = 1 $ TeV. The counterpart for the four-top production is
\begin{align}
   \chi^2_{4t}(c_i) & = 46.957(c_{tt}^1)^2 + 16.085(c_{Qt}^1)^2 + 3.5828(c_{Qt}^8)^2 + 17.209(c_{QQ}^1)^2 + 1.9011(c_{QQ}^8)^2 \nonumber \\
   & \qquad + 11.439c_{QQ}^1c_{QQ}^8  - 32.919c_{QQ}^1c_{Qt}^1 - 10.946c_{QQ}^8c_{Qt}^1  + 15.144c_{QQ}^1c_{Qt}^8 \nonumber \\
   & \qquad + 5.033c_{QQ}^8c_{Qt}^8 - 14.283c_{Qt}^1c_{Qt}^8  + 56.537c_{QQ}^1c_{tt}^1 + 18.793c_{QQ}^8c_{tt}^1 \nonumber \\
   & \qquad - 54.195c_{Qt}^1c_{tt}^1 + 25.511c_{Qt}^8c_{tt}^1.
\end{align}

\bibliographystyle{unsrt}
\bibliography{references.bib}

\end{document}